\documentclass[aps,preprintnumbers,prd,notitlepage,showpacs,superscriptaddress,nofootinbib]{revtex4-2}
\pdfoutput=1
\usepackage[usenames,dvipsnames]{color}
\usepackage{graphicx,slashed,xcolor,multirow,amsmath,amssymb}
\usepackage{tikz}
\usepackage{tikz-feynman}
\usepackage{caption}
\usepackage{subcaption}
\usepackage{adjustbox}
\usepackage[colorlinks=true,linktoc=page,linkcolor=purple,citecolor=purple,urlcolor=magenta]{hyperref}
\usepackage{graphicx}
\usepackage{booktabs}
\usepackage{xparse}
\usepackage{xcolor}
\usepackage{hyperref}

\begin{document}
\title{Exploring Scalar Leptoquarks at Muon Collider via Indirect Signatures and Right-Handed Neutrino-Assisted Decays}

\author{Subham Saha}\email{subham.saha@iopb.res.in}
\affiliation{Institute of Physics, Sachivalaya Marg, Bhubaneswar, Odisha 751005, India}
\affiliation{Homi Bhabha National Institute, BARC Training School Complex, Anushakti Nagar, Mumbai
400094, India}

\author{Arvind Bhaskar}\email{arvind.bhaskar@iopb.res.in}
\affiliation{Institute of Physics, Sachivalaya Marg, Bhubaneswar, Odisha 751005, India}
\affiliation{Homi Bhabha National Institute, BARC Training School Complex, Anushakti Nagar, Mumbai
400094, India}

\author{P. S. Bhupal Dev}\email{bdev@wustl.edu}
\affiliation{Department of Physics and McDonnell Center for the Space Sciences, Washington University, St.~Louis, Missouri 63130, USA}

\author{Manimala Mitra}\email{manimala@iopb.res.in}
\affiliation{Institute of Physics, Sachivalaya Marg, Bhubaneswar, Odisha 751005, India}
\affiliation{Homi Bhabha National Institute, BARC Training School Complex, Anushakti Nagar, Mumbai
400094, India}

\begin{abstract}
\noindent
Scalar leptoquarks (sLQs) appear in a wide range of ultraviolet‑motivated extensions of the Standard Model and provide a natural link between the quark and lepton sectors. In this work, we investigate the discovery potential of an sLQ doublet $\widetilde{R}_{2}({\bf 3},{\bf 2},1/6)$ that couples to light quarks and right‑handed neutrinos (RHNs) at the proposed muon collider. We analyze both indirect probes—arising from $t$‑channel sLQ exchange that affects the high‑$p_T$ behavior of dijet spectra—and direct searches exploiting pair and single production of the sLQs, incorporating the full interplay of kinematic thresholds and decay topologies.  We find that indirect probes at muon colliders deliver remarkably robust sensitivity to the sLQ–quark-muon coupling over a broad mass range. Assuming a sub-$\mathcal{O}(1)$ Yukawa coupling, we achieve a  $5\sigma$ sensitivity up to sLQ masses $ \sim$ $4.0$ TeV ($7.0$ TeV) at $\sqrt{s}=5 \, (10)$ TeV  center-of-mass energy with $\mathcal{L} = 3~\text{ab}^{-1}$ ($10~\text{ab}^{-1}$) integrated luminosity. Direct production channels provide complementary reach: pair production dominates below threshold, while single production, driven by the sLQ–quark–muon/RHN interaction, decisively extends the mass reach well into the multi‑TeV regime. We demonstrate that  with  $\mathcal{O}(1)$ Yukawa couplings, the single production channel can probe sLQ masses up to $3.0$ TeV ($6.0$ TeV) for $\sqrt s=5$ TeV ($10$ TeV). Together, these channels enable a unified exploration of parameter space far beyond the projected capabilities of the HL‑LHC, including regions where conventional charged‑lepton signatures are subdominant.
\end{abstract}
\preprint{IOP/BBSR/2025-04}
\maketitle

\section{Introduction}
\label{sec:1}

Leptoquarks (LQs) are an interesting class of beyond-the-Standard-Model (BSM) particles that can turn quarks into leptons and vice versa~\cite{Dorsner:2016wpm}. They emerge naturally in extensions of the Standard Model (SM) that unify matter, such as the Pati-Salam~\cite{Pati:1974yy}, $SU(5)$~\cite{Georgi:1974sy} and $SO(10)$~\cite{Fritzsch:1974nn} grand unified theories, as well as in supersymmetric theories with $R$-parity violation~\cite{Barbier:2004ez}, technicolor models~\cite{Weinberg:1975gm,Susskind:1978ms}, and radiative neutrino mass models~\cite{Cai:2017jrq}. In recent years, LQs have gained considerable attention in the literature as promising candidates to address several experimental anomalies~\cite{Fischer:2021sqw, Crivellin:2023zui}. They can also play a role in dark matter phenomenology~\cite{Choi:2018stw} and electroweak vacuum stability~\cite{Bandyopadhyay:2016oif}. The LHC has an active search program targeting LQs, with both ATLAS and CMS collaborations investigating pair and single production of LQs in a variety of final states involving leptons (charged or neutrinos) and jets (light-flavor, top, or bottom). The current LHC searches~\cite{ATLAS:2020dsk} exclude scalar leptoquarks (sLQs) up to $1.73$~TeV decaying to $\mu j$ with $100\%$ branching ratio (BR) and vector leptoquarks (vLQs)~\cite{ATLAS:2022wcu} up to $1.98$ TeV decaying with $100\%$ BR to third-generation quarks and second-generation leptons. Indirect constraints on LQ couplings to SM quarks and leptons also arise from high-$p_T$ dilepton and monolepton plus missing energy searches~\cite{Bessaa:2014jya,Mandal:2018kau,ATLAS:2020yat, Babu:2020hun, Bhaskar:2021pml, Angelescu:2021lln,CMS:2023qdw}. Depending on their charges, some LQs can also couple to a SM quark and a right-handed neutrino (RHN) --  a channel that remains largely unconstrained by existing searches. If the RHN is lighter than the LQ, decays such as  $\text{LQ} \rightarrow \text{RHN} + \text{\rm jet}$ become kinematically allowed. In scenarios where this decay mode dominates, the conventional direct LHC bounds get weaker. As a result, a significant portion of the LQ parameter space involving RHNs remains unexplored and unconstrained. For phenomenological studies on enhanced RHN sensitivity via LQ production at the LHC and LHeC, see Refs.~\cite{Das:2017kkm,Mandal:2018qpg,Bhaskar:2020kdr,Cottin:2021tfo}.

A future muon collider~\cite{AlAli:2021let,InternationalMuonCollider:2025sys} offers a unique opportunity to explore BSM physics beyond the LHC, thanks to its clean experimental environment and the possibility of high partonic center-of-mass (C.O.M.) energy. Unlike hadron colliders, collisions of elementary muons allow the full C.O.M. energy to be directly available for new particle production, enabling precise studies with minimal background. Additionally, the larger muon mass significantly reduces synchrotron radiation, allowing higher luminosities and energies to be achieved. These advantages make the muon collider an ideal setting to investigate heavy BSM states such as LQs and RHNs. Several recent studies have demonstrated the potential of a muon collider to discover LQs~\cite{Bandyopadhyay:2021pld,Ghosh:2023xbj,Desai:2023jxh, Varzielas:2023qlb, Han:2025wdy}. In this work, we examine the prospects of probing TeV-scale sLQs at a future high-energy muon collider, focusing on their decays into RHNs in direct search strategy, as well as indirectly through the dijet channel mediated by a $t$-channel sLQ exchange. We specifically consider the weak doublet sLQ $\widetilde{R}_2({\bf 3},{\bf 2},1/6)$ in this work, because it can simultaneously couple to the SM quarks and leptons, as well as to SM quarks and RHNs. 

The RHNs are well-motivated candidates for explaining the observed nonzero but tiny neutrino masses via the tree-level seesaw mechanism~\cite{Minkowski:1977sc, Mohapatra:1979ia, 
Yanagida:1979as,Gell-Mann:1979vob}. Moreover, sLQs provide an alternative to generate neutrino masses via radiative seesaw~\cite{Cai:2017jrq, Babu:2019mfe}. Instead of committing to a particular model for neutrino mass generation, we consider a minimal extension of the SM with a single RHN \( N \) with mass \( M_N > \mathcal{O}(10)~\text{GeV} \), which couples to the sLQ $\widetilde{R}_2$ and mixes with the muon-flavor neutrino \( \nu_\mu \) via an active-sterile mixing angle \( V_{\mu N} \). Our analysis focuses on the parameter space where the RHN is lighter than the sLQs, allowing the sLQ to decay into an RHN and a jet. We consider the RHN to decay promptly into SM particles, such as final-states consisting of a muon and two jets, mediated by the SM \(W\) boson and off-shell sLQ. We focus on the muonic decay channel due to its superior detection efficiency compared to electrons. The sLQ-assisted RHN production mode considered here offers a complementary probe of RHNs at muon colliders, as the conventional RHN production modes~\cite{Mekala:2023diu, Li:2023tbx, Kwok:2023dck} are suppressed by the active-sterile mixing in the minimal seesaw. 
For direct search, we particularly focus on the dimuon and multijet final-state which serves as a powerful and versatile signal topology for probing sLQs and RHNs. The presence of two energetic muons ensures excellent trigger and reconstruction efficiency, while the large jet multiplicity characteristic of $\widetilde{R}_2$ decays provides strong kinematic handles for suppressing the SM backgrounds. Since our analysis does not impose a requirement on the relative charges of the muons, the resulting sensitivity estimates are conservative; imposing a same-sign dimuon requirement would further suppress backgrounds and enhance the sensitivity reach. The indirect search channel, which requires two highly energetic jets in the final-state, provides a complementary sensitivity reach in regions where the direct search is kinematically suppressed.

The rest of the paper is organized as follows. In Section~\ref{sec:2}, we describe the sLQ model and couplings to SM fermions and RHNs. In Section~\ref{sec:3}, we present the analytical expressions for the sLQ decay widths and the BR plots. In Section~\ref{sec:4}, we summarize the current limits on sLQs from various experimental searches. In Section~\ref{sec:5}, we discuss the various sLQ production and decay modes  at a future muon collider. In Section~\ref{sec:6}, we describe our cut-based analysis strategy for studying the sLQ signals. We present our results in Section~\ref{sec:7} and our conclusions in Section~\ref{sec:8}. Appendices~\ref{Appendix_LQ} and \ref{Appendix_n}  provide the analytical expressions for the partial decay widths of the $\widetilde{R}_2$ LQ and the RHN, respectively. 

\section{\label{sec:2} Leptoquark Couplings}

We augment the SM with an RHN $N_R({\bf 1},{\bf 1},0)$ and an sLQ $\widetilde{R}_2({\bf 3},{\bf 2},1/6)$ where the charges shown are under the SM gauge group $SU(3)_c\times SU(2)_L\times U(1)_Y$. The isospin components of the doublet are denoted by $\widetilde{R}_2 = (\widetilde{R}_2^{2/3}, \widetilde{R}_2^{-1/3})$. Following the notation of Refs.~\cite{Dorsner:2016wpm,Buchmuller:1986zs,Padhan:2019dcp}, the renormalizable sLQ Lagrangian can be expressed as 
\begin{align}
-\mathcal{L}_{\rm sLQ} \supset  Y_{ij}\bar{d}_{R}^{i}\widetilde{R}_{2}^{a}\epsilon^{ab}L_{L}^{j,b}+
Z_{ij}\bar{Q}_{L}^{i,a}\widetilde{R}_{2}^{a}N_{R}^{j}+\text{H.c.} \, ,
\label{eq:eq1}
\end{align}
where $i, j = 1, 2, 3$ are the flavor indices, $a, b = 1, 2$ are $SU(2)_L$ indices and $\epsilon^{ab}$ is the $SU(2)$ antisymmetric tensor. Expanding Eq.~\eqref{eq:eq1} in terms of the individual components, we obtain the following Lagrangian:

\begin{align}
&-\mathcal{L}_{\rm sLQ} \supset Y_{ij}\bar{d}_{R}^{i}e_{L}^{j}\widetilde{R}_{2}^{2/3}+(YU_{\text{PMNS}})_{ij}\bar{d}_{R}^{i}\nu_{L}^{j}
\widetilde{R}_{2}^{-1/3} +  (V_{\text{CKM}}Z)_{ij}\bar{u}_{L}^{i}N_R^{j}\widetilde{R}_{2}^{2/3}+Z_{ij}\bar{d}_{L}^{i}
N_{R}^{j}\widetilde{R}_{2}^{-1/3}+\text{H.c.}
\label{eq:eq2}
\end{align}
Here, $U_{\text{PMNS}}$ and $V_{\text{CKM}}$ are the Pontecorvo-Maki-Nakagawa-Sakata (PMNS) and Cabibbo-Kobayashi-Maskawa (CKM) mixing matrices, respectively. The off-diagonal elements of these matrices do not play any distinctive role in the collider analysis that we pursue. 
 
$Y_{ij}$ and $Z_{ij}$ in Eq.~\eqref{eq:eq1} denote the Yukawa couplings, which we assume to be real in the subsequent discussion. In the above equation, $N_R$ denotes the right-chiral component of the RHN field and in the rest of the paper we will use the notation $N$ to denote it in the mass basis. Throughout this paper, we consider the RHN to be of Majorana nature, although taking it to be (pseudo-)Dirac will not change our results significantly. For simplicity, we consider only the first-generation of $N$ while investigating the discovery prospects of $\widetilde{R}_2$ at the proposed muon collider, considering different mass choices for the RHN, by analyzing final-states with muons and light jets. Our choice of Yukawa couplings are thus $Y_{12}$ and $Z_{11}$, where $Y_{12}$ describes the interaction of $\widetilde{R}_2$ to a first-generation quark and a second-generation lepton, and $Z_{11}$ denotes the interaction of $\widetilde{R}_2$ with a first-generation quark and $N$. In addition to the above Yukawa interactions, $N$ also has the following interactions with the SM particles:
\begin{align}
   &-\mathcal{L}_{\mu W N} = \frac{g}{\sqrt{2}} W^-_{\mu} \bar{\mu} \gamma^{\mu} P_L V_{\mu N} N + {\rm H.c.},
\label{eq:WlN} \\
   &-\mathcal{L}_{\nu_{\mu} Z N} = \frac{g}{2\cos\theta_w} Z_{\mu} \bar{\nu}_{\mu} \gamma^{\mu} P_L V_{\mu N} N + {\rm H.c.},
\label{eq:ZnuN} \\
   &-\mathcal{L}_{\nu_{\mu} H N} = \frac{M_N}{v} H \bar{\nu}_{\mu} P_R V_{\mu N} N + {\rm H.c.}
\label{eq:HnuN}
\end{align}
The interactions of the RHN with the SM gauge and Higgs bosons are governed by the active-sterile mixing $V_{\mu N}$ as shown above. Mixing with lepton flavors other than the muon is not directly relevant for our muon collider study;  hence for simplicity, we ignore the mixing elements $V_{\ell N}$ with $\ell=e,\tau$. In the canonical type-I seesaw mechanism, the magnitude of this mixing is related to the light neutrino mass $m_{\nu}$ and the mass of the RHN $M_N$ via the relation $V_{\mu N} \sim \sqrt{\frac{m_{\nu}}{M_N}}$. For a light neutrino mass $m_{\nu}\sim 0.1$ eV and RHN mass $M_N \sim \mathcal{O}(100)$ GeV, the active-sterile mixing is highly  suppressed: $V_{\mu N} \sim \mathcal{O}(10^{-6})$. Due to this strong suppression, the production cross-sections for processes such as $\mu^+ \mu^- \to \nu N$ or $\mu^+ \mu^- \to N N$ are exceedingly small at a muon collider operating with TeV-scale C.O.M. energy, making their experimental observation challenging. As we show in this work, an alternative and more promising production mechanism arises from the decay of $\widetilde{R}_2$. This underscores the crucial role that future collider searches for sLQs can play in probing the existence of RHNs. In the following section, we present a detailed discussion of the viable decay channels of both  $\widetilde{R}_2$ and $N$. 

\section{\label{sec:3} Decay modes}

As discussed in the previous section, in this analysis we consider a simplified scenario where the only nonzero Yukawa couplings associated with $\widetilde{R}_2$ are $Y_{12}$ and $Z_{11}$. The presence of a non-vanishing $Y_{12}$ coupling facilitates the two-body decays of the $\widetilde{R}_2$ components into SM fermions. The $2/3$ and $-1/3$-components of $\widetilde{R}_2$ decay via $\widetilde{R}_2^{2/3}\to \mu d$ and $\widetilde{R}_2^{-1/3}\to {\nu}_{\mu} d$ mode, respectively.

Similarly, a nonzero $Z_{11}$ coupling opens up decay channels involving the RHN. The corresponding decay modes are $\widetilde{R}_2^{2/3}\to N u$ and $\widetilde{R}_2^{-1/3}\to N d$.\footnote{Throughout this paper, decay modes of $\widetilde{R}_2$ are written without explicitly specifying the electric charges of the final-state fermions. All expressions are understood to implicitly include the appropriate charge assignments and, where applicable, the corresponding charge-conjugate processes. This notational convention is adopted purely for brevity and does not imply any restriction on the physical decay channels considered.} The Feynman diagrams corresponding to these decay processes are illustrated in Figs.~\ref{fig:R2tildeupdecay} and \ref{fig:R2tildedowndecay}. For simplicity, we assume mass degeneracy between $\widetilde{R}_2^{2/3}$ and $\widetilde{R}_2^{-1/3}$, which allows us to neglect the decay mode $\widetilde{R}_2^{2/3}\rightarrow \widetilde{R}_2^{-1/3} {W^*}$. Furthermore, our analysis is conducted under the assumption that $M_{\widetilde{R}_2} > M_N$, ensuring that the sLQ decays exclusively through two-body channels into a quark and a SM lepton/RHN. The analytical expressions for the partial decay widths of the $\widetilde{R}_2$ components are given in Appendix~\ref{Appendix_LQ}. 
    
\begin{figure}[htb!]
    \centering
    \begin{subfigure}{0.20\textwidth}
        \centering
        \begin{tikzpicture}
            \begin{feynman}
				\vertex (a);
				\vertex[right=of a](b);
				\vertex[above right=of b](c);
				\vertex[below right=of b] (f);
				
				\diagram*  {
					(a) -- [edge label=$\widetilde{R}^{2/3}_2 / \widetilde{R}^{-1/3}_2$, scalar] (b),
					(c) -- [edge label=$N$,  draw] (b),
					(b) -- [edge label=$u/d$, fermion] (f),
				};
			\end{feynman}
            \label{fig:R2tildeupdecay}
        \end{tikzpicture}
        \caption{}
        \label{fig:R2tildeupdecay}
    \end{subfigure}
    \hfill
    \begin{subfigure}{0.20\textwidth}
        \centering
        \begin{tikzpicture}
           \begin{feynman}
				\vertex (a);
				\vertex[right=of a](b);
				\vertex[above right=of b](c);
				\vertex[below right=of b] (f);
				
				\diagram*  {
					(a) -- [edge label=$\widetilde{R}^{2/3}_2 / \widetilde{R}^{-1/3}_2$, scalar] (b),
					(b) -- [edge label=$d$, fermion] (c),
					(f) -- [edge label=$\mu^{+} / \bar{\nu_{\mu}}$, fermion] (b),
				};
			\end{feynman}
            \label{fig:R2tildedowndecay}
        \end{tikzpicture}
        \caption{}
        \label{fig:R2tildedowndecay}
    \end{subfigure}
    \hfill
    \begin{subfigure}{0.20\textwidth}
        \centering
        \begin{tikzpicture}
            \begin{feynman}
				\vertex (a);
				\vertex[right=of a](b);
				\vertex[above right=of b](c);
				\vertex[below right=of b] (f);
				\vertex[above right=of f] (g);
				\vertex[below right=of f] (h);
				
				\diagram*  {
					(a) -- [edge label=$N$, draw] (b),
					(b) -- [edge label=${u} / d$, fermion] (c),
					(b) -- [scalar, edge label'={\(\widetilde{R}^{-2/3}_2/ \widetilde{R}^{-1/3}_2\)}, sloped] (f),
					(g) -- [edge label'=$\bar{d}$, fermion] (f),
					(f) -- [edge label=$\mu^{-}/ \nu_{\mu}$, fermion] (h),
				};
			\end{feynman}
        \end{tikzpicture}
        \caption{}
        \label{fig:NR2tildedecay}
    \end{subfigure}
    \hfill
    \begin{subfigure}{0.20\textwidth}
        \centering
        \begin{tikzpicture}
            \begin{feynman}
				\vertex (a);
				\vertex[right=of a](b);
				\vertex[above right=of b](c);
				\vertex[below right=of b] (f);
				\vertex[above right=of f] (g);
				\vertex[below right=of f] (h);
				
				\diagram*  {
					(a) -- [edge label=$N$, draw] (b),
					(b) -- [edge label=$\mu^{-}$, fermion] (c),
					(b) -- [edge label=$W^{+}$, boson] (f),
					(g) -- [edge label=$\bar{d}/\nu_{\ell}$, fermion] (f),
					(f) -- [edge label=${u}/ \ell^+$, fermion] (h),
				};
			\end{feynman}
        \end{tikzpicture}
        \caption{}
        \label{fig:NWdecay}
        \end{subfigure}
\captionsetup{justification=raggedright,singlelinecheck=false}   

\caption{Representative Feynman diagrams for various decay modes of $\widetilde{R}^{2/3}_2(\widetilde{R}^{-1/3}_2) \rightarrow N u(Nd)$ and $\widetilde{R}^{2/3}_2(\widetilde{R}^{-1/3}_2) \rightarrow \mu d (\nu_{\mu} d)$ are shown in Figs. (a) and (b), respectively. In Figs. (c) and (d), we show the different decay modes of $N$. $N$ can decay to muon/neutrino and light quarks via an off-shell $\widetilde{R}_2$ or $W$ boson. In addition, $N$ can also decay to charged leptons and neutrinos. There are similar decay modes of $N$ mediated by $Z$ and $H$ bosons which we do not show here. However, all decay modes have been considered in the computation of BRs.}

\label{Fig:NR_production}
\end{figure}
\begin{figure}[hbt!]
\centering
\captionsetup[subfigure]{labelformat=empty}
\subfloat[\quad\quad(a)]{\includegraphics[width=0.45\textwidth,height=0.35\textwidth]{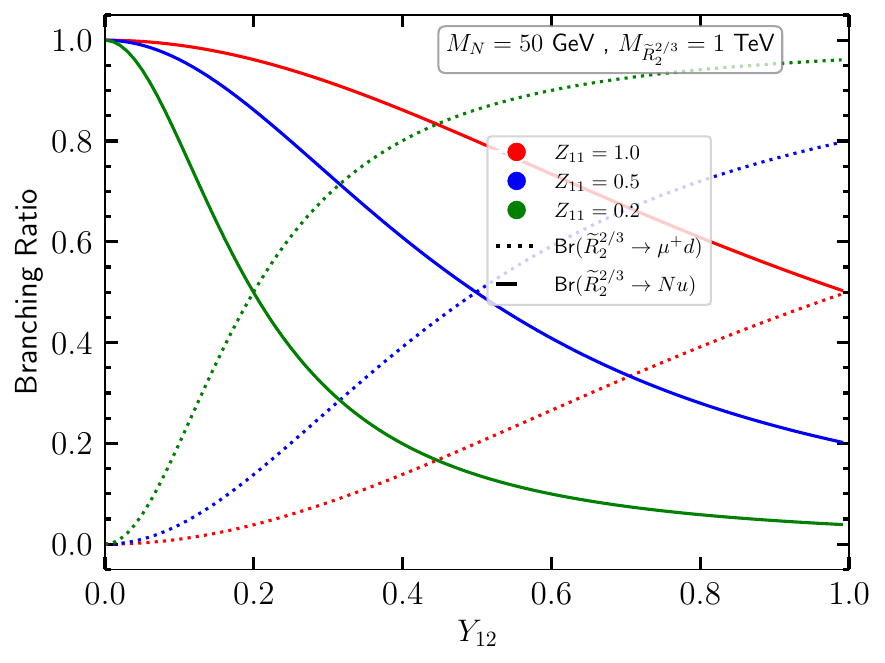}
\label{fig:BR_PLOT_LQ_50}}
\subfloat[\quad\quad(b)]{\includegraphics[width=0.45\textwidth,height=0.36\textwidth]{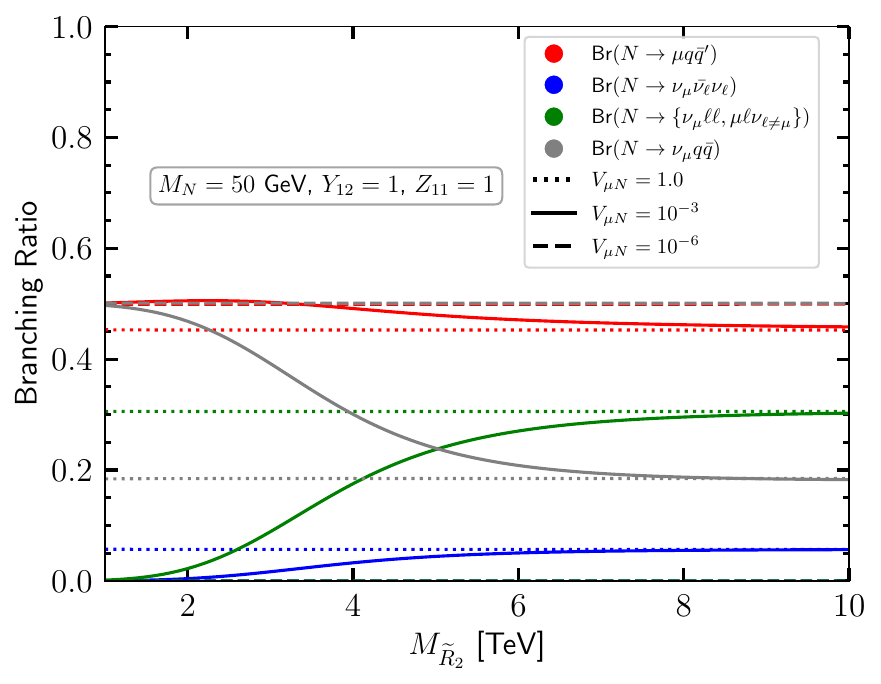}
\label{fig:BR_plot_RHN_mn_50}}\\
\subfloat[\quad\quad(c)]{\includegraphics[width=0.45\textwidth,height=0.35\textwidth]{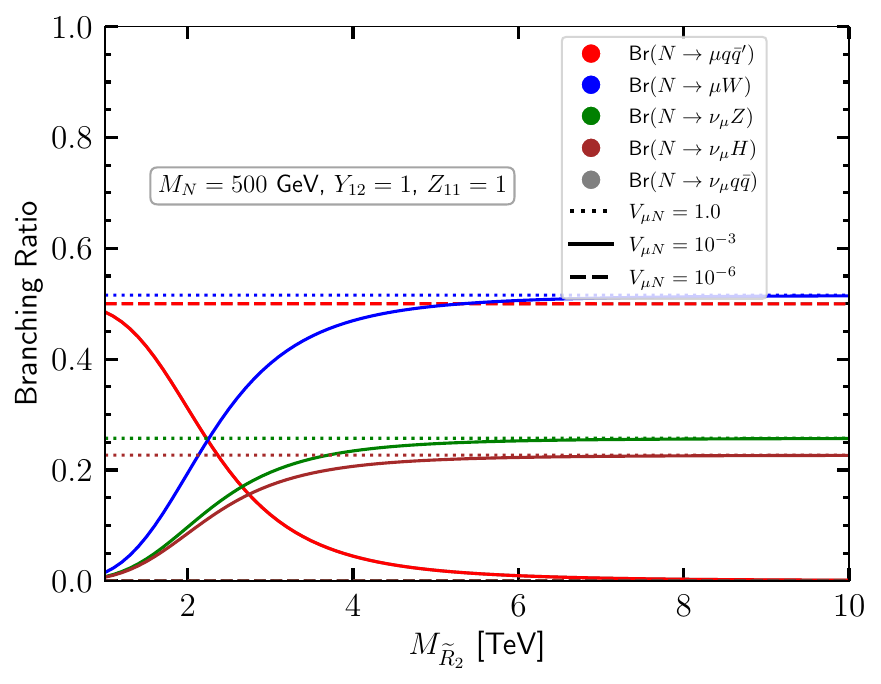}\label{fig:BR_plot_RHN_mn_500}}
\subfloat[\quad\quad(d)]{\includegraphics[width=0.45\textwidth,height=0.35\textwidth]{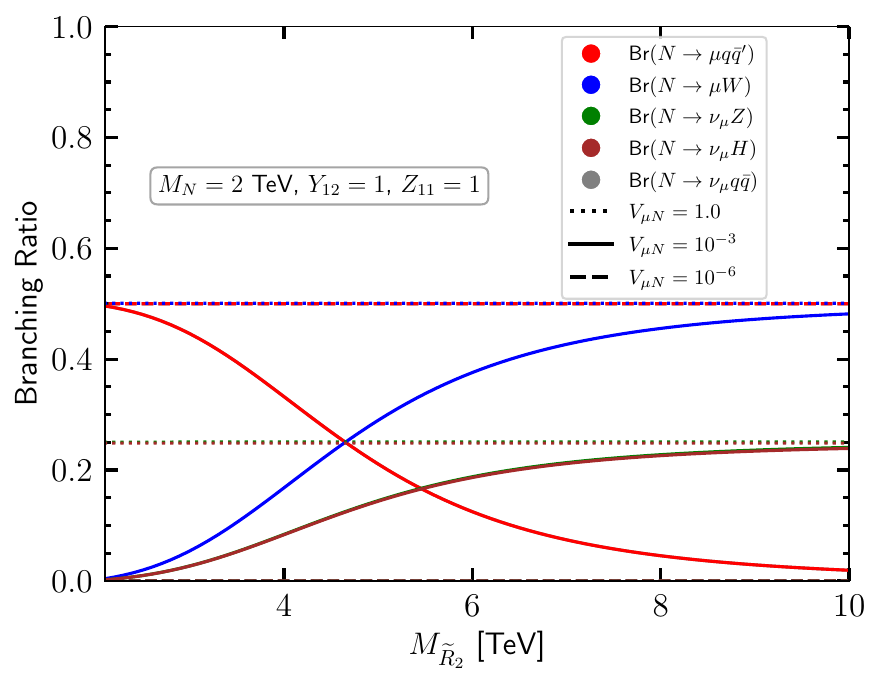}\label{fig:BR_plot_RHN_mn_2TeV}}
\captionsetup{justification=raggedright,singlelinecheck=false} 
\caption{(a) BRs of $\widetilde{R}^{2/3}_2 \rightarrow \mu d$ and $\widetilde{R}^{2/3}_2\rightarrow N u$ as functions of the coupling $Y_{12}$, evaluated for three benchmark values of $Z_{11}$ ($Z_{11}=0.2, 0.5, 1.0$), keeping $M_N = 50$ GeV and $M_{\widetilde{R}^{2/3}_2} = 1.0$ TeV.  Panels (b), (c), and (d) show the BRs of different decay modes of $N$ for $M_N = 50~\text{GeV}$, $500~\text{GeV}$, and $2.0~\text{TeV}$, respectively, as functions of the mass of $\widetilde{R}_2$, for three benchmark values of the active–sterile mixing angle $V_{\mu N}$ ($V_{\mu N} = 1.0$, $10^{-3}$, and $10^{-6}$).}
\label{fig:branchingratio}
\end{figure}

The decay modes of the RHN depend on the mass difference between $M_N$ and the masses of the SM gauge and Higgs bosons. For an RHN heavier than the {$W$, $Z$, or $H$}, the corresponding two-body decay modes, such as $N\to W\mu,~Z \nu_{\mu},~H\nu_{\mu}$ are kinematically possible. These $V_{\mu N}$-dependent decay modes are sub-dominant for our choice of a small  $V_{\mu N} \sim 10^{-6}$.
 
The three-body decay modes of the RHN can arise through two distinct mechanisms. Assuming $M_{\widetilde{R}_2} > M_N$ and our choice of Yukawa couplings, the RHN can undergo three-body decays such as $N\to \mu u\bar{d}$ ($N\to \nu_{\mu} d\bar{d}$) via an off-shell $\widetilde{R}^{2/3}_2$ ($\widetilde{R}^{-1/3}_2$). This is illustrated in Fig.~\ref{fig:NR2tildedecay}. Furthermore, if the RHN is lighter than the $W$ boson ($M_N < M_W$), we can have the decay mode $N \to W^*\mu \to (q\bar{q}^{\prime}, \ell \nu_{\ell})\mu$, as shown in Fig.~\ref{fig:NWdecay}. Similarly, for $M_N<M_Z$ and $M_N<M_H$, the decays $N \to Z^*\nu_{\mu} \to (q\bar q^{\prime}, b\bar b, \ell \ell, \bar{\nu_{\ell}}\nu_{\ell}) \nu_{\mu}$ and $N \to H^*\nu_{\mu} \to (q\bar{q},\ b\bar{b},\ \ell\ell) \nu_{\mu} $ (where $\ell = e,\mu,\tau$ and $q^{\prime}, q$ are first- and second-generation quarks) become possible. These decay modes depend on $V_{\mu N}$, and our choice of small $V_{\mu N}$, which we use to pursue the collider analysis, ensures that these contributions are nominal. 

The decay widths of all possible channels of $N$ have been calculated and are presented in Appendix~\ref{Appendix_n}. We have even included the sub-dominant channels to ensure an accurate determination of the BRs. Although in the subsequent sections and in the collider analysis, we consider $V_{\mu N}=10^{-6}$, for illustrative purpose, below we demonstrate the BRs for widely different choices of $V_{\mu N}$. 

In Fig.~\ref{fig:BR_PLOT_LQ_50}, we show that the BR of $\widetilde{R}^{2/3}_2$ decaying to $u N$ and $\mu d$ as a function of the Yukawa coupling $Y_{12}$. We consider three different values of $Z_{11}$: $Z_{11} = 0.2$, $0.5$, and $1.0$. We set $M_{\widetilde{R}^{2/3}_2} = 1.0$ TeV and $M_N = 50$ GeV. Neglecting the SM fermion masses, the BR for $\widetilde{R}^{2/3}_2 \rightarrow \mu d$ is approximately given as $Y^2_{12}/(Y^2_{12}+Z^2_{11})$. Thus, decreasing \( Z_{11} \) (while keeping \( Y_{12} \) constant) or increasing \( Y_{12} \) (while keeping \( Z_{11} \) constant) enhances the BR of decay \( \widetilde{R}^{2/3}_2 \rightarrow \mu d \), while the BR of \( \widetilde{R}^{2/3}_2 \rightarrow uN \) exhibits the opposite behavior. The BRs for $\widetilde{R}^{-1/3}_2 \to ~{\nu}_\mu d, N d$ show a similar behavior, and hence we do not show them explicitly in the plot.

Fig.~\ref{fig:BR_plot_RHN_mn_50} represents the variation of the BR for different three-body decay modes of $N$ as a function of the mass of $\widetilde{R}_{2}$. We set $Y_{12}=Z_{11}=1.0$, $M_N=50$ GeV, and present results for three different values of $V_{\mu N}$: $V_{\mu N} = 10^{-6}$, $10^{-3}$ and $1.0$. Although $V_{\mu N} = 1.0$ is disallowed by the electroweak precision constraints~\cite{Antusch:2014woa}, we show this only for illustration purposes.  For $V_{\mu N}=1.0$, the decays occur mainly via the off-shell $W$, $Z$ and $H$ boson (shown as dotted lines), resulting in constant BRs. This occurs because despite having a large $\mathcal{O}(1)$ Yukawa coupling, the contribution from heavy off-shell $\widetilde{R}_2$ is suppressed, and thus, the BRs do not vary with respect to $M_{\widetilde{R}_2}$. For $V_{\mu N} = 10^{-6}$, the BRs for the decay modes $N \rightarrow \mu q \bar{q}'$ and $N \rightarrow \nu_{\mu} q \bar{q}$ (depicted by the coinciding red and grey dashed lines) are $0.5$ each. This is because, for small values of $V_{\mu N}$, the dominant contributions arise from off-shell $\widetilde{R}_{2}$-mediated diagrams and depend on large $\mathcal{O}(1)$ Yukawa couplings. The pure leptonic decay modes $N\rightarrow \nu_\mu \bar{\nu}_{\ell} \nu_{\ell}$ and $N \rightarrow \nu_\mu  {{\ell}} {\ell},\ \mu \ell \bar\nu_{\ell\neq{\mu}}$  (represented by the blue and green dashed lines) for $V_{\mu N}=10^{-6}$ originate from the off-shell $W$, $Z$, and $H$ bosons and are suppressed. A key point to note here is that, the decay process $N\to \mu\mu\nu_{\mu}$ can arise from the off-shell decay of $W$, $Z$, and $H$ bosons as well as the interference between these processes. These contributions are considered in the decay width expression in Appendix~\ref{Appendix_n}. The solid lines represent BRs of different decay modes for a moderate value of $V_{\mu N}=10^{-3}$, and show significant variation with respect to $M_{\widetilde{R}_2}$ before becoming constant for a higher sLQ mass. For $M_{\widetilde{R}_2} > 8.0$ TeV,  the contributions from the off-shell $W$, $Z$, and $H$ bosons to the respective partial decay widths of different channels become the most dominant as the sLQ-mediated ones get off-shell mediator suppression, leaving the BR independent of the mass of $\widetilde{R}_2$.

In Fig.~\ref{fig:BR_plot_RHN_mn_500}, we consider $M_N = 500$ GeV. This allows the two-body decays of $N$ to be kinematically possible. The BRs for the two-body decays $N \rightarrow W \mu$, $N \rightarrow \nu_{\mu} Z$, and $N \rightarrow \nu_{\mu} H$ are shown in blue, green, and brown, respectively. The relevant three-body decay channels are $N \rightarrow \mu q \bar{q}^{\prime}$ and $N \rightarrow \nu_\mu  q \bar{q}$. For $V_{\mu N} = 1.0$, the two-body decays exhibit constant BRs, as they are independent of the mass of $\widetilde{R}_2$. Unlike the previous scenario, the three-body decays here are mediated solely by an off-shell $\widetilde{R}_2$ and due to the large mass of $\widetilde{R}_2$, these decay modes have small branching ratios. For $V_{\mu N} = 10^{-6}$, the two-body decay modes are highly suppressed, and the corresponding BRs are close to zero, as is evident from the figure. For $V_{\mu N} = 10^{-3}$, the BRs of the three-body decay modes follow the same qualitative trend as the previous case. The BRs of the two-body decays initially increase with $M_{\widetilde{R}_2}$ and eventually saturate at higher masses. This behavior arises because, at low $M_{\widetilde{R}_2}$, the three-body decays contribute significantly to the total width, thereby reducing the BRs of the two-body channels. As $M_{\widetilde{R}_2}$ increases, the contribution from three-body decays to the total decay width becomes negligible, and two-body modes begin to dominate, making the BRs increase with increasing  sLQ mass, eventually becoming independent for large values of $M_{\widetilde{R}_2}$.

In Fig.~\ref{fig:BR_plot_RHN_mn_2TeV}, we consider  a very heavy RHN with  $M_N = 2.0$ TeV and obtain similar BR plots as in Fig.~\ref{fig:BR_plot_RHN_mn_500}. The different lines corresponding to the different decay modes of $N$ follow a similar trend, and the explanations for $V_{\mu N}=1.0$ and $V_{\mu N}=10^{-6}$ remain the same. For $V_{\mu N} = 10^{-3}$, the BRs for decays $N \rightarrow \nu_{\mu} Z$ and $N \rightarrow \nu_{\mu} H$ overlap. This occurs because, at $M_N = 2.0$ TeV, these decay channels exhibit  similar behavior, as the masses of the $Z$ and Higgs bosons are of similar order compared to the heavy $N$. The three-body decay lines representing $N\rightarrow \mu q \bar{q}^{\prime}$ and $N\rightarrow \nu_{\mu}  q \bar{q}$ also overlap and follow a similar behavior as seen in the previous figure.\\

\vspace{2.5cm}
\noindent
In our analysis, we fix the value of $V_{\mu N} = 10^{-6}$ as expected from the canonical type-I seesaw mechanism, and we focus mainly on the decay $N \rightarrow \mu^- u \bar{d}$.\footnote{Since $N$ is Majorana in nature, we can also have the decay $N \rightarrow \mu^+ \bar{u} d$.} Thus, for this small value of $V_{\mu N}$, the contribution from the off-shell $W$-mediated diagram becomes negligible and the decay proceeds dominantly via the off-shell sLQ mediated channel for our choice of sLQ mass. The BR of $N \rightarrow\mu u \bar{d}$ is approximately $0.5$ for benchmark values of $M_N = 50,\ 500, \ 2000$ GeV and remains the same even for a smaller value of $Y_{12}$.

\section{Limits on leptoquarks from direct and indirect searches}
\label{sec:4}
\noindent
The parameter space for the sLQ is constrained by various direct and indirect searches at the LHC. The direct search for sLQ in the pair production channel with a subsequent  decay to a pair of muon and dijet ($pp\rightarrow {\rm sLQ}+ \overline{\rm sLQ}\rightarrow \mu j\ \mu j$) has been performed by the ATLAS collaboration~\cite{ATLAS:2020dsk}. Assuming a BR of $100\%$ for the decay mode ${\rm sLQ} \rightarrow \mu j$, this search excludes sLQ mass up to $1730$ GeV. 

A search for pair-produced sLQs in the $\mu \nu j j$ final-state excludes masses up to $1175~\mathrm{GeV}$ under the assumption of a $50\%$ BR for each of the decay channels $s\mathrm{LQ} \to j \nu$ and $s\mathrm{LQ} \to \mu j$~\cite{CMS:2018lab}. In our $\widetilde{R}_2$ model, the direct search limits are modified due to the presence of additional Yukawa interactions that alter both the pair production cross-section and the decay BRs. In particular, for nonzero values of $ Z_{11}$, the interaction $\widetilde{R}^{2/3}_2 \bar{u} N$ induces additional $t$-channel diagrams with heavy neutrino exchange, leading to a modification of the total production rate. The presence of a non-zero $Y_{12}$ also introduces additional $t$-channel diagrams mediated via muon which contributes to the sLQ pair production.  The sLQ $\widetilde{R}^{2/3}_2$ couples not only to $\mu d$ through the Yukawa coupling $Y_{12}$ but also to $N u$ via $Z_{11}$, so a nonzero $Z_{11}$ necessarily affects the ${\rm BR}(\widetilde{R}^{2/3}_2 \rightarrow \mu j)$. As a consequence of these combined effects, the exclusion limits from direct searches can differ significantly from the canonical sLQ scenario. 

\begin{figure}[t!]
\centering
\captionsetup[subfigure]{labelformat=empty}
\subfloat[\quad\quad\quad(a)]{

\includegraphics[width=0.45\textwidth,height=0.35\textwidth]{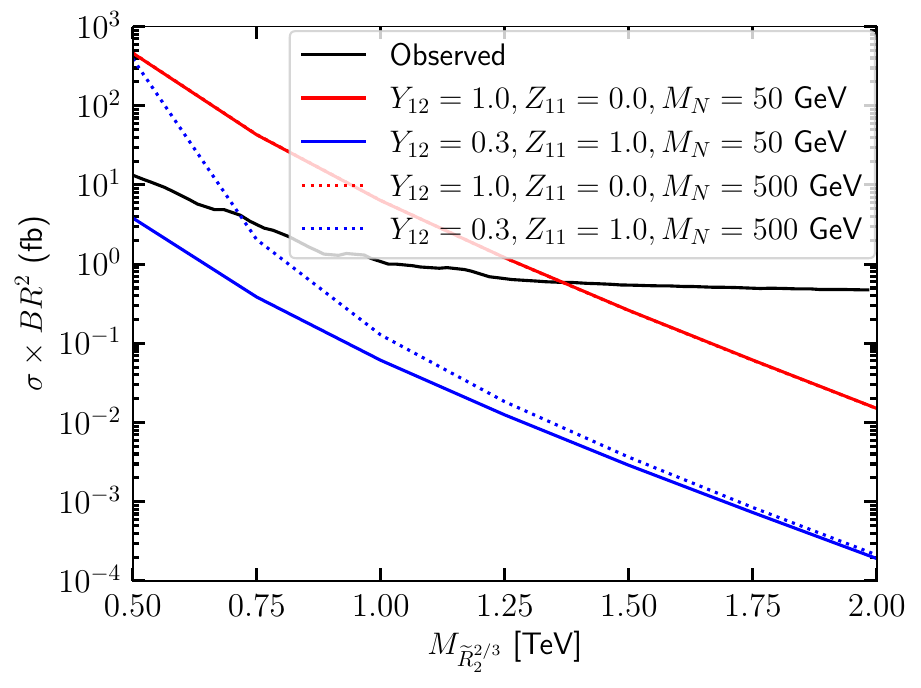}
\label{fig:lqlq_limits}}
\subfloat[\quad\quad\quad(b)]{

\includegraphics[width=0.45\textwidth,height=0.35\textwidth]{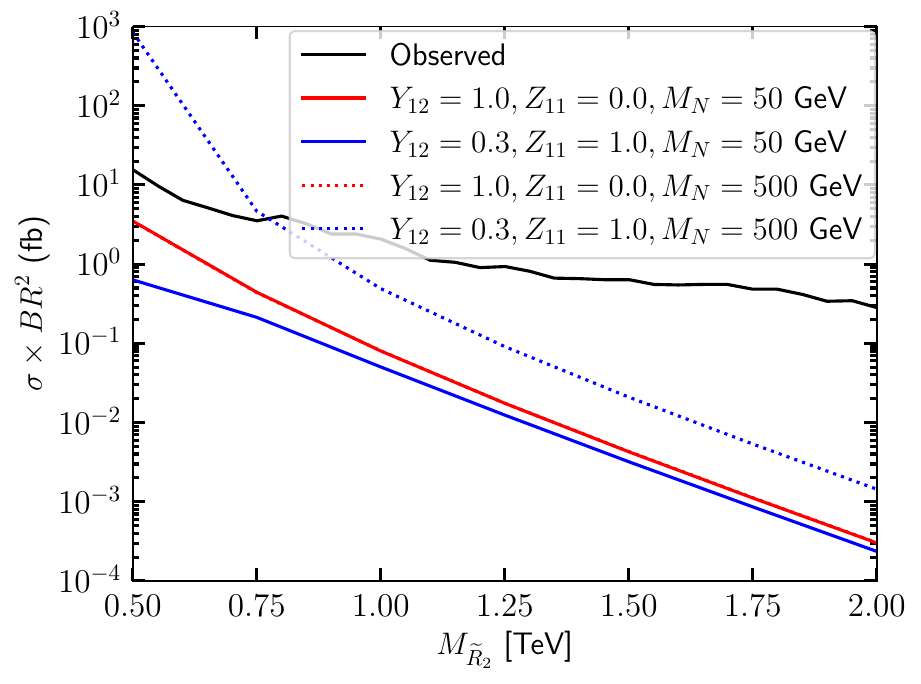}

\label{fig:lqlq_CMS_limits_1}
}
\captionsetup{justification=raggedright,singlelinecheck=false} 
\caption{Variation of sLQ pair production cross-section times ${\rm BR}^2$ versus sLQ mass. Panels (a) and (b) show the results for $\mu\mu jj$ and $\mu\nu jj$ channels, respectively. The observed limits are indicated by the solid black lines, from the ATLAS search for $pp \to \mu \mu jj$~\cite{ATLAS:2020dsk} and from the CMS search for $\mu\nu jj$~\cite{CMS:2018lab}, respectively. The blue and red lines (dotted and solid) represent the theoretical cross-sections times ${\rm BR}^2$ for various benchmark values of $M_N$, $Y_{12}$ and $Z_{11}$.}
\label{fig:LQlimits}
\end{figure}

In Fig.~\ref{fig:lqlq_limits} we display the observed ATLAS limit (solid black curve)~\cite{ATLAS:2020dsk} together with the theoretical predictions in the $pp\to \mu\mu jj$ channel for a few different choices of couplings and RHN masses. Note that this channel only involves the pair production of the $\widetilde{R}^{2/3}_2$ components. The red solid (dotted) curve corresponds to \(Y_{12} = 1.0\) and \(Z_{11} = 0.0\) with \(M_{N} = 50~\mathrm{GeV}\) (\(M_{N} = 500~\mathrm{GeV}\)). The perfect overlap between the solid and dotted curves indicates the absence of heavy neutrino effects in this benchmark scenario.
 The solid and  dotted blue curves represent the cross-section for $Y_{12} = 0.3$ and $Z_{11} = 1.0$ with $M_{N} = 50~{\rm GeV}$ and $M_{N} = 500~{\rm GeV}$, respectively. These curves clearly illustrate the interplay between the enhanced branching ratio to $\mu j$ and the altered pair production rate induced by the $t$-channel $N$ exchange. For these benchmark choices, the enhancement in the BR dominates over the moderate reduction in the production cross-section when comparing the $M_{N}=500~{\rm GeV}$ case to the $M_{N}=50~{\rm GeV}$. This leads to a more stringent bound for heavier $N$. However, the region $M_{\widetilde{R}_2}\gtrsim 720~{\rm GeV}$ remains unconstrained by the present $\mu\mu jj$ search for both heavy neutrino masses. 
 
 Fig.~\ref{fig:lqlq_CMS_limits_1} shows the exclusion limits from the CMS search for pair-produced sLQs decaying into the $\mu\nu jj$ final-state~\cite{CMS:2018lab}. For the choice of couplings $Y_{12}=0.3$ and $Z_{11}=1.0$, this analysis excludes sLQ masses below $M_{\widetilde{R}_2}\simeq 775~\text{GeV}$. Note that the $\mu\nu jj$ channel involves contributions from both $\widetilde{R}_2^{2/3}$ and $\widetilde{R}_2^{-1/3}$ components, and we have used a common mass $M_{\widetilde{R}_2}$ for them. The choice of  adopted $Y_{12}$  in the above figures is guided by the complementary indirect constraints. In particular, the CMS search for $t$-channel sLQ exchange in the high-mass dimuon spectrum yields $2\sigma$ exclusion limits in the sLQ mass–coupling plane~\cite{CMS:2024bej}. Translating their notation to ours, $M_{S_{\mu d}}\to M_{\widetilde{R}^{2/3}_2}$ and $Y_{d\mu}\to Y_{12}$, we identify the region of parameter space consistent with these bounds. For $M_{\widetilde{R}^{2/3}_2}\gtrsim 1.0~\text{TeV}$, a Yukawa coupling $Y_{12}=0.3$ remains allowed and is therefore adopted as a representative benchmark. We further fix $Z_{11}=1.0$, as no dedicated direct experimental constraints currently exist on this coupling.

 For \(Y_{12}=0.3\) and \(Z_{11}=1.0\), the dominant decay mode of
\(\widetilde{R}_2^{2/3}\) is \(\widetilde{R}_2^{2/3}\to Nj\). Hence, the mass range $M_{\widetilde{R}_2}\geq 1.0 $ TeV for the above choice of couplings remains unconstrained from LHC direct searches, as can also be seen from Fig.~\ref{fig:lqlq_limits} and Fig.~\ref{fig:lqlq_CMS_limits_1}.  

We note that  a fully consistent interpretation of the ATLAS search would require incorporating single production modes such as $\widetilde{R}_2 N j,\,  \widetilde{R}_2 \mu j$, as well as decay modes $(\mu j)$ and $(N j)$, 
 and performing a dedicated recast of the experimental search.
Moreover, for small mass ratios \(M_N/M_{\widetilde{R}_2^{2/3}}\), the \(Nj\) channel can lead
to non-isolated muons, which would be rejected by the isolation requirements employed in
Ref.~\cite{ATLAS:2020dsk}. Given these complications, we adopt a simplified approach and derive the limits as has been discussed above.
A complete recast including these effects is left for future work.

\begin{figure}[t!]
	\centering

	\begin{subfigure}[h]{0.30\textwidth}
		\centering
		\begin{tikzpicture}
			\begin{feynman}
				\vertex (a) at (0,0);
				\vertex (b) at (-1.5, 1.5);
				\vertex (c) at (-1.5, -1.5);
				\vertex (f) at (2.0, 0);
				\vertex (d) at (3.5, 1.5);
				\vertex (e) at (3.5, -1.5);
				
				\diagram*  {
					(a) -- [edge label=$\mu^{+}$, fermion] (b),
					(c) -- [edge label=$\mu^{-}$, fermion] (a),
					(a) -- [edge label=$Z / \gamma$, boson] (f),
					(d) -- [edge label=$\widetilde{R}^{2/3}_2/ \widetilde{R}^{-1/3}_2$, scalar] (f),
					(f) -- [edge label=$\widetilde{R}^{-2/3}_2 / \widetilde{R}^{1/3}_2$, scalar] (e),
				};
			\end{feynman}
		\end{tikzpicture}
		\caption{}
		\label{fig:R2pairQED}
	\end{subfigure}
\begin{subfigure}[h]{0.30\textwidth}
	\centering
	\begin{tikzpicture}
		\begin{feynman}
			\vertex (a) at (0,0);
			\vertex (b) at (-1.0, 0.8);
			\vertex (c) at (1.0, 0.8);
			\vertex (f) at (0,-1.2);
			\vertex (d) at (1.0, -2.0);
			\vertex (e) at (-1.0, -2.0);
			
			\diagram* {
				(a) -- [edge label=$\mu^{+}$, fermion] (b),
				(c) -- [edge label=$\widetilde{R}^{2/3}_2$, scalar] (a),
				(f) -- [edge label=$d$, fermion] (a),
				(f) -- [edge label=$\widetilde{R}^{-2/3}_2$, scalar] (d),
				(e) -- [edge label=$\mu^{-}$, fermion] (f),
			};
		\end{feynman}
	\end{tikzpicture}
	\caption{}
	\label{fig:R2pairNP}    
\end{subfigure}
    \begin{subfigure}[h]{0.30\textwidth}
	\centering
	\begin{tikzpicture}[baseline={(current bounding box.center)}, scale=0.9]
		\begin{feynman}
			\vertex (a) at (0,0);
			\vertex (b) at (-1.2, 1.0);
			\vertex (c) at (1.2, 1.0);
			\vertex (f) at (0,-1.0);
			\vertex (d) at (1.2, -2.0);
			\vertex (e) at (-1.2, -2.0);
			\vertex (m) at (2.0, 0.2);
			\vertex (n) at (2.0, 1.4);
			
			\diagram* {
				(a) -- [edge label=$\mu^{+}$, fermion] (b),
				(c) -- [edge label=$\bar{\nu_{\mu}}$, fermion] (a),
				(f) -- [edge label=$W^{+}$, boson] (a),
				(f) -- [edge label=$N$, draw] (d),
				(e) -- [edge label=$\mu^{-}$, fermion] (f),
				(m) -- [edge label=$\bar{d}$, fermion] (c),
				(c) -- [edge label=$\widetilde{R}_{2}^{-1/3}$, scalar] (n),
			};
		\end{feynman}
	\end{tikzpicture}
	\caption{}
	\label{fig:R2single6}
\end{subfigure}
\begin{subfigure}[h]{0.30\textwidth}
	\centering
	\begin{tikzpicture}
		\begin{feynman}
			\vertex (a) at (0,0);
			\vertex (b) at (-1.0, 0.8);
			\vertex (c) at (1.0, 0.8);
			\vertex (f) at (0,-1.2);
			\vertex (d) at (1.0, -2.0);
			\vertex (e) at (-1.0, -2.0);
			\vertex (m) at (1.8, 0.1);     
			\vertex (n) at (1.6, 1.6);     
			
			\diagram* {
				(a) -- [edge label=$\mu^{+}$, fermion] (b),
				(a) -- [edge label=${\widetilde{R}_2}^{2/3*}$, scalar] (c),
				(f) -- [edge label=$\bar{d}$, fermion] (a),
				(f) -- [edge label=$\widetilde{R}^{-2/3}_2$, scalar] (d),
				(e) -- [edge label=$\mu^{-}$, fermion] (f),
				(c) -- [edge label'=$u/d$, fermion] (m),
				(n) -- [edge label=$N/\mu^{+}$, fermion] (c),
			};
		\end{feynman}
	\end{tikzpicture}
	\caption{}
	\label{fig:R2single1}
\end{subfigure}
\begin{subfigure}[h]{0.30\textwidth}
	\centering
	\begin{tikzpicture}
		\begin{feynman}
			\vertex (a) at (0,0);
			\vertex (b) at (-1.2, 1.0);
			\vertex (c) at (1.2, 1.0);
			\vertex (g) at (0,-1.0);
			\vertex (k) at (1.2,-1.0);
			\vertex (f) at (0,-2.0);
			\vertex (d) at (1.2,-3.0);
			\vertex (e) at (-1.2,-3.0);
			
			\diagram* {
				(b) -- [edge label=$\mu^{-}$, fermion] (a),
				(a) -- [edge label=$N$, draw] (c),
				(a) -- [edge label=$W^{+}$, boson] (g),
				(g) -- [edge label=$\bar{d}$, fermion] (f),
				(k) -- [edge label=$\bar{u}$, fermion] (g),                
				(f) -- [edge label=$\widetilde{R}^{2/3}_2$, scalar] (d),
				(f) -- [edge label'=$\mu^{+}$, fermion] (e),
			};
		\end{feynman}
	\end{tikzpicture}
	\caption{}
	\label{fig:R2single2}
\end{subfigure}
\begin{subfigure}[h]{0.30\textwidth}
	\centering
	\begin{tikzpicture}
		\begin{feynman}
			\vertex (a) at (0,0);
			\vertex (b) at (-1.2, 1.0);
			\vertex (c) at (1.2, 1.0);
			\vertex (g) at (0,-1.0);
			\vertex (k) at (1.2,-1.0);
			\vertex (f) at (0,-2.0);
			\vertex (d) at (1.2,-3.0);
			\vertex (e) at (-1.2,-3.0);
			
			\diagram* {
				(b) -- [edge label=$\mu^{-}$, fermion] (a),
				(a) -- [edge label=$\mu^{-}$, fermion] (c),
				(a) -- [edge label=$\gamma/Z/H$, boson] (g),
				(g) -- [edge label=$\bar{d}$, fermion] (f),
				(k) -- [edge label=$\bar{d}$, fermion] (g),                
				(f) -- [edge label'=$\mu^{+}$, fermion] (e),
				(f) -- [edge label=$\widetilde{R}^{2/3}_2$, scalar] (d),
			};
		\end{feynman}
	\end{tikzpicture}
	\caption{}
	\label{fig:R2single3}
\end{subfigure}
\begin{subfigure}[h]{0.30\textwidth}
	\centering
	\begin{tikzpicture}[baseline={(current bounding box.center)}, scale=0.8]
		\begin{feynman}
			\vertex (a) at (0,0);
			\vertex (b) at (-1.2, 1.2);
			\vertex (c) at (-1.2, -1.2);
			\vertex (f) at (1.5, 0);
			\vertex (d) at (2.8, 1.2);
			\vertex (e) at (2.8, -1.2);
			\vertex (m) at (4.5, -2.2);
			\vertex (n) at (4.2, 0.0);
			
			\diagram*  {
				(a) -- [edge label=$\mu^{+}$, fermion] (b),
				(c) -- [edge label=$\mu^{-}$, fermion] (a),
				(a) -- [edge label=$Z / \gamma$, boson] (f),
				(d) -- [edge label=$\bar{d}$, fermion] (f),
				(f) -- [edge label=$d$, fermion] (e),
				(e) -- [edge label=${\widetilde{R}}^{-1/3}_2$, scalar] (m),
				(e) -- [edge label=$N$, draw] (n),
			};
		\end{feynman}
	\end{tikzpicture}
	\caption{}
	\label{fig:R2single4}
\end{subfigure}
\begin{subfigure}[h]{0.30\textwidth}
	\centering
	\begin{tikzpicture}[baseline={(current bounding box.center)}, scale=0.9]
		\begin{feynman}
			\vertex (a) at (0,0);
			\vertex (b) at (-1.2, 1.2);
			\vertex (c) at (1.2, 1.2);
			\vertex (g) at (0.2,-1.0);
			\vertex (k) at (1.3,-0.7);
			\vertex (f) at (0,-1.8);
			\vertex (d) at (1.2,-2.6);
			\vertex (e) at (-1.2,-2.6);
			
			\diagram* {
				(a) -- [edge label=$\mu^{+}$, fermion] (b),
				(c) -- [edge label=$\bar{d}$, fermion] (a),
				(a) -- [edge label=$\widetilde{R}_{2}^{2/3}$, scalar] (g),
				(g) -- [edge label=$W^{+}$, boson] (f),
				(k) -- [edge label=$\widetilde{R}_{2}^{-1/3}$, scalar] (g),                
				(f) -- [edge label=$N$, draw] (d),
				(e) -- [edge label'=$\mu^{-}$, fermion] (f),
			};
		\end{feynman}
	\end{tikzpicture}
	\caption{}
	\label{fig:R2single5}
\end{subfigure}
\begin{subfigure}[h]{0.30\textwidth}
	\centering
	\begin{tikzpicture}
		\begin{feynman}
			\vertex (a);
			\vertex[above left=of a](b);
			\vertex[above right=of a](c);
			\vertex[below=of a] (f);
			\vertex[below right=of f](d);
			\vertex[below left=of f] (e);
			
			\diagram*  {
				(a) -- [edge label=$\mu^{+}$, fermion] (b),
				(c) -- [edge label=$\bar{d}$, fermion] (a),
				(f) -- [edge label=$\widetilde{R}^{2/3}_2$, scalar] (a),
				(f) -- [edge label=$d$, fermion] (d),
				(e) -- [edge label=$\mu^{-}$, fermion] (f),
			};
		\end{feynman}
	\end{tikzpicture}
	\caption{}
	\label{fig:mumujjfeynman}    
\end{subfigure}
\captionsetup{justification=raggedright,singlelinecheck=false}     
\caption{Representative Feynman diagrams illustrating the pair and single production of $\widetilde{R}_2$. The last diagram represents dijet production at a muon collider, mediated via an sLQ.}
\label{Fig:LQ_production}
\end{figure}

\begin{figure}[t!]
\centering
\captionsetup[subfigure]{labelformat=empty}

\makebox[\textwidth][l]{
    \hspace{0.8cm} 
    \begin{minipage}{0.49\textwidth}
        \centering
        \includegraphics[width=\textwidth,height=0.30\textwidth]{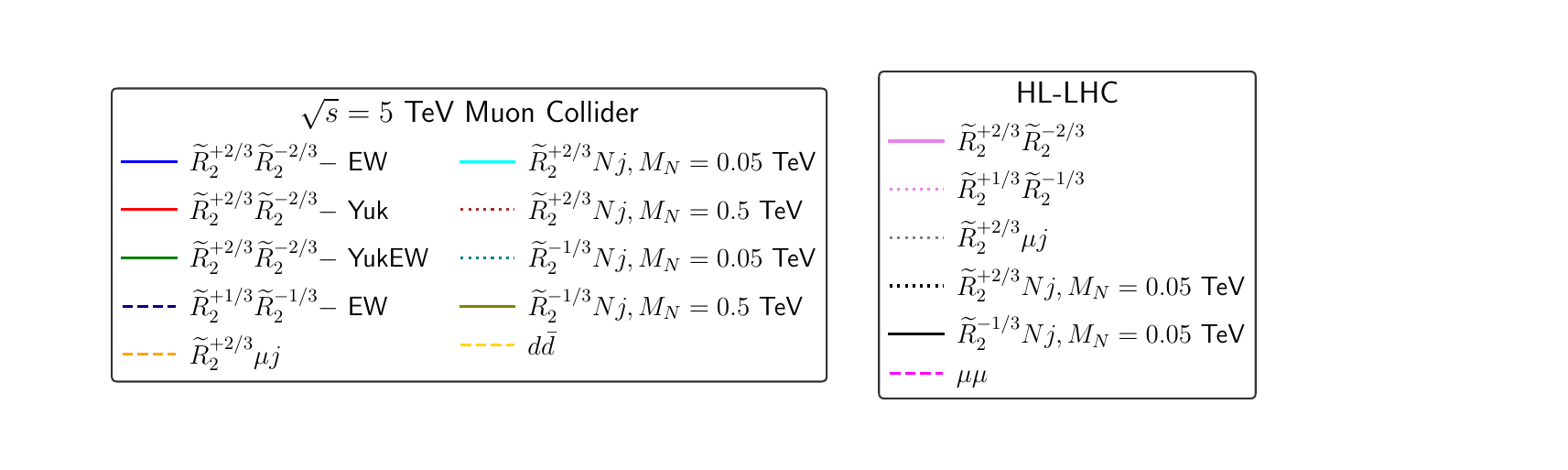}
    \end{minipage}
    \hfill
    \begin{minipage}{0.49\textwidth}
        \centering
        \includegraphics[width=\textwidth,height=0.30\textwidth]{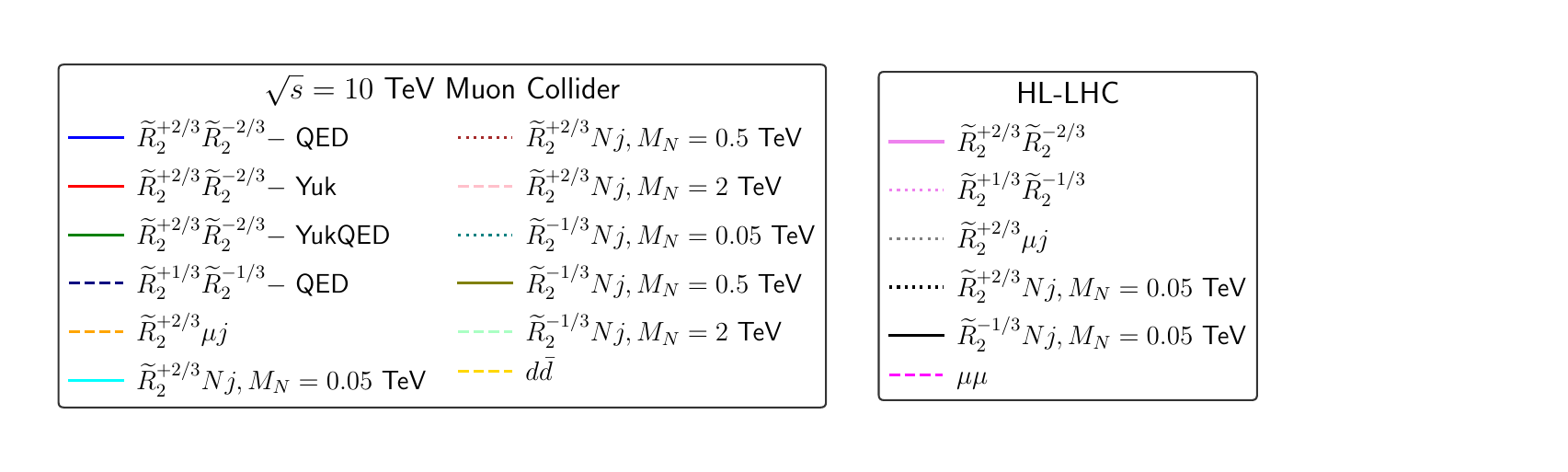}
    \end{minipage}
}

\vspace{0.3cm}

\captionsetup[subfigure]{labelformat=simple}

\subfloat[]{
    \includegraphics[width=0.48\textwidth,height=0.36\textwidth]
    {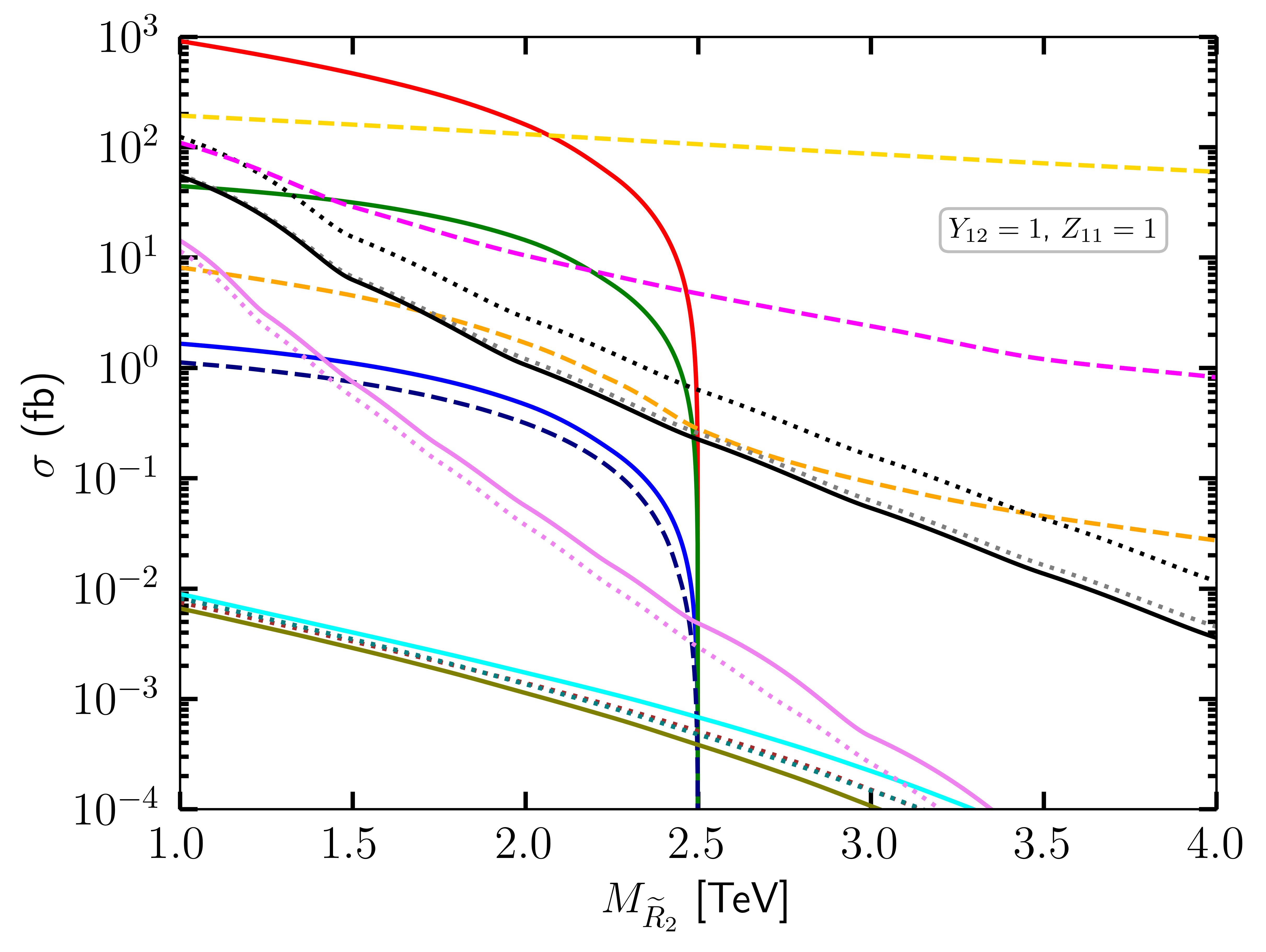}
    \label{fig:MR2tilde_CS_5TeV_1}
}
\hfill
\subfloat[]{
    \includegraphics[width=0.48\textwidth,height=0.36\textwidth]
    {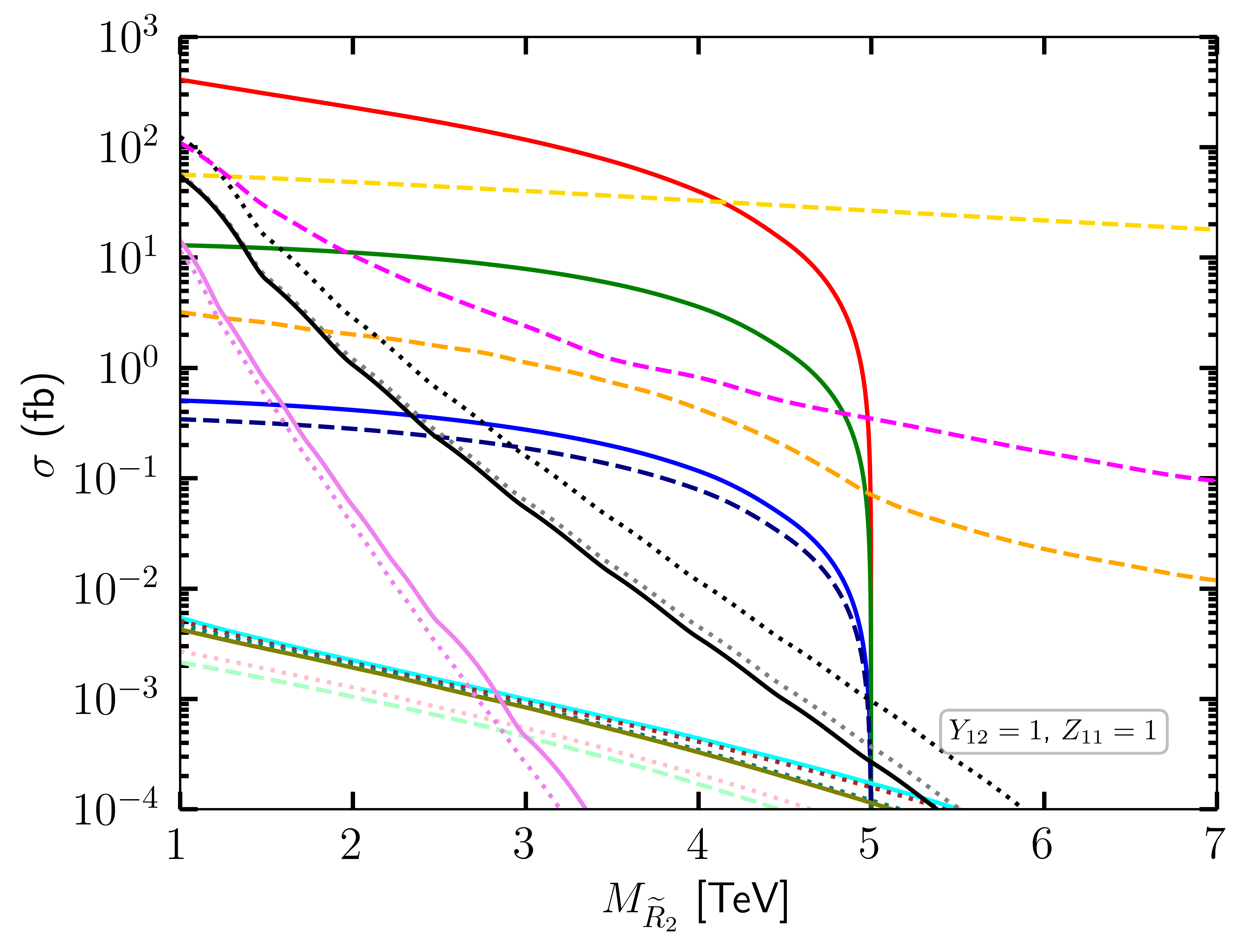}
    \label{fig:MR2tilde_CS_10TeV_1}
}

\captionsetup{justification=raggedright,singlelinecheck=false} 
\caption{Variation of cross-section for different $\widetilde{R}_2$ production modes as a function of $M_{\widetilde{R}_2}$. Left and right panel correspond to the muon collider C.O.M. energy $\sqrt{s} = 5$ TeV and $10$ TeV, respectively. The Yukawa couplings are set to $Z_{11}=Y_{12}=1$. We explain the labels in detail in the text.}
\label{fig:crossSectionvsmass_1}

\end{figure}

\section{Production of $\widetilde{R}_2$ at a muon collider}
\label{sec:5}

The sLQ $\widetilde{R}_2$  can be produced at the muon collider either in pairs or singly along with a light quark and a SM lepton/RHN. For the single production mode, $\widetilde{R}^{2/3}_2$ is produced along with $N u$ or $\mu d$, and $\widetilde{R}^{1/3}_2$ is produced along with $Nd$ or $\nu_{\mu} d$. As discussed in Section~\ref{sec:3}, $\widetilde{R}^{2/3}_2$ ($\widetilde{R}^{-1/3}_2$) subsequently decays to $Nu, \mu d$ ($Nd, \nu_\mu d$). As we are interested in dimuon final-state, we consider the decay mode of $N \to \mu u \bar{d}$ only. Depending on the final decay products, we categorize the entire production and decay chain of sLQs into two different scenarios: symmetric and asymmetric modes. In the symmetric mode, both the sLQs in the pair decay to the same  states, i.e., for pair-produced $\widetilde{R}^{2/3}_2$, both decay to $N u$ or $\mu d$, and for $\widetilde{R}^{-1/3}_2$ both decay to $N d$ or $\nu_\mu d$. In the case of symmetric single production where $\widetilde{R}^{2/3}_2$ ($\widetilde{R}^{1/3}_2$) is produced alongside $N u$ or $\mu d$ ($N d$ or $\nu_\mu d$), the sLQ decaying to $N u$ or $\mu d$ ($N d$ or $\nu_\mu d$) leads to the same decay products as the symmetric pair production mode.
In the case of asymmetric mode and for pair production of sLQ, $\widetilde{R}^{2/3}_2$ and $\widetilde{R}^{-2/3}_2$  decay via different decay chains. For single production mode such as $\widetilde{R}^{2/3}_2 \mu d$, in the asymmetric mode $\widetilde{R}^{2/3}_2$ decay to $Nu$ is chosen so that identical final-states $\mu d \mu \bar{d}$ as in  symmetric single production can be avoided. Similar conclusion holds for other asymmetric single production modes $\widetilde{R}^{2/3}_2 u N, \widetilde{R}^{-1/3}_2 N d, \widetilde{R}^{-1/3}_2 \nu d$. For clarity, we explicitly illustrate the symmetric and asymmetric modes for the pair and single production processes below.

\begin{center}
    \textbf{Symmetric mode: Pair and single production }
\end{center}
\begin{equation}
\label{eq:pairRD1A}
\mu^+\mu^-\to\left\{\begin{array}{lclcl}
  \widetilde{R}^{+2/3}_2 \widetilde{R}^{-2/3}_2 &\rightarrow& (u N) \, ({u} N)  &\rightarrow& j(\mu j j) j(\mu jj) \equiv \mu \mu + N_{\rm jet} \ge 6 \\
  \widetilde{R}^{+2/3}_2 \widetilde{R}^{-2/3}_2 &\rightarrow& (\mu {d}) \, (\mu d) &\equiv& \mu \mu + N_{\rm jet} \ge 2 \\
  \widetilde{R}^{+1/3}_2 \widetilde{R}^{-1/3}_2 &\rightarrow& (N {d}) \, (N d)   &\rightarrow& j(\mu j j) j(\mu jj) \equiv \mu \mu + N_{\rm jet} \ge 6 \\
  \widetilde{R}^{+1/3}_2 \widetilde{R}^{-1/3}_2 &\rightarrow& (\nu {d}) \, ({\nu} d)  &\equiv& \slashed{E}_T + N_{\rm jet} \ge 2 \\
  
\end{array}\right.
\end{equation}
\begin{equation}
\label{eq:singleRD1A}
\mu^+\mu^-\to\left\{\begin{array}{lclcl}
  \widetilde{R}^{2/3}_2 {u} N & \rightarrow & (u N)\ {u} N &\rightarrow& j(\mu j j)\  j(\mu jj) \equiv \mu \mu + N_{\rm jet} \ge 6 \\
  \widetilde{R}^{2/3}_2 \mu {d} & \rightarrow & (\mu d)\ \mu {d} & \equiv & \mu \mu + N_{\rm jet} \ge 2 \\
  \widetilde{R}^{-1/3}_2 d N  & \rightarrow & ({d} N)\ d N  &\rightarrow& j(\mu j j)\  j(\mu jj) \equiv \mu \mu + N_{\rm jet} \ge 6 \\
  \widetilde{R}^{-1/3}_2 {\nu} d & \rightarrow & (\nu {d})\ {\nu} d &\equiv& \slashed{E}_T + N_{\rm jet} \ge 2 \\

\end{array}\right.
\end{equation}
     \begin{center}
     \textbf{Asymmetric mode: Pair and single production}
     \end{center}

\begin{equation}
\label{eq:pairRD1B}
\mu^+\mu^-\to\left\{\begin{array}{lclcl}
\widetilde{R}^{+2/3}_2 \widetilde{R}^{-2/3}_2 &\rightarrow& (u N) \, ({d} \mu) &\rightarrow& j (\mu j j)\ j \mu \equiv \mu \mu + N_{\rm jet} \ge 4 \\
\widetilde{R}^{+1/3}_2 \widetilde{R}^{-1/3}_2 &\rightarrow& ({d} N) \, (d \nu) &\rightarrow& j (\mu j j)\ j \nu \equiv \mu \slashed{E}_T + N_{\rm jet} \ge 4 \\

\end{array}\right.
\end{equation}

\begin{equation}
\label{eq:singleRD1B}
\mu^+\mu^-\to\left\{\begin{array}{lclcl}
\widetilde{R}^{2/3}_2 \mu {d} &\rightarrow& (u N)\, \mu {d} &\rightarrow& j (\mu j j)\ j \mu \equiv \mu \mu + N_{\rm jet} \ge 4 \\
\widetilde{R}^{2/3}_2 {u} N &\rightarrow& (d \mu) \ {u} N  &\rightarrow& j \mu j (\mu j j)\ \equiv \mu \mu + N_{\rm jet} \ge 4 \\
\widetilde{R}^{-1/3}_2 d N &\rightarrow& (\nu {d}) \, d N &\rightarrow& \nu j \ j (\mu j j) \equiv \mu + \slashed{E}_T + N_{\rm jet} \ge 4 \\
\widetilde{R}^{-1/3}_2 d \nu &\rightarrow& (N{d}) \, d \nu &\rightarrow& (\mu j j) j (j \nu) \equiv \mu + \slashed{E}_T + N_{\rm jet} \ge 4 
\end{array}\right. \, .
\end{equation}
In the above, $N_{\rm jet}$ denotes the number of jets and, as can be seen, the typical jet multiplicity is large with $N_{\rm jet}\ge 2-6$. Among the processes mentioned above, we particularly focus on the modes in which at least one RHN is generated from $\widetilde{R}^{2/3}$/$\widetilde{R}^{-1/3}$ decay {or at least one is produced alongside $\widetilde{R}^{2/3}$/$\widetilde{R}^{-1/3}$} and in the final-state we have dimuons associated with multi-jets. These modes serve as viable RHN production modes at the muon collider and can probe both sLQ and RHN. These channels have large cross-sections and due to the presence of dimuon and at least four jets in the final-state, they encounter a very small SM background. Note that, from the above, we ignore the channels with missing transverse energy ($\slashed{E}_T$), as they are not accompanied with dimuons and hence they are not relevant for our study. Furthermore we also neglect the contribution from $\mu \mu d d$ final-state, as reconstruction of $N$ can make this contribution negligible.
Therefore, the specific $\widetilde{R}_2$ production channels considered in our analysis for symmetric and asymmetric modes are the following: \\

\begin{equation}
\label{eq:finalsym}
\mu^+\mu^-\to\left\{\begin{array}{lclcl}
  \widetilde{R}^{+2/3}_2 \widetilde{R}^{-2/3}_2 &\rightarrow& (u N) \, (\bar{u} N)  &\rightarrow& j(\mu j j) j(\mu jj) \equiv \mu \mu + n_{\rm jet} \ge 6\\
  \widetilde{R}^{+1/3}_2 \widetilde{R}^{-1/3}_2 &\rightarrow& (N d)  \,  (N \bar{d}) &\rightarrow& j(\mu j j) j(\mu jj) \equiv \mu \mu + n_{\rm jet} \ge 6 \\  
 \widetilde{R}^{+2/3}_2 \bar{u} N,\ \widetilde{R}^{-2/3}_2 u N & \rightarrow & (u N)\ \bar{u} N, (\bar{u} N)\ u N &\rightarrow& j(\mu j j)\  j(\mu jj) \equiv \mu \mu + n_{\rm jet} \ge 6\\
 \widetilde{R}^{+1/3}_2 d N,\ \widetilde{R}^{-1/3}_2 \bar{d} N  & \rightarrow & (\bar{d} N)\ d N,\ (d N)\ \bar{d} N &\rightarrow& j(\mu j j)\  j(\mu jj) \equiv \mu \mu  + n_{\rm jet} \ge 6 \\
\end{array}\right\} {\rm:  Symmetric\,  mode}
\end{equation}

\begin{equation}
\label{eq:finalasymm}
 \mu^+\mu^-\to\left\{\begin{array}{lclcl}
 \widetilde{R}^{+2/3}_2 \widetilde{R}^{-2/3}_2 &\rightarrow& (u N) \, (\bar{d} \mu) &\rightarrow& j (\mu j j)\ j \mu \equiv \mu  \mu  + n_{\rm jet} \ge 4 \\
 \widetilde{R}^{+2/3}_2 \mu \bar{d},\ \widetilde{R}^{-2/3}_2 \mu d &\rightarrow& (u N)\, \bar{d} \mu,\ (\bar{u} N)\, \mu d &\rightarrow& j (\mu j j)\ j \mu \equiv \mu \mu  + n_{\rm jet} \ge 4 \\
 \widetilde{R}^{+2/3}_2 \bar{u} N,\ \widetilde{R}^{-2/3}_2 u N &\rightarrow& (d \mu) \ \bar{u} N, \ (\bar{d} \mu) \ u N &\rightarrow& j \mu  j\ (\mu j j)\ \equiv \mu \mu  + n_{\rm jet} \ge 4 \\
\end{array}\right\}: {\rm Asymmetric\,  mode}
\end{equation}
\noindent
In the above equations, $\mu\mu$ denotes both the opposite-charge ($\mu^{\pm}\mu^{\mp}$) and same-charge ($\mu^{\pm}\mu^{\pm}$) dimuon pairs. Since our analysis does not impose any requirement on the relative charges of the muons, we remain agnostic to the charge combination. If one demands same-sign dimuon in the final-state as the signal, the sensitivity will improve since typically the SM background for such signal is extremely small. Consequently, our sensitivity estimates are conservative.
{Although there is a difference in jet multiplicity between symmetric and asymmetric modes, as shown in Eqs.~\eqref{eq:finalsym} and \eqref{eq:finalasymm}, our search strategy discussed in Section~\ref{sec:6} is  applicable for both the scenarios, and hence more generic.}

In Figs.~\ref{fig:MR2tilde_CS_5TeV_1} and \ref{fig:MR2tilde_CS_10TeV_1} we show the production cross-sections of $\widetilde{R}_{2}$ as a function of its mass for  C.O.M. energies of $5$ and $10$ TeV, respectively. We assume Yukawa couplings $Z_{11}=Y_{12}=1$. In the following, we explain the notation used to label the different $\widetilde{R}_{2}$ production modes in the figures mentioned above, with the corresponding Feynman diagrams shown in Fig.~\ref{Fig:LQ_production}. 

\begin{itemize}
    \item [--] In Fig.~\ref{fig:R2pairQED}, we illustrate the pair production process $\mu^+ \mu^- \rightarrow \widetilde{R}^{+2/3}_{2} \widetilde{R}^{-2/3}_{2}$ or $\widetilde{R}^{+1/3}_{2} \widetilde{R}^{-1/3}_{2}$ mediated by the electroweak (EW) $\gamma/ Z^*$ boson; this contribution is labeled as $\widetilde{R}^{+2/3}_{2}\ \widetilde{R}^{-2/3}_{2}-$EW ($\widetilde{R}^{+1/3}_{2}\ \widetilde{R}^{-1/3}_{2}-$EW) which is denoted by the solid blue line (dashed blue line) in Figs.~\ref{fig:MR2tilde_CS_5TeV_1} and \ref{fig:MR2tilde_CS_10TeV_1}. Similarly, Fig.~\ref{fig:R2pairNP} illustrates the pair production of $\widetilde{R}^{2/3}_{2}$ through the $t$-channel quark. Here, the cross-section contribution scales as $Y^4_{12}$ and is denoted by $\widetilde{R}^{+2/3}_{2}\ \widetilde{R}^{-2/3}_{2}-$Yuk (solid red line) in both cross-section plots. $\widetilde{R}^{+2/3}_{2}\ \widetilde{R}^{-2/3}_{2}-$YukEW (solid green line) represents the cross-section resulting from interference between Figs.~\ref{fig:R2pairQED} and \ref{fig:R2pairNP} and its contribution scales as $Y^2_{12}$. As we show the cross-sections of different processes for $Y_{12}=1$, and give the dependency with respect to $Y_{12}$, hence for any other smaller values of $Y_{12}$ the cross-section can be scaled trivially.

    \item[--] The Feynman diagrams for few of the representative single production modes of $\widetilde{R}^{2/3}_{2}$ ($\widetilde{R}^{-1/3}_{2}$) are shown in Figs.~\ref{fig:R2single1}, \ref{fig:R2single2} and \ref{fig:R2single3} (Figs.~\ref{fig:R2single6}, \ref{fig:R2single4}, and \ref{fig:R2single5}). In Figs.~\ref{fig:MR2tilde_CS_5TeV_1} and \ref{fig:MR2tilde_CS_10TeV_1}, the corresponding single production cross-sections are labeled $\widetilde{R}^{+2/3}_{2} \mu j$, $\widetilde{R}^{+2/3}_{2} N j$ and $\widetilde{R}^{-1/3}_{2} N j$, and are shown for different benchmark masses of $M_N=0.05,\  0.5,\ 2.0$ TeV. The inclusion of single production processes is important because their contribution becomes significant, or even dominant, at higher $\widetilde{R}_2$ masses where pair production is phase-space suppressed.
    
    \item[--] In addition to direct production of sLQ, we also show the cross-section of the $\widetilde{R}^{2/3}_{2}$-mediated $t$-channel process $\mu^+ \mu^- \to d \bar{d}$ (yellow dashed line). The corresponding Feynman diagram for this indirect production is shown in Fig.~\ref{fig:mumujjfeynman}. 
\end{itemize}

For comparison, we show the cross-section of the pair and single production of $\widetilde{R}^{2/3}_2$ and $\widetilde{R}^{-1/3}_2$ at the High-Luminosity LHC (HL-LHC) in Figs.~\ref{fig:MR2tilde_CS_5TeV_1} and \ref{fig:MR2tilde_CS_10TeV_1}. To avoid confusion, we have shown separate legends for the production modes at the muon collider and HL-LHC.
\begin{itemize}
    \item [--] The pair production cross-sections for $\widetilde{R}^{+ 2/3}_2 \widetilde{R}^{- 2/3}_2$ (solid pink) and $\widetilde{R}^{+ 1/3}_2 \widetilde{R}^{- 1/3}_2$ (dotted light pink) at the HL-LHC are shown for comparison with those at the muon collider. The EW–mediated pair production at the muon collider begins to exceed the corresponding HL-LHC production for $M_{\widetilde{R}_2} \gtrapprox 1.5~\text{TeV}$. Furthermore, the Yukawa-induced production of $\widetilde{R}^{2/3}_2$ (denoted “Yuk”), together with its interference with the EW-mediated channel (“YukEW”), yields a total cross-section that consistently remains above the HL-LHC prediction across the entire mass range considered.
    
    \item[--] In analogy to the muon collider setup, we have also generated single production processes ($\widetilde{R}^{2/3}_2 \mu j$, $\widetilde{R}^{2/3}_2 N j$, and $\widetilde{R}^{-1/3}_2 N j$) at the HL-LHC. Among these, the single production modes $\widetilde{R}^{2/3}_2 N j$ and $\widetilde{R}^{-1/3}_2 N j$ at the HL-LHC are larger than their muon collider counterparts at both C.O.M energies and for all benchmark $M_N$ values considered.

    \item[--] In Fig.~\ref{fig:MR2tilde_CS_5TeV_1}, for a $\sqrt{s}=5$ TeV muon collider, the $\widetilde{R}^{2/3}_2 \mu j$ single production channel remains sub-dominant compared to the HL-LHC single production modes ($\widetilde{R}^{2/3}_2 \mu j$, $\widetilde{R}^{2/3}_2 N j$, and $\widetilde{R}^{-1/3}_2 N j$) for $M_{\widetilde{R}_2} < 1.75~\text{TeV}$. For $M_{\widetilde{R}_2} > 1.75~\text{TeV}$, the production at muon collider becomes more significant than the HL-LHC single production modes of $\widetilde{R}^{-1/3}_2 N j$, and $\widetilde{R}^{2/3}_2 \mu j$, but still remains lower than $\widetilde{R}^{2/3}_2 N j$ process for $M_{\widetilde{R}_2} < 3.5~\text{TeV}$. However, for a $\sqrt{s}=10~\text{TeV}$ (Fig.~\ref{fig:MR2tilde_CS_10TeV_1}) muon collider, the same process exceeds all the HL-LHC single production modes for $M_{\widetilde{R}_2} \gtrsim 2.2~\text{TeV}$ demonstrating that a muon collider can probe significantly heavier masses compared to the HL-LHC.

    \item[--] The dimuon production at the HL-LHC via a $t$-channel sLQ is denoted by the dashed pink line. It is sub-dominant compared to the dijet production (dashed yellow line) via an off-shell sLQ at the muon collider.
   
\end{itemize}

The major contribution to the signal comes from the pair production of $\widetilde{R}_{2}$ and the single production mode $\widetilde{R}^{2/3}_2 \mu j$. Thus, it is evident from Figs.~\ref{fig:MR2tilde_CS_5TeV_1} and \ref{fig:MR2tilde_CS_10TeV_1} that a muon collider can probe heavier masses compared to the HL-LHC, emphasizing its importance in the search for heavier BSM particles.

\section{Search strategy for $\widetilde{R}_{2}$ at the muon collider} 
\label{sec:6}

We begin by outlining the software tools utilized in our analysis. The Lagrangian presented in Eq.~\eqref{eq:eq2} is implemented using the \texttt{FeynRules}~\cite{Alloul:2013bka} package, which is then used to generate the corresponding Universal FeynRules Output (UFO) model file. Signal and background events are simulated at leading order using the Monte Carlo event generator \texttt{MadGraph5\_aMC@NLO-v3.5.3}~\cite{Alwall:2014hca}. The generated events are passed to \texttt{Pythia8}~\cite{Sjostrand:2006za} for parton showering and hadronization. Detector effects are modeled using~\texttt{Delphes3}~\cite{deFavereau:2013fsa} with the \texttt{Delphes ILD} detector card. Jet clustering is performed using the anti-$k_T$ algorithm~\cite{Soyez:2008pq}, with a radius parameter of $R = 0.4$, as implemented in \texttt{FastJet}~\cite{Cacciari:2011ma}. Both signal and background samples are generated in \texttt{MadGraph} with minimal generation-level cuts. Specifically, we impose a transverse momentum cut of $p_T > 10$~GeV on final-state muons and $p_T > 20$~GeV on final-state jets. In the Delphes simulation, muons are selected with the generic cuts of $|\eta| < 2.4$ on the pseudorapidity and $p_T > 10$~GeV, while jets are clustered with a transverse momentum threshold of $p_T > 20$~GeV.\\

In the following subsections, we present the search strategies used to assess the discovery potential of the $\widetilde{R}_2$ at a muon collider. We first examine the indirect search mode of $\widetilde{R}_2^{2/3}$, followed by the direct production channels of both components of $\widetilde{R}_2$. For each case, we discuss the signal topology, the relevant SM background processes, kinematic distributions, and the selection cuts applied to enhance signal significance.

\subsection{ Indirect search mode of $\widetilde{R}^{2/3}_2$} 
\label{sec:6A}

As discussed in the previous sections, $\widetilde{R}^{2/3}_2$ can mediate a $t$-channel dijet production process at a muon collider. The cross-section for this process scales with the fourth power of the coupling, i.e. $\sigma \propto Y_{12}^4$. 
Note that $\widetilde{R}^{-1/3}_2$ does not contribute to this process as it does not couple to a muon and a light quark. We list the relevant SM background processes below:
 
\begin{enumerate}
\item Dijet Production: This background arises from the production of dijets in the final-state through an $s$-channel mediated $\gamma$ or $Z$ boson.
\item Single gauge boson production ($V+X$): This category includes SM gauge boson production along with dijet/two charged leptons/two SM neutrinos or one SM neutrino and one lepton. The produced gauge boson decaying  to jets can mimic the dijet topology. The relevant backgrounds of this type are $
W_h \ell \nu_{\ell},\ 
Z_h  \ell \ell, \ 
Z_h \nu_{\ell} \bar{\nu}_{\ell} \ 
$. While generating these backgrounds, we add an invariant mass ($m_{\ell \ell}/m_{\ell \nu}$) cut on dileptons to ensure that these backgrounds are not arising from diboson background. Here, the subscripts $h$ indicate the hadronic ($W/Z\rightarrow j j$) decay modes of the gauge bosons. Implementing a lepton-veto can substantially reduce this background. 

\item Diboson production ($VV$): This category includes various SM diboson production channels whose decay products can mimic the dijet topology. The relevant diboson backgrounds are $W_h W_{\ell}, Z_\ell Z_{h}$ and $Z_h Z_h$. Here, $\ell$ denotes the leptonic decay ($W\rightarrow \ell \nu_\ell$ or $Z\rightarrow \ell^- \ell^+$) modes of the gauge bosons. Similarly, implementing a lepton-veto can substantially reduce this background. 

\item Top quark pair production ($t\bar{t}$): The pair-produced top quark $t_h t_{\ell}$, and $t_h t_h$, with the subsequent decays of top $t_h \to b W \to b jj$, $t_{\ell} \to b W \to b \ell \nu_{\ell}$ can mimic the signal. Implementing a lepton and $b$-veto can substantially reduce this background. 
\end{enumerate}

In our analysis, the \( V+X \) and \( VV \) background processes have been combined into a single category. The cross-sections of all background processes discussed above are listed in Table~\ref{tab:cross_sections_backgrounddijet}. Among these, the combined \( V+X \) and \( VV \) processes constitute the dominant background contribution. For the $jj$ and $t\bar t$ backgrounds, which proceed dominantly via $s$-channel $\gamma/Z$ exchange, the cross-section scales as $\sigma \propto 1/s$; consequently, away from the resonance region, the $\sqrt{s}=10\,\mathrm{TeV}$ collider yields a smaller cross-section than the $\sqrt{s}=5\,\mathrm{TeV}$ machine.

\begin{table}[h!]
\centering 
\begin{tabular}{|l|l|c|c|}
\hline
\textbf{Process} & \textbf{Description} & \textbf{$\sigma$ (pb) at $\sqrt{s} = 10$ TeV} & \textbf{$\sigma$ (pb) at $\sqrt{s} = 5$ TeV} \\ 
\hline
$VV, V+X$   & $\mu^+ \mu^- \to W_h W_{\ell}, Z_\ell Z_{h}, Z_h Z_h, W_h \ell \nu_{\ell}, Z_h \ell \ell, Z_h \nu_{\ell} \bar{\nu}_{\ell}$ & $1.58$                 & $1.40$ \\ 
\hline
$j j$       & $\mu^+ \mu^- \to j j $ via $Z/ \gamma$ & $5.28 \times 10^{-3}$  & $2.11 \times 10^{-2}$ \\ 
\hline
$t\bar{t}$  & $\mu^+ \mu^- \to t_h t_{\ell}, t_h t_h $ & $2.01 \times 10^{-4}$  & $1.22 \times 10^{-3}$ \\ 
\hline
\end{tabular}
\captionsetup{justification=raggedright,singlelinecheck=false} 
\caption{Cross-sections of the relevant SM background processes considered in the $\widetilde{R}^{2/3}_2$ indirect search. 
}
\label{tab:cross_sections_backgrounddijet}
\end{table}

\subsubsection{Kinematic distributions of signal and background for indirect probe and selection criterion}

\begin{figure*}[ht]
    \centering
    \captionsetup[subfigure]{labelformat=parens, labelfont=bf}

    \subfloat[]{
        \includegraphics[width=0.48\textwidth, height=0.35\textwidth]{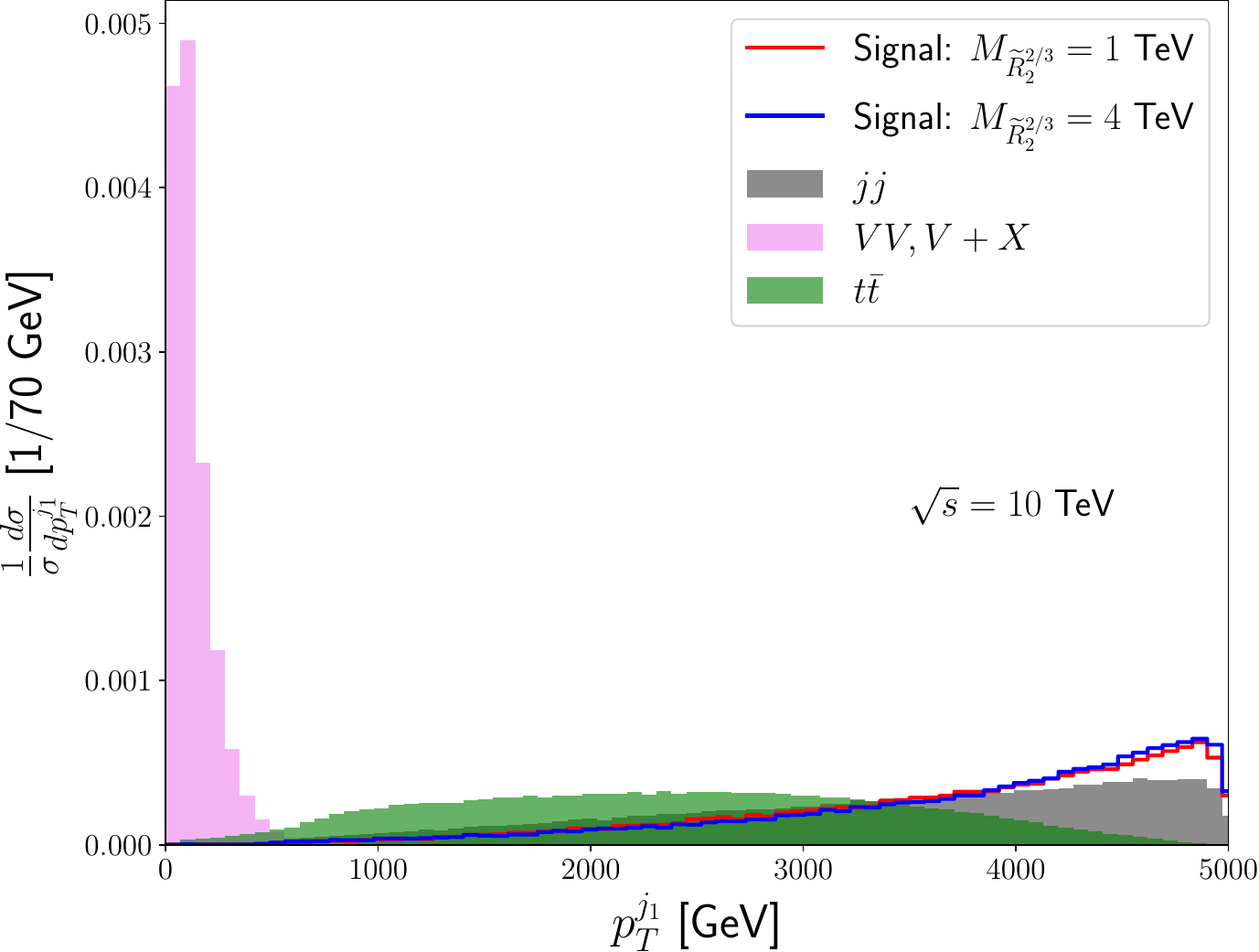}
        
    \label{fig:PTj1_10T_mumujj_28May}
    }
    \hfill
    \subfloat[]{
        \includegraphics[width=0.48\textwidth, height=0.35\textwidth]{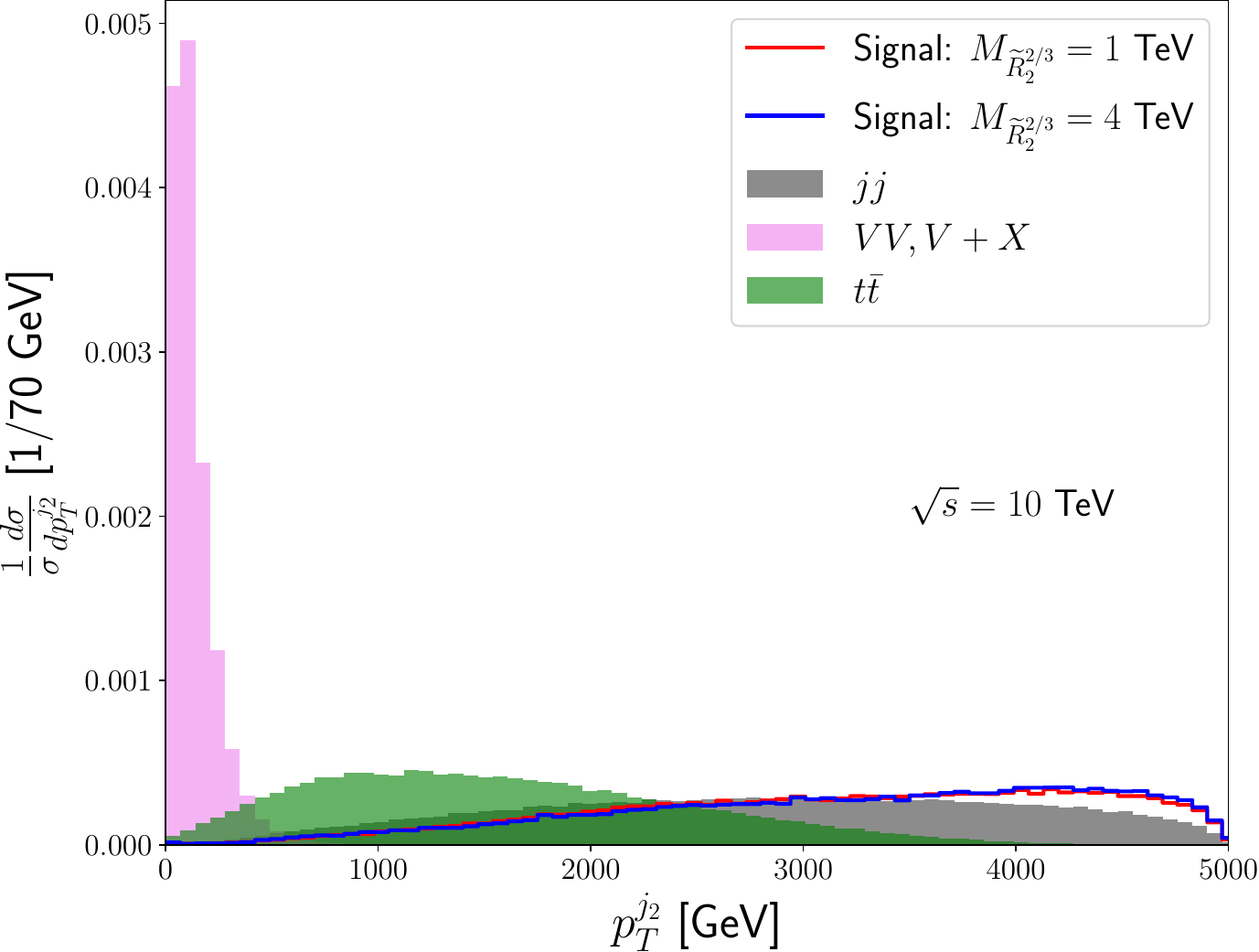}
       
    \label{fig:PTj2_10T_mumujj_28May}
    }
\captionsetup{justification=raggedright,singlelinecheck=false} 
    \caption{Distributions of the transverse momentum of the  (a) leading jet ($p_{T}^{j_{1}}$) and (b) the sub-leading jet ($p_{T}^{j_{2}}$) for the indirect search mode of $\widetilde{R}_{2}^{2/3}$, shown for two different benchmark values of $M_{\widetilde{R}_{2}^{2/3}}$ at  C.O.M. energy $\sqrt{s} = 10$~TeV. }
    \label{fig:kinematicdistribution_indirect}
\end{figure*}

We show the distributions of the kinematic variables that are essential to separate the signal from the SM backgrounds in Fig.~\ref{fig:kinematicdistribution_indirect}. In particular, we show the transverse momentum distributions of the leading ($p_T^{j_1}$) and sub-leading ($p_T^{j_2}$)  jets for the signal and background processes for the  C.O.M. energy $\sqrt{s}=10$ TeV, and benchmark sLQ masses: $M_{\widetilde{R}^{2/3}_2} = 1.0~\rm{TeV}$ (red solid line) and $M_{\widetilde{R}^{2/3}_2} = 4.0~\rm{TeV}$ (blue solid line). From Fig.~\ref{fig:PTj1_10T_mumujj_28May}, it is evident that the transverse momentum distribution of the leading jet ($p_T^{j_1}$) peaks around $5$ TeV for both sLQ masses.  A similar pattern is observed for the sub-leading jet ($p_T^{j_2}$) distribution shown in Fig.~\ref{fig:PTj2_10T_mumujj_28May}, where the peak occurs for $p_T < 5$ TeV. Overall, the multi-TeV transverse momentum of the leading and sub-leading jets for the signal occurs, as the dijet is produced through the $t$-channel $\widetilde{R}^{2/3}_2$ exchange, and thus the $\sqrt{s} = 10$ TeV collider energy is effectively distributed almost equally among the final-state jets. For a  C.O.M. energy of $\sqrt{s} = 5$~TeV, the overall shape of the distributions remains similar to those in the $\sqrt{s} = 10$~TeV case, except that the peak shifts towards lower $p_T$ and the tail of the distribution extends up to around $2.5$~TeV. 

Since $p_T^{j_{1}}$ and $p_T^{j_{2}}$ show a clear distinction between signal and SM backgrounds, we implement the following set of selection cuts.  

\begin{itemize}
\item 
Number of jets $N_j\ge 2$.
\item 
The transverse momentum of the leading jet, $p^{j_1}_T >3.5$ TeV ($2.0$ TeV) for a $\sqrt{s}=10$ TeV ($\sqrt{s}=5$ TeV) muon collider.
\item 
The transverse momentum of the sub-leading jet, $p^{j_2}_T >3.0$ TeV ($1.5$ TeV) for a $\sqrt{s}=10$ TeV ($\sqrt{s}=5$ TeV) muon collider.
\item 
A $b$ veto: number of $b$-jets $N_b=0$ to suppress the $t \bar{t}$ background.
\item Lepton veto: number of leptons $N_l=0$  in the final-state.
\end{itemize}

The cut-flow for the two benchmark mass points corresponding to $Y_{12} = 1.0\, (0.3)$, after applying the aforementioned cuts, is presented in 
Table~\ref{tab:cut_flow_indirect}. 
After these selection cuts, the effective signal cross-sections are found to be $18.49$ fb ($0.15$ fb) and $10.57$ fb ($0.085$ fb) for sLQ masses of $1.0~\text{TeV}$ and $4.0~\text{TeV}$, respectively, corresponding to couplings $Y_{12} = 1.0\, (0.3)$, at a $\sqrt{s} = 10~\text{TeV}$ muon collider. The effective background cross-section is found to be $1.56$ fb. Following the prescription mentioned in Section~\ref{sec:7}, we evaluate the significance of the signal. To achieve a $5\sigma$ sensitivity, the required coupling values $Y_{12}$ are $0.23$ and $0.34$ for the benchmark masses $M_{\widetilde{R}_{2}^{2/3}} = 1.0$ TeV and $4.0$ TeV, respectively.
In the subsequent Section~\ref{sec:7}, we will show the results for a large variation of $M_{\widetilde{R}_{2}^{2/3}}$.

\begin{table*}[hbt!]
	\centering
	 
	\begin{adjustbox}{width=17.80cm, height=1.4cm} 
		\addtolength{\tabcolsep}{-1pt}
		\begin{tabular}{||c|c|c|c|c|c|c||}
			\hline 
			& $N_b=0$  & $N_l=0$ &  $N_j\ge 2$ & $p^{j_1}_T >3.5$ TeV  & $p^{j_2}_T >3.0$ TeV & ${\sigma}_{\text{eff}}$[fb]   \\
			\hline
			$M_{\widetilde{R}_2^{2/3}} = 1.0$ TeV, $Y_{12} = 1.0$($Y_{12} = 0.3$) [$53.28$ ($4.31 \times 10^{-1}$) fb]&  35.21 ($2.85 \times 10^{-1}$)  & 35.18 ($2.84 \times 10^{-1}$) & 35.13 ($2.84 \times 10^{-1}$) & 22.62 ($1.83 \times 10^{-1}$)& 18.49 ($1.50 \times 10^{-1}$) &18.49 ($1.50 \times 10^{-1}$) \\
			
            $M_{\widetilde{R}_2^{2/3}} = 4.0$ TeV, $Y_{12} = 1.0$($Y_{12} = 0.3$) [$29.28$ ($2.37 \times 10^{-1}$) fb] &  19.25 ($1.56 \times 10^{-1}$)  & 19.22 ($1.55 \times 10^{-1}$)& 19.20 ($1.55 \times 10^{-1}$)& 12.85 ($1.04 \times 10^{-1}$) &  10.57 ($8.56 \times 10^{-2}$)& 10.57 ($8.56 \times 10^{-2}$)\\
			\hline \hline
			$VV, V+X$  [$1586$ fb]     & 1475.72 & 1440.19 & 924.54 & $6.47 \times 10^{-1}$ & $3.17 \times 10^{-2}$ & $3.17 \times 10^{-2}$ \\
			$jj$  [$5.28$ fb]      & 3.66 & 3.66 & 3.65 & 1.93  & 1.52 & 1.52    \\
			$t\bar{t}$  [$2.01 \times 10^{-1}$ fb]  & $9.25 \times 10^{-2}$ & $8.64 \times 10^{-2}$ & $8.61 \times 10^{-2}$ & $1.88 \times 10^{-2}$ & $6.85 \times 10^{-3}$ & $6.85 \times 10^{-3}$  \\
			\hline 
		\end{tabular}
	\end{adjustbox}\captionsetup{justification=raggedright,singlelinecheck=false} 
	\caption{Cut flow using the selection cuts mentioned in Section~\ref{sec:6A} for $\sqrt{s} = 10$ TeV muon collider.}
	\label{tab:cut_flow_indirect}
\end{table*}       

\subsection{Direct production of $\widetilde{R}_2$}
\label{sec:6B}

As discussed in Section~\ref{sec:5}, the direct production of $\widetilde{R}_2$, i.e., pair or single production of $\widetilde{R}_{2}^{2/3}$ and $\widetilde{R}_{2}^{-1/3}$ states, and their subsequent decays can lead to dimuon and multi-jet final-state. As discussed previously, in our analysis we demand a high jet multiplicity $N_j \geq 4$. We consider signals that receive contributions  both from pair and single production channels of $\widetilde{R}_2$ in  the symmetric and asymmetric modes; see Eqs.~\eqref{eq:finalsym} and~\eqref{eq:finalasymm}. The dominant contribution to the signal arises from the channels $\widetilde{R}_{2}^{+2/3} \widetilde{R}_{2}^{-2/3}$, $\widetilde{R}_{2}^{-1/3} \widetilde{R}_{2}^{+1/3}$, and $\widetilde{R}_{2}^{ 2/3} \mu d$ followed by the subsequent decay of at least one of the $\widetilde{R}_2$ states to a $N$ and a light quark. Additional channels such as $\widetilde{R}_{2}^{2/3} Nu$ followed by the decay modes $\widetilde{R}_{2}^{2/3} \to \mu d/u N$ and $N \to \mu u d$ give sub-dominant contributions over a large portion of the assumed parameter space, except near the kinematic/collider threshold. This occurs due to suppression in  production cross-section of such configurations and the corresponding BRs.

\subsubsection{Potential Backgrounds for direct production of $\widetilde{R}_2$}
 
A number of  SM  processes including $VV+X, \mu \mu+ \rm{jets}$ can mimic the signal.
\begin{enumerate}
\item Dimuon production $(\mu\mu) +$ jets: This background comprises of  dimuon plus additional jets in the final-state. We combine contributions from the following possible modes to obtain such a final-state. $\mu^+\mu^- \rightarrow Z_h + \mu^+\mu^-$, $Z_{\ell}+\rm{jets}$, $Z_{\ell}+Z_{h}$ and ${\mu^+\mu^-}_{\text{excl}} + j j$. Here, $Z_{h}$ and $Z_{\ell}$ denote the hadronic ($Z\rightarrow j j$) and leptonic ($Z\rightarrow\mu^+\mu^-$) decays of the $Z$ boson, respectively. In the case of the ${\mu^+\mu^-}_{\text{excl}} + jj$ final-state, we generate the process $\mu^+\mu^- \rightarrow {\mu^+\mu^-} + jj$ exclusively with an invariant mass cut of $M_{\mu\mu} > 100$~GeV and $M_{jj} > 100$~GeV at the generation level, in order to suppress contributions from on-shell $Z \to \mu^+ \mu^-$ and $Z \to jj$ decays to avoid double counting.

\item Diboson production $VV + X $ ($V = W, Z$): The diboson production in association with jets/muons and their decay to the SM final-state can contribute as a background. The different background processes are $W_{\ell}W_{\ell} +$ jets, $W_{h}W_{h} + \mu\mu $, $Z_{\ell}Z_{h} +$ jets, $Z_{h}Z_{h} + \mu \mu$, and $Z_{\ell}W_{h} +$ jets, with subsequent decays of $W_{\ell}, Z_{\ell}$ into leptonic and $W_{h}, Z_{h}$ into hadronic final-states.

In addition to the above processes, the other sub-dominant backgrounds are as follows: The production of a $W$ boson in association with jets is a potential background, where the leptonically decaying $W$ boson provides one muon, and a second muon could arise from a jet being misidentified. However, given the inherently low probability that a jet is misidentified as a muon, the contribution from this background channel is negligible. {Background events featuring leptonically decaying top-quark pairs ($t_{\ell}\bar{t}_{\ell}$) in association with jets also represent a potential source of background, as the ditop system naturally produces a final-state dimuon signature through leptonic decays. However, the corresponding cross-section is relatively low, with values of $4.7 \times 10^{-5}$~pb ($1.10 \times 10^{-4}$~pb) at a  C.O.M. energy of $\sqrt{s} = 10$~TeV ($5$~TeV), respectively. Furthermore, applying a \( b \)-veto effectively suppresses this background to a negligible level, allowing it to be safely ignored in the analysis. Similarly, the $t_h\bar{t}_h Z_{\ell}$ ($t_\ell\bar{t}_\ell Z_{h}$) process, involving a hadronically (leptonically) decaying top-quark pair and a leptonically (hadronically) decaying $Z$ boson, could potentially mimic the signal topology. This process also has a relatively low cross-section of $6.01 \times 10^{-6}$~pb ($1.65 \times 10^{-5}$~pb) at $\sqrt{s} = 10$~TeV ($5$~TeV). Similar  to the $t_{\ell}\bar{t}_{\ell}+$ jets scenario, implementing $b$ veto  ensures that the contribution from this channel is also rendered negligible.
}

\end{enumerate}
 
In Table~\ref{tab:cross_sections_background}, we list all the dominant background processes relevant to our analysis and have skipped the processes that contribute negligibly. As evident from the table, the dominant background contributions arise from the $(\mu\mu$ + jets) and $(W_h W_h$ + $\mu\mu)$ channels.

\begin{table}[h!]
\centering

\begin{tabular}{|l|l|c|c|}
\hline
\textbf{Process} & \textbf{Description} & \textbf{$\sigma$ (pb) at $\sqrt{s} = 10$ TeV} & \textbf{$\sigma$ (pb) at $\sqrt{s} = 5$ TeV} \\
\hline

$\mu\mu$ + jets & $\mu^+ \mu^- \to (\mu\mu)_{\text{excl}} jj$, $Z_{\ell}$ + jets, $Z_{h}$ + $\mu\mu$, $Z_{\ell}+ Z_{h}$ & $6.52 \times 10^{-3}$ & $1.51 \times 10^{-2}$ \\
\hline
$W_{h}W_h$ + $\mu \mu$ & $\mu^+ \mu^- \to W^+ W^- \mu^+ \mu^-$, $(W \to jj)$ & $3.73 \times 10^{-3}$ & $6.1 \times 10^{-3}$ \\
\hline
$W_hZ_{\ell}$ + jets & $\mu^+ \mu^- \to W Z jj$, $(Z \to \mu^+ \mu^-)$, $(W \to jj)$ & $1.17 \times 10^{-4}$ & $3.21  \times 10^{-6}$ \\
\hline
$W_{\ell}W_{\ell}$ + jets & $\mu^+ \mu^- \to W^+ W^- jj$, $(W \to \mu \nu_\mu)$ & $3.33 \times 10^{-5}$ & $8.27 \times 10^{-5}$ \\
\hline
$Z_hZ_h$ + $\mu \mu$ & $\mu^+ \mu^- \to ZZ \mu^+ \mu^-$, $(Z \to jj)$ & $1.98 \times 10^{-5}$ & $2.82 \times 10^{-5}$ \\
\hline
$Z_{\ell}Z_{h}$ + jets & $\mu^+ \mu^- \to ZZ jj$, $(Z \to \mu^+ \mu^-)$, $(Z \to jj)$ & $2.57 \times 10^{-6}$ & $7.76 \times 10^{-6}$ \\
\hline

\end{tabular}
\captionsetup{justification=raggedright,singlelinecheck=false} 
\caption{Cross-sections of different SM backgrounds which can mimic dimuon+multi-jet signal in direct production mode search.} 
\label{tab:cross_sections_background}
\end{table}

\subsubsection{Kinematic distributions of signal and background for direct production}

\begin{figure*}
\centering
\captionsetup[subfigure]{labelformat=empty}

\subfloat[\quad\quad\quad(a)]{\includegraphics[width=0.5\textwidth,height=0.35\textwidth]{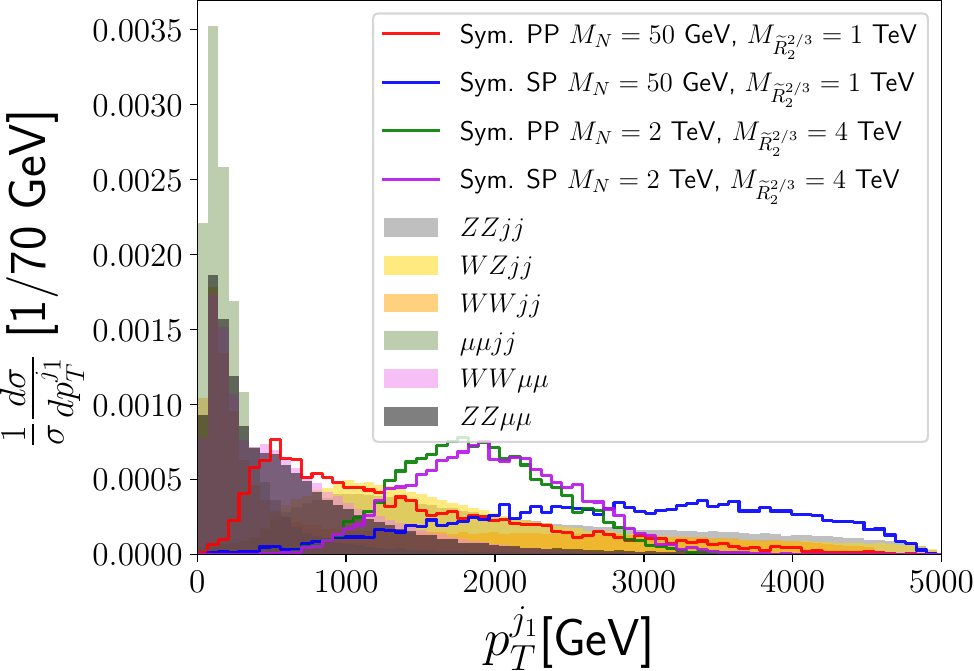}\label{fig:PTj1_sym}}
\subfloat[\quad\quad\quad(b)]{\includegraphics[width=0.5\textwidth,height=0.35\textwidth]{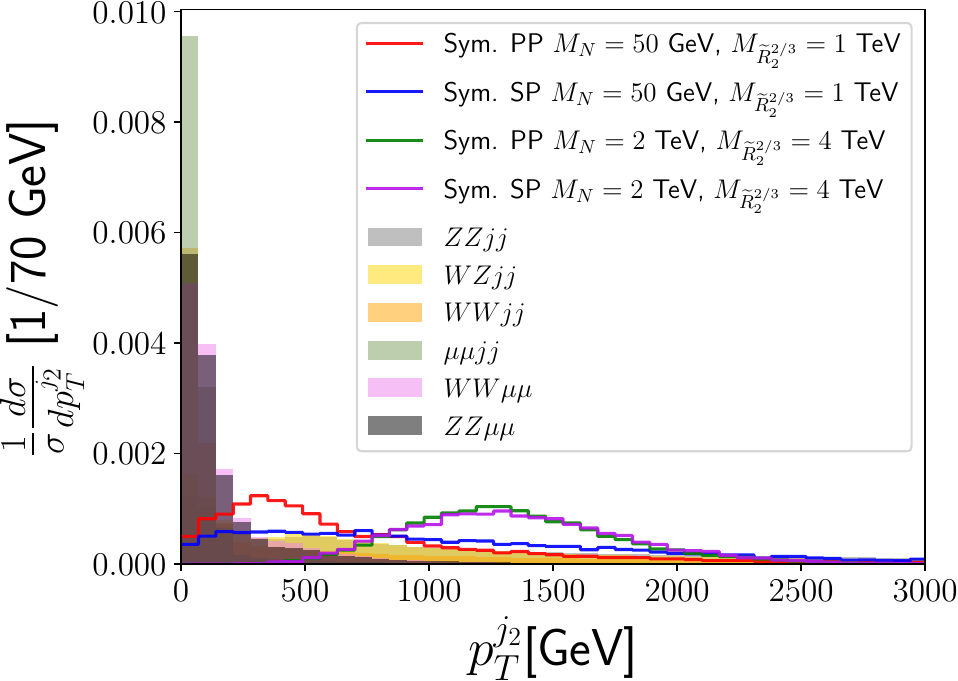}\label{fig:PTj2_sym}}

\subfloat[\quad\quad\quad(c)]{\includegraphics[width=0.5\textwidth,height=0.35\textwidth]{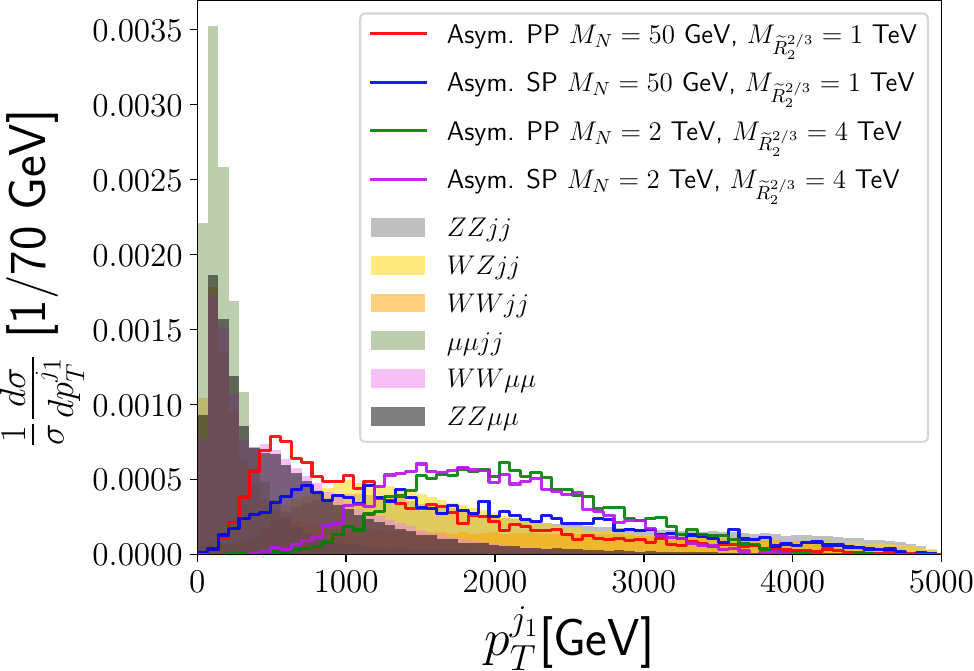}\label{fig:PTj1_asym}}
\subfloat[\quad\quad\quad(d)]{\includegraphics[width=0.5\textwidth,height=0.35\textwidth]{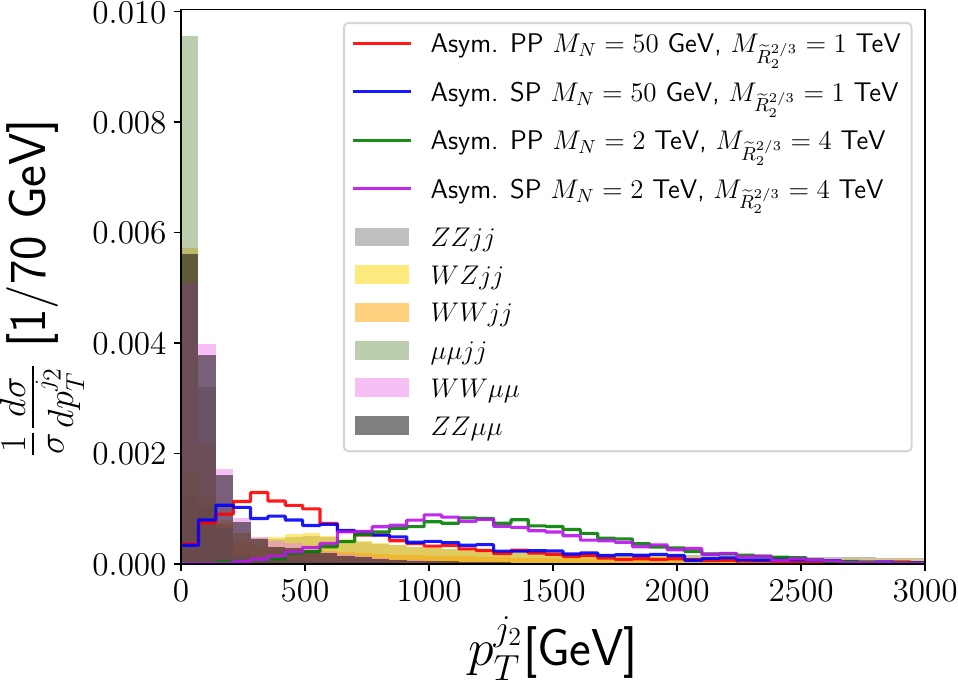}\label{fig:PTj2_asym}}
\captionsetup{justification=raggedright,singlelinecheck=false} 
\caption{Kinematic distributions for the symmetric (a, b) and asymmetric (c, d) modes for different choices of $M_{\widetilde{R}^{2/3}}$ and $M_N$. Figures (a) and (c) correspond to the leading $p_T^{j_1}$ distributions, while figures (b) and (d) display the sub-leading $p_T^{j_2}$ distributions.}
\label{fig:kinematicdistribution}
\end{figure*}

In Fig.~\ref{fig:kinematicdistribution}, we present the distributions of the relevant kinematic variables for the signal and background processes, which are the transverse momentum of the leading ($p_T^{j_1}$) and sub-leading ($p_T^{j_2}$) jets. The choice of these kinematic variables is motivated by the clear separation between signal and background event distributions.  
We present the kinematic distributions for the symmetric (pair) and asymmetric (single) production modes of sLQ separately, for the following benchmark mass points at a C.O.M. energy of $\sqrt{s} = 10$~TeV:
\begin{itemize}
    \item [--] $M_{\widetilde{R}^{2/3}_2} = 1.0~\text{TeV}$ and $M_{N} = 50~\text{GeV}$,
    \item [--] $M_{\widetilde{R}^{2/3}_2} = 4.0~\text{TeV}$ and $M_{N} = 2.0~\text{TeV}$.
\end{itemize}

The kinematic distributions of $\widetilde{R}^{-1/3}_2$ are largely similar to those of $\widetilde{R}^{2/3}_2$ except for asymmetric production where it does not contribute, and therefore, we do not present them separately.
~ 
Figs.~\ref{fig:PTj1_sym},   \ref{fig:PTj2_sym} present symmetric mode distributions and Figs.~\ref{fig:PTj1_asym}, \ref{fig:PTj2_asym} present asymmetric mode. In all of these figures, the red (blue) and green (violet) curves correspond to pair (single) production for the first and second mass points, respectively.
For Fig.~\ref{fig:PTj1_sym}, since the leading jet ($j_1$) for symmetric pair production originates from the decay of $\widetilde{R}_{2}^{2/3}$ with a mass of $1.0$ TeV and $4.0$ TeV, its momentum distribution peaks around $500$~GeV and $2.0$ TeV, respectively. A similar behavior is observed in the pair production channel for the asymmetric mode in Fig.~\ref{fig:PTj1_asym}.
 
For sLQ mass $1.0$ TeV, the momentum distribution for the symmetric-single production mode $\mu^+\mu^- \to \widetilde{R}_2^{\pm 2/3} N \bar{u}/N {u} $ peaks near $3.3$~TeV and extends up to approximately $5$~TeV. Notably, this distribution is broader and flatter relative to the others, reflecting a less sharply defined peak and a wider spread in momenta.
This occurs because the jet produced alongside $\widetilde{R}_{2}^{\pm 2/3}$ tends to carry away a larger momentum, as mass of the RHN is only 50 GeV. For the other heavier mass point, sLQ mass of $4.0$ TeV and RHN mass of 2.0 TeV, the $p_T$ distribution of the leading jet represented by the magenta line on the contrary has a peak at about $1.9$ TeV, resulting in less momentum being shared with the jet. 
The asymmetric production mode $\mu^+ \mu^- \to \widetilde{R}_2^{\pm 2/3} \mu d/\mu \bar{d} $ also exhibits a similar behavior with a long tail. 
For the sub-leading jet, the overall pattern of the  distributions in Fig.~\ref{fig:PTj2_sym} and Fig.~\ref{fig:PTj2_asym} mimic the distributions of the leading jet.

\subsubsection{Selection criteria for the direct production of $\widetilde{R}_2$}
 
{After analyzing the kinematic distributions shown in Fig.~\ref{fig:kinematicdistribution}, we implement the following set of cuts, which provide effective discrimination between signal and background processes across the entire range of $M_{\widetilde{R}_{2}}$ and $M_{N}$. }

\begin{itemize}
    \item The number of muons and jets: $N_{\text{muon}} \geq 2$ and  $N_{\text{\rm jet}} \geq 4$. Since for relatively light $M_N$, the produced muon will be closer to the jets originated from the RHN, we do not demand any specific isolation criterion for the muon. 
    \item Among the jets, $N_{\text{jets}}(p_T^j > 400~\text{GeV},~\Delta R_{\mu j} > 0.4) \geq 2$.

\end{itemize}
Table~\ref{tab:cut_flow_direct} presents the cut-flow for the considered mass points, for which the after-cut effective signal cross-sections are $47.88$ fb ($0.38$ fb) and $5.64$ fb ($0.041$ fb) for the first and second benchmarks, respectively, at a $\sqrt{s}=10$~TeV muon collider with $Y_{12}=1.0\,(0.3)$.
 The effective background cross-section is $0.026$ fb. Following the prescription mentioned in Section~\ref{sec:7}, we evaluate the significance of the signal. The required values of the coupling $Y_{12}$ to achieve a $5\sigma$ discovery are found to be below $10^{-2}$ for both illustrative mass points. The selection strategy is generic and effectively probes both symmetric and asymmetric production modes of $\widetilde{R}_{2}$ in almost the entire parameter space. While our approach is most generic, further improvement can be achieved by the reconstruction of  $N$, which would enhance signal sensitivity.

\begin{table*}[hbt!]
	\centering
	 
	\begin{adjustbox}{width=18.0cm, height=2.8cm} 
		\addtolength{\tabcolsep}{-1pt}
		\begin{tabular}{||c|c|c|c|c||}
			\hline 
			& $N_{\text{muon}} \geq 2$,$N_{\text{\rm jet}} \geq 4$   & $N_{\text{jets}}(p_T^j > 400~\text{GeV},~\Delta R_{\mu j} > 0.4) \geq 2$ &  ${\sigma}_{\text{eff}}$[fb]   \\
			\hline
			Sym PP: $M_{\widetilde{R}_2} = 1.0$ TeV,$M_{N} = 50$ GeV, $Y_{12} = 1.0$($Y_{12} = 0.3$) [$24.60$ ($6.32 \times 10^{-1}$) fb]&  18.99 ($4.88 \times 10^{-1}$)  & 11.78 ($3.02 \times 10^{-1}$) & 11.78 ($3.02 \times 10^{-1}$)  \\
			
            Sym PP: $M_{\widetilde{R}_2} = 4.0$ TeV,$M_{N} = 2.0$ TeV, $Y_{12} = 1.0$($Y_{12} = 0.3$) [$1.18$ ($3.61 \times 10^{-2}$) fb] &  $9.27 \times 10^{-1}$ ($2.84 \times 10^{-2}$)  & $9.25 \times 10^{-1}$ ($2.83 \times 10^{-2}$)& $9.25 \times 10^{-1}$ ($2.83 \times 10^{-2}$) \\
            Asym PP: $M_{\widetilde{R}_2} = 1.0$ TeV,$M_{N} = 50$ GeV, $Y_{12} = 1.0$($Y_{12} = 0.3$) [$99.52$ ($2.04 \times 10^{-1}$) fb] &  55.75 ($1.14 \times 10^{-1}$)  & 35.49 ($7.27 \times 10^{-2}$)& 35.49 ($7.27 \times 10^{-2}$) \\
            Asym PP: $M_{\widetilde{R}_2} = 4.0$ TeV,$M_{N} = 2.0$ TeV, $Y_{12} = 1.0$($Y_{12} = 0.3$) [$8.39$ ($1.41 \times 10^{-2}$) fb] &  4.80 ($8.06 \times 10^{-3}$)  & 4.72 ($7.93 \times 10^{-3}$) & 4.72 ($7.93 \times 10^{-3}$) \\          

			Sym SP: $M_{\widetilde{R}_2} = 1.0$ TeV,$M_{N} = 50$ GeV, $Y_{12} = 1.0$($Y_{12} = 0.3$) [$2.48 \times 10^{-3}$ ($4.31\times 10^{-3}$) fb]&  $1.97 \times 10^{-3}$ ($3.43 \times 10^{-3}$)  & $1.53 \times 10^{-3}$ ($2.67 \times 10^{-3}$) & $1.53 \times 10^{-3}$ ($2.67 \times 10^{-3}$)  \\
			
            Sym SP: $M_{\widetilde{R}_2} = 4.0$ TeV,$M_{N} = 2.0$ TeV, $Y_{12} = 1.0$($Y_{12} = 0.3$) [$6.74 \times 10^{-5}$ ($1.02 \times 10^{-4}$) fb] &  $5.34 \times 10^{-5}$ ($8.09 \times 10^{-5}$)  & $5.33 \times 10^{-5}$ ($8.07 \times 10^{-5}$)& $5.33 \times 10^{-5}$ ($8.07 \times 10^{-5}$) \\
            Asym SP: $M_{\widetilde{R}_2} = 1.0$ TeV,$M_{N} = 50$ GeV, $Y_{12} = 1.0$($Y_{12} = 0.3$) [$1.16$ ($7.12 \times 10^{-2}$) fb] &  $5.01 \times 10^{-1}$ ($3.08 \times 10^{-2}$)  & $3.35 \times 10^{-1}$ ($2.06 \times 10^{-2}$) & $3.35 \times 10^{-1}$ ($2.06 \times 10^{-2}$) \\
            Asym SP: $M_{\widetilde{R}_2} = 4.0$ TeV,$M_{N} = 2.0$ TeV, $Y_{12} = 1.0$($Y_{12} = 0.3$) [$1.55 \times 10^{-1}$ ($4.29 \times 10^{-3}$) fb] &  $1.06 \times 10^{-1}$ ($2.95 \times 10^{-3}$)  & $1.04 \times 10^{-1}$ ($2.90 \times 10^{-3}$) & $1.04 \times 10^{-1}$ ($2.90 \times 10^{-3}$) \\            
            
			\hline \hline
			$\mu\mu$ + jets  [ 6.52 fb]     & $2.20 \times 10^{-2}$ &  $1.25 \times 10^{-2}$ & $1.25 \times 10^{-2}$  \\
			$W_{h}W_h$ + $\mu \mu$  [ 3.73 fb]      & $4.18 \times 10^{-1}$ & $1.02 \times 10^{-2}$  &  $1.02 \times 10^{-2}$       \\
			$W_hZ_{\ell}$ + jets  [ $1.17 \times 10^{-1}$ fb]  &$3.50 \times 10^{-3}$ & $2.06 \times 10^{-3}$ & $2.06 \times 10^{-3}$       \\
           $W_{\ell}W_{\ell}$ + jets  [ $3.33 \times 10^{-2}$ fb]  & $1.27 \times 10^{-3}$& $8.97 \times 10^{-4}$ &  $8.97 \times 10^{-4}$    \\
           $Z_hZ_h$ + $\mu \mu$  [ $1.98 \times 10^{-2}$  fb]  & $1.45 \times 10^{-3}$ & $5.47 \times 10^{-5}$ & $5.47 \times 10^{-5}$      \\
             $Z_{\ell}Z_{h}$ + jets  [ $2.57 \times 10^{-3}$ fb]  & $5.15 \times 10^{-4}$ & $4.14 \times 10^{-4}$ & $4.14 \times 10^{-4}$       \\
			\hline 
		\end{tabular}
	\end{adjustbox}
    \captionsetup{justification=raggedright,singlelinecheck=false} 
	\caption{Cut flow using the selection cuts mentioned in Section~\ref{sec:6B} for $\sqrt{s} = 10$ TeV muon collider in direct production mode search.}
	\label{tab:cut_flow_direct}
\end{table*}

\section{Methodology} 
\label{sec:7}
\noindent
In this section, we describe the methodology used to calculate the significance and sensitivity of the direct and indirect search mode separately. For each case, the statistical significance $\mathcal{Z}$ is calculated using the following expression outlined in Ref.~\cite{Cowan:2010js}:
\begin{align}
\mathcal{Z} = \sqrt{2\left(N_S+N_B\right)\ln\left(\frac{N_S+N_B}{N_B}\right)-2N_S}\, ,
\label{eq:sig}
\end{align} 
Here, $N_S$ and $N_B$ are the number of signal and background events, respectively.
The total number of background events are computed as follows:
\begin{align}
\label{eq:Numevents_bkg}
N_B = \left(\sum_{i}\sigma_{B}^{i} \times \epsilon_{B}^{i}\right) \times \mathcal{L},    
\end{align}
\noindent
Here, $\sigma_{B}^{i}$ and $\epsilon_{B}^{i}$ denote the cross-section and cut efficiency of the $i^{\rm th}$ background process, respectively and $\mathcal{L}$ is the integrated luminosity for which we use 3 ab$^{-1}$ and 10 ab$^{-1}$ at $\sqrt s=5$ TeV and 10 TeV muon collider, respectively. 

As for the signals, we first compute the number of signal events for the indirect search mode:
\begin{align}
    N^{\rm ind}_S = Y^4_{12} \times \sigma_{\rm ind} \times \epsilon_{\rm ind} \times \mathcal{L}.
\end{align}
Here, $\sigma_{\rm ind}$  denotes the cross-section of the  channel $\mu^+ \mu^- \to j j$   for $Y_{12}=1.0$  and  $\epsilon_{\rm ind}$ represents the efficiency obtained after applying the selection cuts described in Section~\ref{sec:6A}.

The number of signal events for the direct production mode of $\tilde{R}_2$ depends on both symmetric and asymmetric modes, as well as pair and single production. This can be expressed as follows:
\begin{align}
\label{eq:Numevents_direct}
N^{\rm dir}_S &= \left( N_{\text{pair}}^{\text{sym}} + N_{\text{single}}^{\text{sym}} + N_{\text{pair}}^{\text{asym}} + N_{\text{single}}^{\text{asym}} \right) \, .
\end{align}
Here, the individual contributions $N_{\text{pair}}^{\text{sym}}$, $N_{\text{single}}^{\text{sym}}$, $N_{\text{pair}}^{\text{asym}}$, and $N_{\text{single}}^{\text{asym}}$ represent the number of events for the symmetric pair production mode, symmetric single production mode, asymmetric pair production mode, and asymmetric single production mode, respectively. Their expressions are given below:
\begin{align}
N_{\text{pair}}^{\text{sym}} &= \sigma_{\text{pair}}^{\text{sym}}(\mu^+ \mu^- \to \widetilde{R}_2 \widetilde{R}_2 ) \times {\rm BR}^2(\widetilde{R}_2 \rightarrow Nj) \times {\rm BR}^2(N \rightarrow \mu jj) \times \epsilon_{\text{pair}}^{\text{sym}}\times \mathcal{L} \,, \\
N_{\text{pair}}^{\text{asym}} &= \sigma_{\text{pair}}^{\text{asym}} (\mu^+ \mu^- \to \widetilde{R}_2^{+2/3} \widetilde{R}_2^{-2/3})\times {\rm BR}(\widetilde{R}_2^{+2/3} \rightarrow Nj/\mu j) \times {\rm BR}(\widetilde{R}_2^{-2/3} \rightarrow \mu j/Nj) \times {\rm BR}(N \rightarrow \mu jj) \nonumber \\
& \hspace{12cm} \times \epsilon_{\text{pair}}^{\text{asym}} \times \mathcal{L}\,, \\
N_{\text{single}}^{\text{sym}} &= \sigma_{\text{single}}^{\text{sym}} (\mu^+ \mu^- \to \widetilde{R}_2 j N)\times {\rm BR}(\widetilde{R}_2 \rightarrow Nj) \times {\rm BR}^2(N \rightarrow \mu jj) \times \epsilon_{\text{single}}^{\text{sym}} \times \mathcal{L}\,, \\
N_{\text{single}}^{\text{asym}} &= \sigma_{\text{single}}^{\text{asym}} (\mu^+ \mu^- \to \widetilde{R}_2^{2/3}\mu j, \widetilde{R}_2^{2/3}j N)\times {\rm BR}(\widetilde{R}_2^{2/3} \rightarrow Nj/\mu j) \times {\rm BR}(N \rightarrow \mu jj) \times \epsilon_{\text{single}}^{\text{asym}} \times \mathcal{L}\,.
\end{align}

\begin{figure}[h]
\centering
\subfloat[]{\includegraphics[width=0.49\textwidth,height=0.40\textwidth]{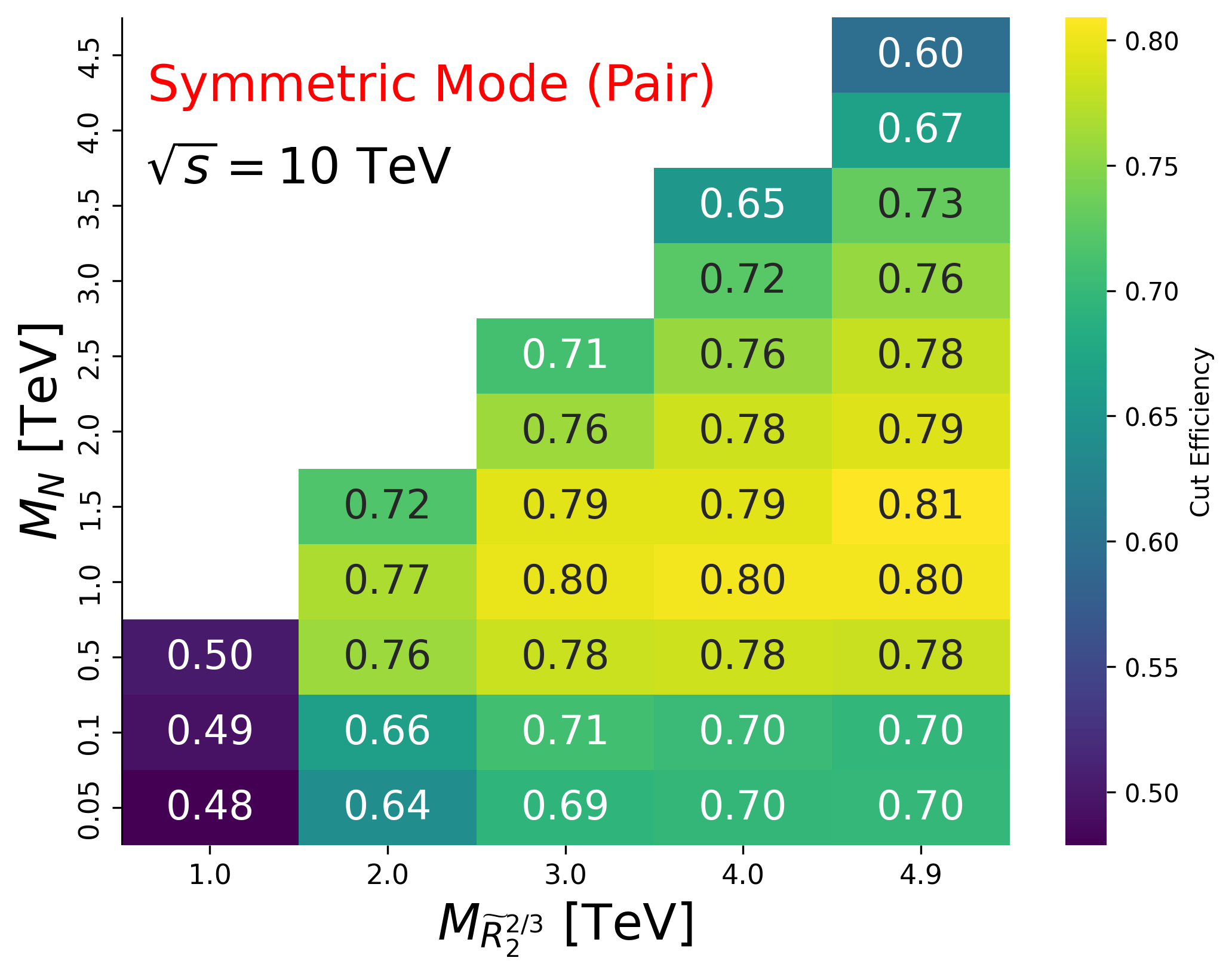}
\label{fig:HeatMap_MN_MLQ_PP_Sym_cut_eff_10TeV_collider_30May}}
\subfloat[]{\includegraphics[width=0.49\textwidth,height=0.40\textwidth]{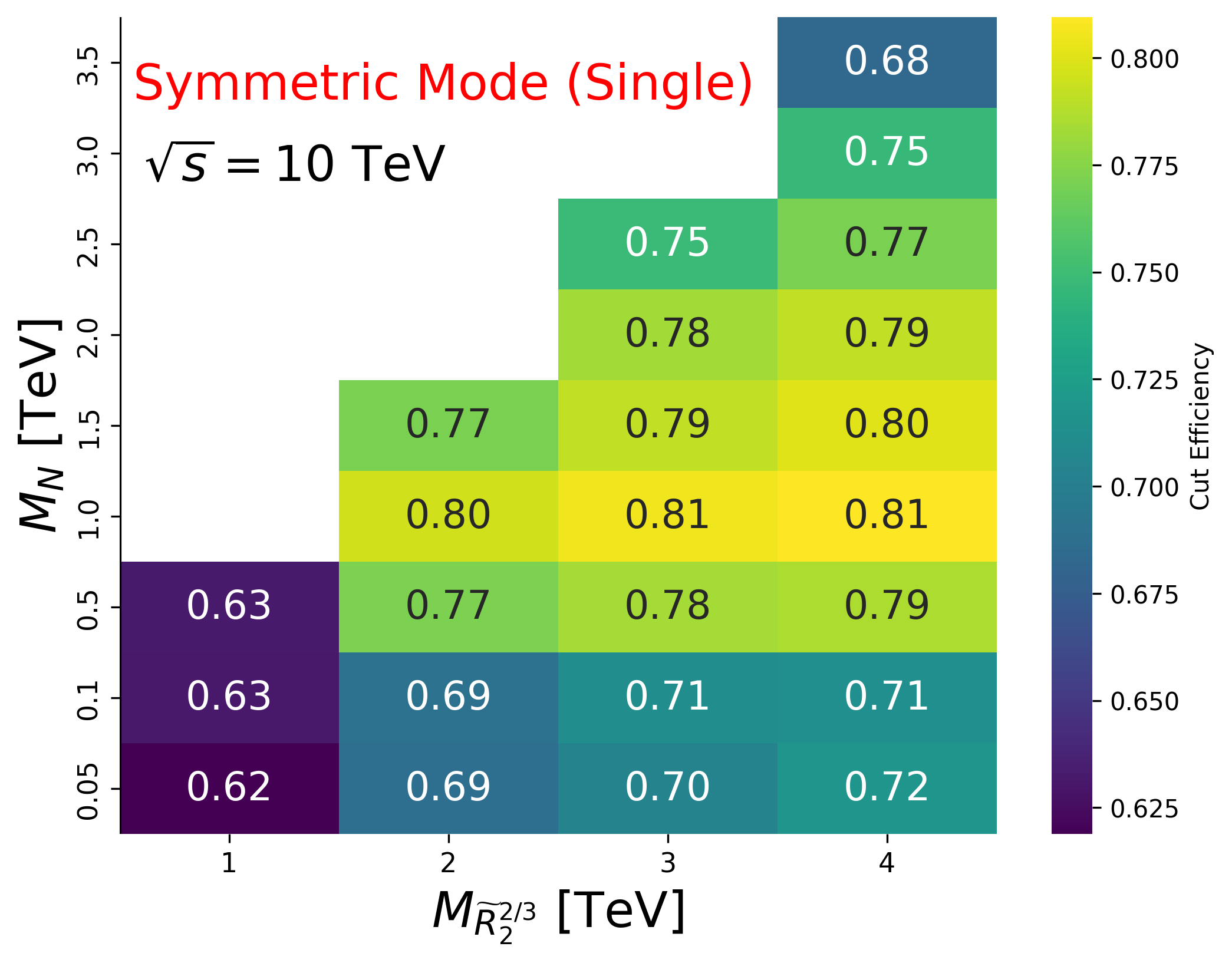}
\label{fig:HeatMap_MN_MLQ_SP_Sym_cut_eff_10TeV_collider_30May}}  \\
\subfloat[]{\includegraphics[width=0.49\textwidth,height=0.40\textwidth]{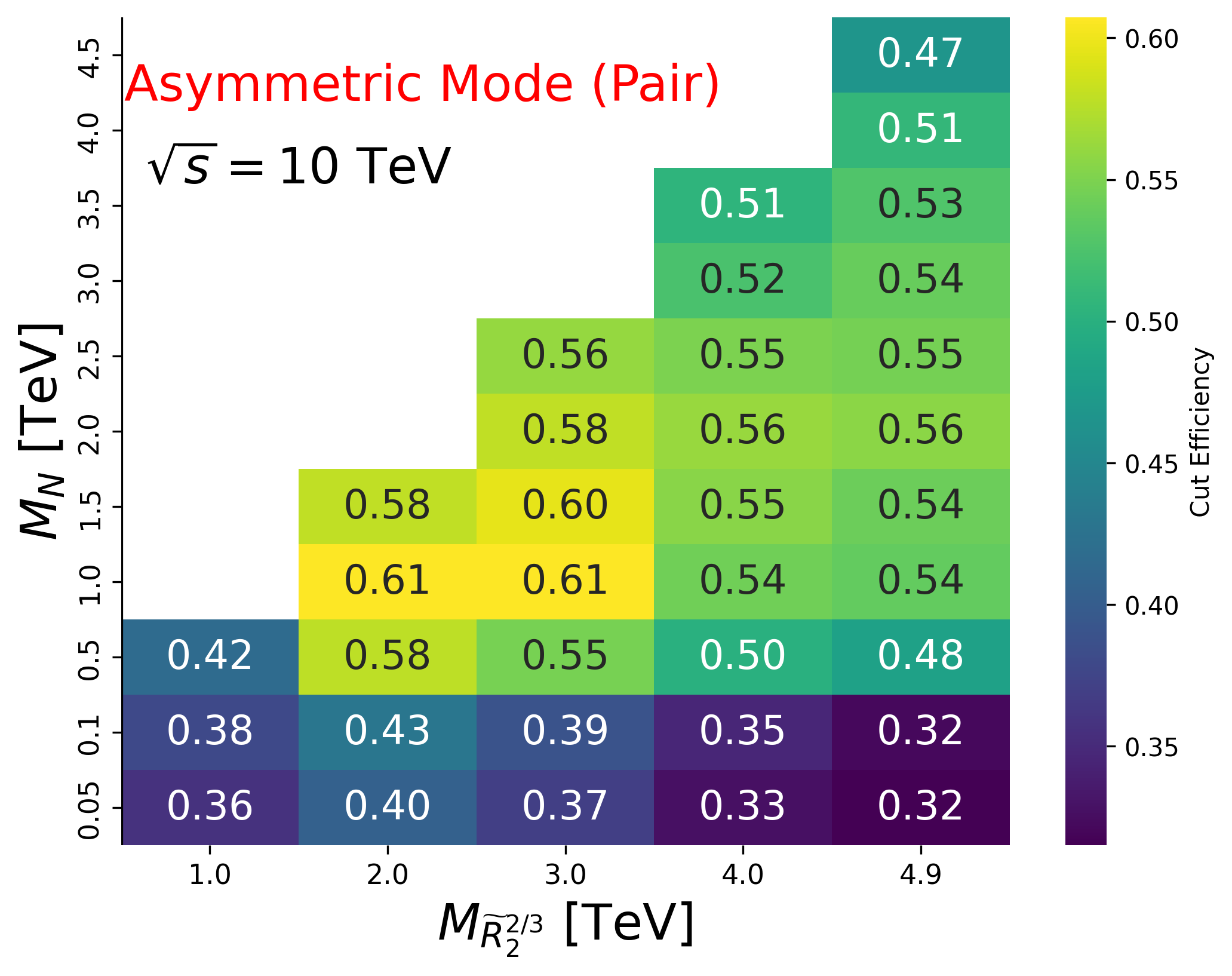}
\label{fig:HeatMap_MN_MLQ_PP_Asym_cut_eff_10TeV_collider_30May}}
\subfloat[]{\includegraphics[width=0.49\textwidth,height=0.40\textwidth]{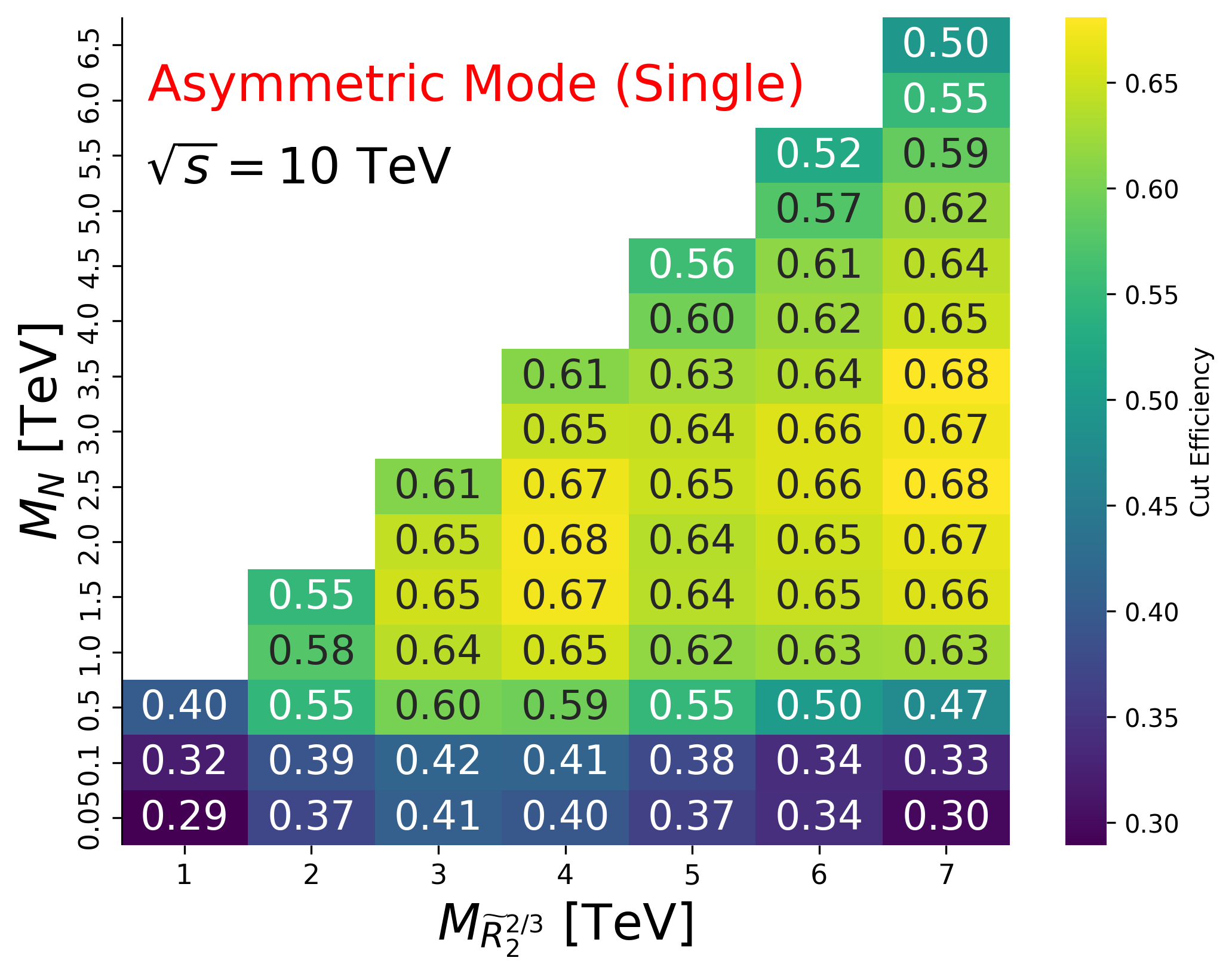}
\label{fig:HeatMap_MN_MLQ_SP_ASym_cut_eff_10TeV_collider_30May}}
\captionsetup{justification=raggedright,singlelinecheck=false} 
\caption{Cut efficiencies for the different production modes in the $M_{\widetilde{R}_2^{2/3}}-M_N$ plane for a $\sqrt{s} = 10~\text{TeV}$ muon collider in the (a) symmetric pair production mode, (b) symmetric single production mode, (c) asymmetric pair production mode, and (d) asymmetric single production mode.}  
\label{fig:heatmap_eff_10Tev_col}
\end{figure}
The partonic cross-sections for symmetric and asymmetric pair production can be written as
\begin{align}
\label{eq:sigma_pair_sym}
\sigma_{\text{pair}}^{\text{sym}} &= 
\sigma^{\widetilde{R}_2^{2/3}}_{\text{EW}} 
+ \sigma^{\widetilde{R}_2^{-1/3}}_{\text{EW}} 
- Y_{12}^2 \, \sigma^{\widetilde{R}_2^{2/3}}_{\text{YukEW}} 
+ Y_{12}^4 \, \sigma^{\widetilde{R}_2^{2/3}}_{\text{Yuk}} \,, \\
\label{eq:sigma_pair_asym}
\sigma_{\text{pair}}^{\text{asym}} &= 
\sigma^{\widetilde{R}_2^{2/3}}_{\text{EW}} 
- Y_{12}^2 \, \sigma^{\widetilde{R}_2^{2/3}}_{\text{YukEW}} 
+ Y_{12}^4 \, \sigma^{\widetilde{R}_2^{2/3}}_{\text{Yuk}} \,.
\end{align}
\noindent
Here, $\sigma^{\widetilde{R}_2^{2/3}}_{\rm EW}$ and $\sigma^{\widetilde{R}_2^{-1/3}}_{\rm EW}$ denote the $Z^*/\gamma$-mediated electroweak pair production cross-sections for the $\widetilde{R}_2^{2/3}$ and $\widetilde{R}_2^{-1/3}$ states, respectively. The quantity $\sigma^{\widetilde{R}_2^{2/3}}_{\rm Yuk}$ corresponds to the $t$-channel quark-exchange contribution evaluated for $Y_{12}=1$, while $\sigma^{\widetilde{R}_2^{2/3}}_{\rm YukEW}$ represents the interference between the electroweak and Yukawa-mediated amplitudes. The negative sign reflects the destructive nature of this interference. In the symmetric production mode, both charge states of $\widetilde{R}_2$ contribute, although the $\widetilde{R}_2^{-1/3}$ component enters only through the electroweak channel. In contrast, the asymmetric mode receives contributions exclusively from the $\widetilde{R}_2^{2/3}$ state. For single production, the symmetric channel involves both charge states of $\widetilde{R}_2$, whereas the asymmetric channel contains only the $\widetilde{R}_2^{2/3}$ component. In all cases, the observable event rates are obtained by multiplying the production cross-sections with the relevant branching ratios and the selection efficiencies ($\epsilon$), after applying the cuts described in Section~\ref{sec:6B}.

Figure~\ref{fig:heatmap_eff_10Tev_col} shows the cut efficiencies for the direct production mode of $\widetilde{R}_2^{2/3}$ at a C.O.M. energy of $\sqrt{s}=10~\text{TeV}$, evaluated for different choices of sLQ and RHN masses. These efficiencies are obtained after applying the event selection criteria described in Section~\ref{sec:6B}. As illustrated in Fig.~\ref{fig:heatmap_eff_10Tev_col}, the symmetric production mode consistently exhibits a higher cut efficiency than the asymmetric mode across the full range of $M_N$ and $M_{\widetilde{R}_2^{2/3}}$ considered. This behavior is primarily driven by the requirement of at least four reconstructed jets ($N_{\text{\rm jet}} \geq 4$) in our event selection. While the symmetric mode produces at least six quarks, the asymmetric mode yields only four at the partonic level. Moreover, jets originating from the decay of the RHN can be highly collimated, leading to jet merging during reconstruction. This effect is particularly pronounced in the asymmetric mode, where the reconstructed jet multiplicity can drop to three, causing such events to fail the $N_{\text{\rm jet}} \geq 4$ requirement and thus reducing the overall efficiency relative to the symmetric case. We also observe that the efficiency initially increases and subsequently decreases as the RHN mass $M_N$ increases. This trend is strongly influenced by the selection requirement on the muon multiplicity, $N_{\mu} > 2$. Variations in $M_N$ modify the kinematics of the RHN decay products, which in turn affect the muon transverse momentum spectrum and tagging efficiency. All efficiencies shown here are computed assuming the benchmark choice $Y_{12}=Z_{11}=1$. If one instead considers the hierarchy $Y_{12} > Z_{11}$, additional signal contributions arise from $\widetilde{R}_2$ pair production in which both sLQs decay directly into $\mu j$ final-states with large BRs. Since our cut-based analysis is agnostic to the origin of high-$p_T$ jets, this channel would also contribute  to the signal yield. In the present analysis, however, our primary goal is to perform a sensitivity study for  $N$ and $\widetilde{R}_2$. Consequently, we restrict our signal definition to production and decay topologies in which at least one RHN appears in the production or decay chain of the sLQ. As discussed in Section~\ref{sec:6B}, we deliberately keep the event selection generic. If  instead RHN reconstruction strategy were employed, the contribution from purely $\mu j$ decay modes of the sLQ would become sub-dominant. The results presented here is a conservative estimate, and the inclusion of  additional channels would further enhance the overall sensitivity.

\begin{figure}[h]
\centering
  
\subfloat[]{\includegraphics[width=0.49\textwidth,height=0.40\textwidth]{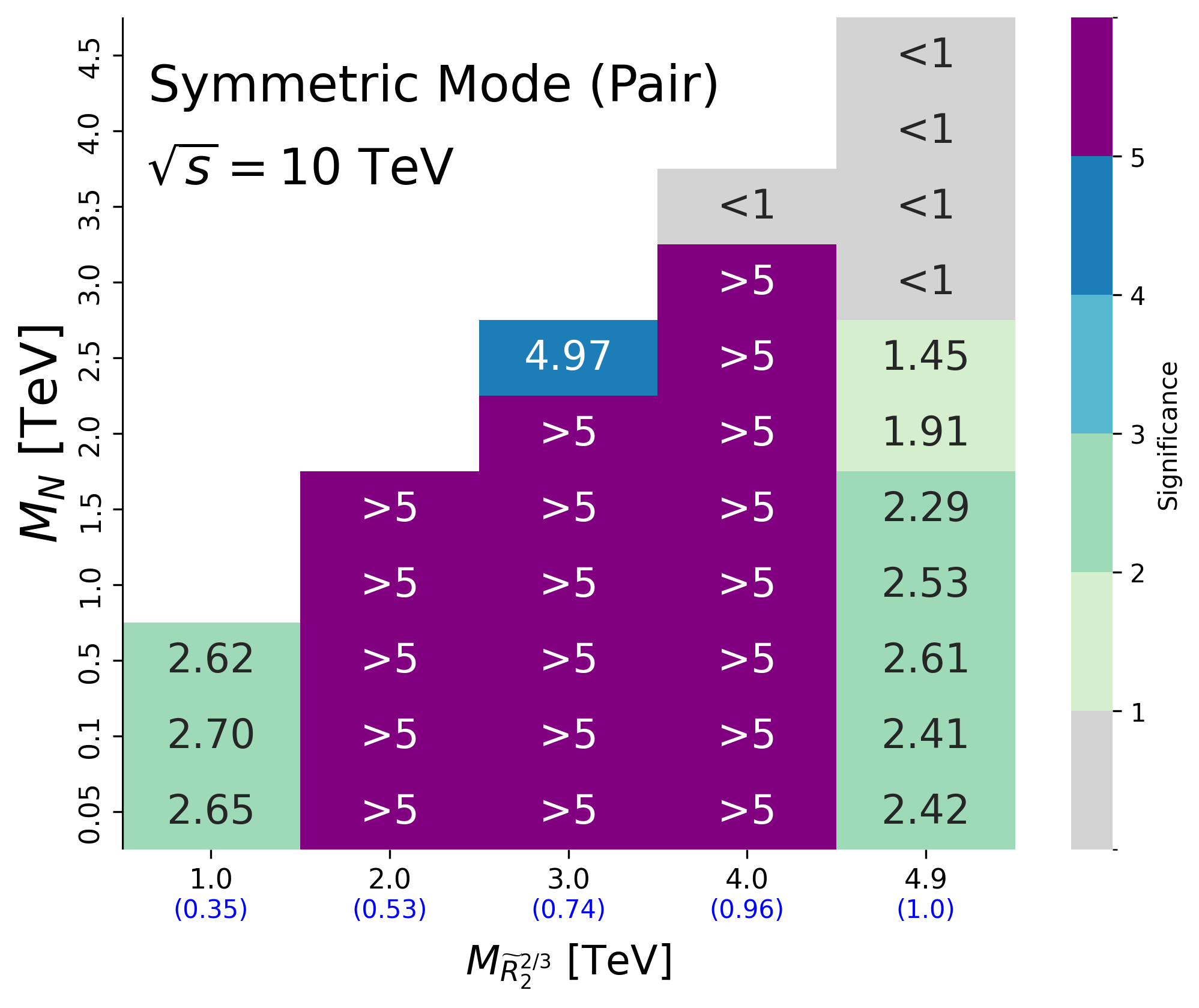}
\label{fig:HeatMap_MN_MLQ_PP_Sym_sig_10TeV_collider}}

\subfloat[]{\includegraphics[width=0.49\textwidth,height=0.40\textwidth]{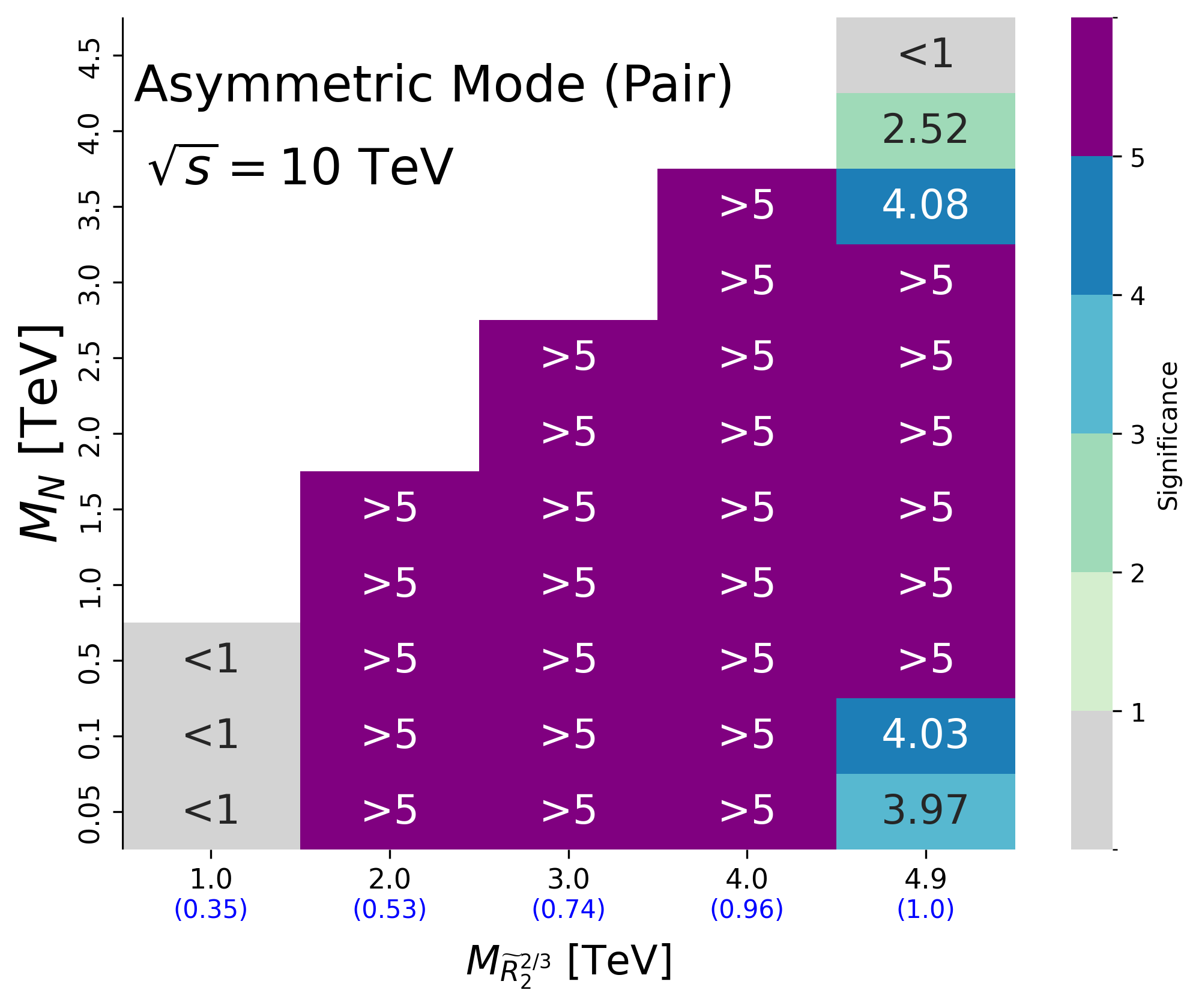}
\label{fig:HeatMap_MN_MLQ_PP_ASym_sig_10TeV_collider}}
\subfloat[]{\includegraphics[width=0.49\textwidth,height=0.40\textwidth]{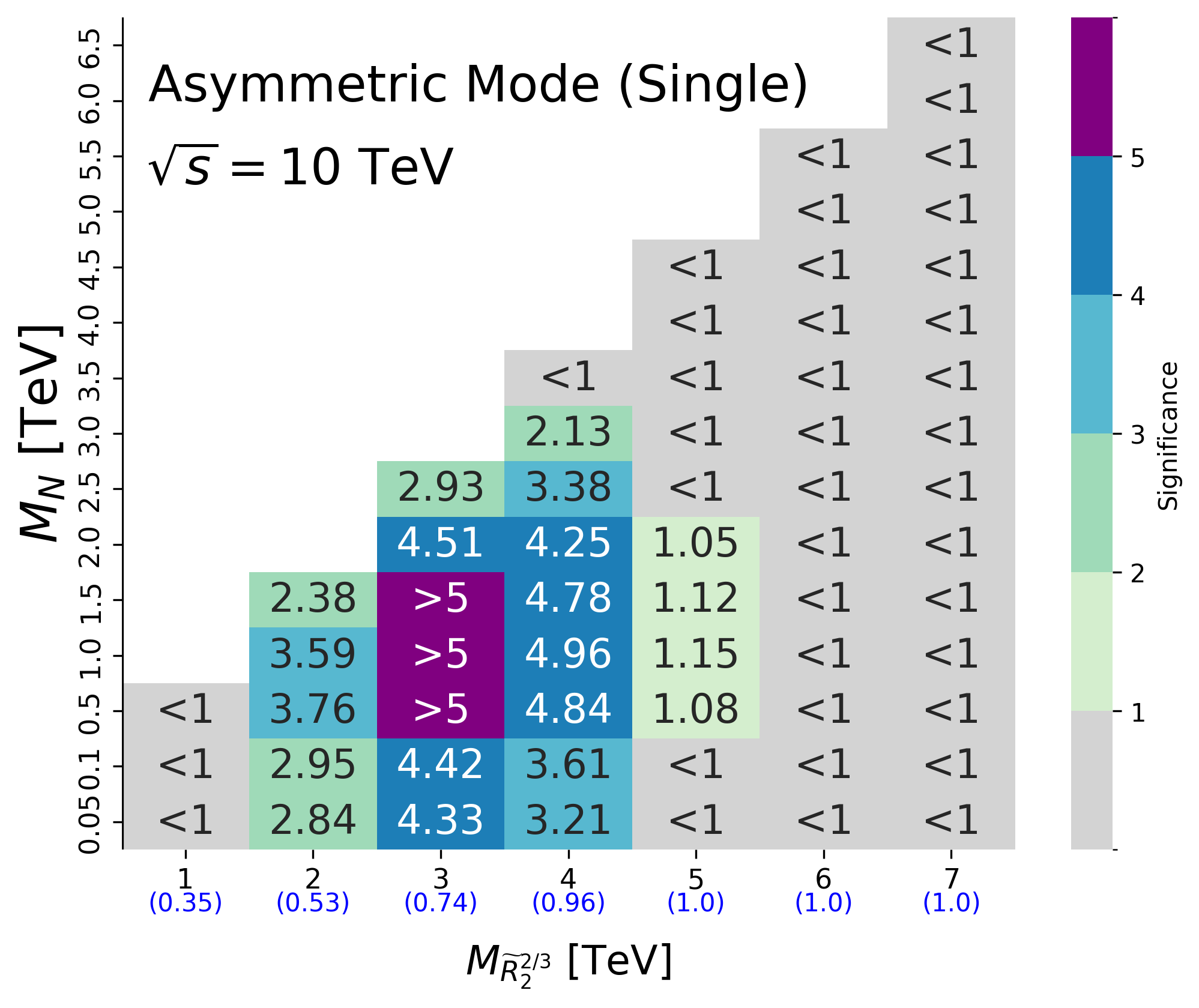}
\label{fig:HeatMap_MN_MLQ_SP_ASym_sig_10TeV_collider}}
\captionsetup{justification=raggedright,singlelinecheck=false} 
\caption{Signal significance in the $M_{\widetilde{R}_2^{2/3}}-M_N$ plane for a $\sqrt{s} = 10~\text{TeV}$ muon collider with an integrated luminosity of $\mathcal{L} = 100~\text{fb}^{-1}$, taking $Z_{11} = 1.0$ and the maximum value of $Y_{12}$ allowed by indirect searches (indicated below each $M_{\widetilde{R}_{2}^{2/3}}$ in blue brackets). The panels correspond to: (a) symmetric pair production mode, (b) asymmetric pair production mode, and (c) asymmetric single production mode.}

\label{fig:heatmap_SIG_10Tev_col}
\end{figure}
Similar to Fig.~\ref{fig:heatmap_eff_10Tev_col}, the corresponding significance distributions are shown in Fig.~\ref{fig:heatmap_SIG_10Tev_col}, assuming a modest  integrated luminosity of $\mathcal{L}=100~\text{fb}^{-1}$, with $Z_{11}=1.0$ and $Y_{12}$ fixed to its maximum value allowed by the indirect LHC constraint from $pp\to\mu\mu$ searches. In these heatmaps, regions with significances above $5\sigma$ are indicated by ``$>5$'', while those below $1\sigma$ are denoted by ``$<1$''. The symmetric single production mode, owing to its comparatively small cross-section, yields significances below $1\sigma$ across the scanned parameter space, and is therefore not shown. As expected, the production cross-section decreases with increasing $M_{\widetilde{R}_2^{2/3}}$, leading to a corresponding reduction in signal significance at higher masses. Since pair production generally yields a substantially larger cross-section than single production, it exhibits a correspondingly stronger statistical reach. In both the symmetric and asymmetric pair production channels, the significance exceeds the $5\sigma$ threshold over nearly the full set of mass points considered. In contrast, the single production channels display markedly different behavior.  The asymmetric single production mode, however, benefits from an enhanced production rate and achieves significances above $2\sigma$ across most of the mass range.

\section{Results and Discussion} 
\label{sec:8}
\begin{figure}[t]
\centering
\subfloat[]{\includegraphics[scale=0.25]{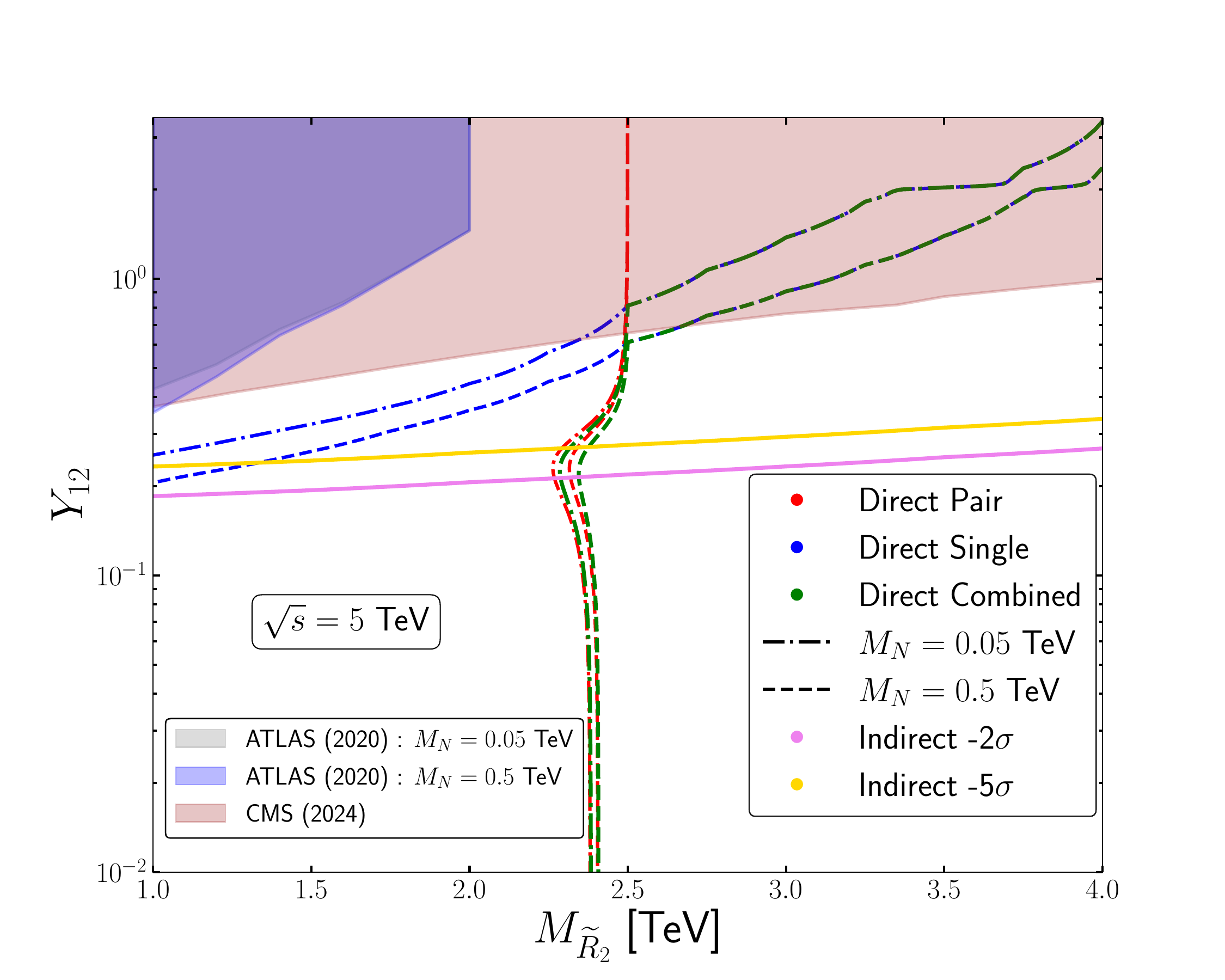}\label{fig:Y12_vs_MLQ_all_5_TeV_FINAL_17May}}
\subfloat[]{

\includegraphics[scale=0.25]{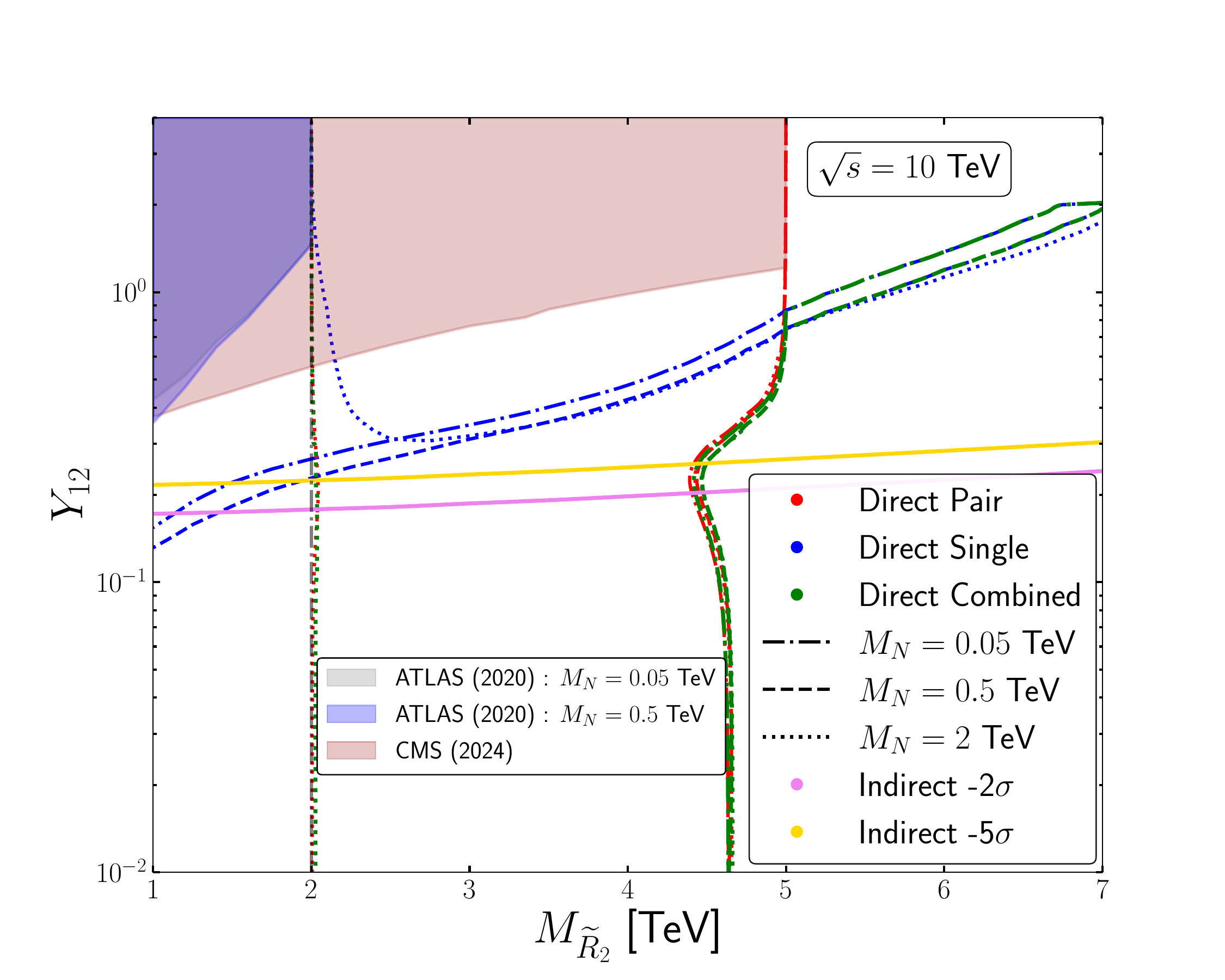}
\label{fig:Y12_vs_MLQ_all_10_TeV_FINAL_17May}}
\captionsetup{justification=raggedright,singlelinecheck=false} 
\caption{Projected sensitivity in the $Y_{12}$–$M_{\widetilde{R}_2}$ plane for direct and indirect sLQ search at the muon collider. Left and right panel correspond to C.O.M. energies $5$ and $10$ TeV, respectively. The pink and yellow solid lines represent $2\sigma$ and $5\sigma$ significance for indirect search, while the results for direct search are shown by red, green and blue lines. See text for further details. The shaded regions indicate the exclusion bounds from existing ATLAS and CMS searches, as specified.}
\label{Fig:Ydl_MR2_final}
\end{figure}

\noindent
In this section, we present a detailed discussion of the important results of our work. In Figs.~\ref{fig:Y12_vs_MLQ_all_5_TeV_FINAL_17May} and \ref{fig:Y12_vs_MLQ_all_10_TeV_FINAL_17May}, we show sensitivity reach in the $M_{\widetilde{R}_2}$ -- $Y_{12}$ plane for direct and indirect searches of $\widetilde{R}_2$ at the proposed muon collider operating at C.O.M $\sqrt{s}=$ $5$ TeV with luminosity $3~\text{ab}^{-1}$ and $\sqrt{s}=$ $10$ TeV with luminosity $10~\text{ab}^{-1}$, respectively, for Yukawa coupling $Z_{11}=1$. We also show  the limits from the current LHC searches in the same plot. The grey and light purple-shaded regions indicate parameter space excluded at $95\%$ confidence limit (CL) by the  ATLAS direct sLQ search~\cite{ATLAS:2020dsk} for $M_{N} = 50$ and $500$ GeV, respectively, and the red-shaded area shows the CMS bound from the indirect search for dimuon production $pp \to \mu^+ \mu^-$~\cite{CMS:2024bej}. In order to re-interpret the direct search limits, we consider all the relevant pair production modes of $\widetilde{R}_{2}^{2/3}$. At the hadron collider, the dominant mode of $\widetilde{R}_{2}^{2/3}$ pair production occurs via the gluon-mediated channel. In addition to this, we also include contributions from the quark-initiated and $t$-channel $\mu/N$-mediated diagrams. Once produced, we consider the decay $\widetilde{R}_{2}^{2/3}\to \mu j$. The inclusion of these additional channels do not alter cut-efficiencies significantly. Hence, to derive the limit, we compare the resulting signal cross-section with the observed upper limit on cross-section reported  in Ref.~\cite{ATLAS:2020dsk}. Among different sLQ production modes at the LHC, pair production via a $t$-channel muon exchange  scales as $Y_{12}^{4}$, while the pair production of sLQ via $t$-channel $N$ exchange scales as $Z_{11}^{4}$. The presence of the latter contribution renders the resulting constraints sensitive to $M_{N}$, as is evident from the figures. In case of limits from indirect search (red-shaded area), only $p p \to \mu^+ \mu^-$ mediated by a $t$-channel $\tilde{R}^{2/3}_2$ in our model contributes to the cross-section and thus it is independent of $Z_{11}$ and $M_N$. Since $\tilde{R}^{2/3}_2$ contributes as an off-shell mediator, the exclusion limit spans a larger $\tilde{R}^{2/3}_2$ mass range. 

Similar to the indirect search at the LHC, $\tilde{R}_2$ can also be probed in the muon collider via $\mu^+ \mu^- \to j j$ mode, where the process is mediated via a $t$-channel $\tilde{R}^{2/3}_2$ exchange and is dependent on the $Y_{12}$ coupling. The pink and yellow solid lines in Fig.~\ref{fig:Y12_vs_MLQ_all_5_TeV_FINAL_17May} and Fig.~\ref{fig:Y12_vs_MLQ_all_10_TeV_FINAL_17May} indicate the $2\sigma$ and $5\sigma$ sensitivity projection for this process. The indirect search into visible final-states are sensitive only to $M_{\widetilde{R}_2^{2/3}}$, which provides an enhanced sensitivity reach for muon collider. The indirect search results indicate that our choice of selection cuts is largely insensitive to $M_{\widetilde{R}_2^{2/3}}$ at both C.O.M. energies.

The direct production of $\widetilde{R}_2$ at the muon collider offer a complimentary measure. The red, blue and green lines in Fig.~\ref{fig:Y12_vs_MLQ_all_5_TeV_FINAL_17May} and Fig.~\ref{fig:Y12_vs_MLQ_all_10_TeV_FINAL_17May} denote the $5\sigma$  projection corresponding  to the pair, single and combined production modes, respectively. As discussed in Section~\ref{sec:7}, the direct production modes include contributions both from $\widetilde{R}_2^{2/3}$ and $\widetilde{R}_2^{ 1/3}$, in their symmetric and asymmetric production channels. These direct production contours are shown for benchmark RHN masses: $M_N=50$ GeV (dash-dotted lines) and $M_N=500$ GeV (dashed lines) at both  $5$ and 10 TeV colliders, and $M_N=2.0$ TeV (dotted lines) at the $10$ TeV collider.  

We observe that up to $M_{\widetilde{R}_2}\sim \sqrt{s}/2$, the $5\sigma$ projection contours for  pair  (red) and  combined (green) production mode overlaps, as the cross-section in this region is mostly dominated by  pair production. For $M_{\widetilde{R}_2}\ge \sqrt{s}/2$,  the single production becomes the dominant contribution. Hence the combined limit follows the single production contour. This transition highlights a crucial result: the vital inclusion of the single production of $\widetilde{R}_2$ is essential to probe heavier sLQs with $\mathcal{O}(1)$ Yukawa coupling, especially probing parameter space beyond the kinematic capability of pair production alone. Note that, the pair production contour develops a bump near \(M_{\widetilde R_2} \simeq 2.3~\text{TeV}\) (left panel) and \(4.5~\text{TeV}\) (right panel), resulting from destructive interference between the EW diagram (Fig.~\ref{fig:R2pairQED}) and the $t$-channel quark-exchange diagram (Fig.~\ref{fig:R2pairNP}). A similar pattern was observed in an earlier work~\cite{Han:2025wdy}. When \(M_{\widetilde R_2} \approx \sqrt{s}/2\) and Yukawa $Y_{12} \sim \mathcal{O}{(0.1-0.6)}$, the destructive interference becomes comparable in magnitude to the EW contribution~\footnote{In this region of the parameter space the contribution from the $t$-channel diagram is sub-dominant.}, reducing the overall pair production cross-section. To maintain a \(5\sigma\) significance, \(M_{\widetilde R_2}\) must be lowered so that the EW contribution dominates, leading to the bump observed on the left side of the contour.

\begin{figure}[h!]
\centering
\subfloat[]{
\includegraphics[scale=0.25]{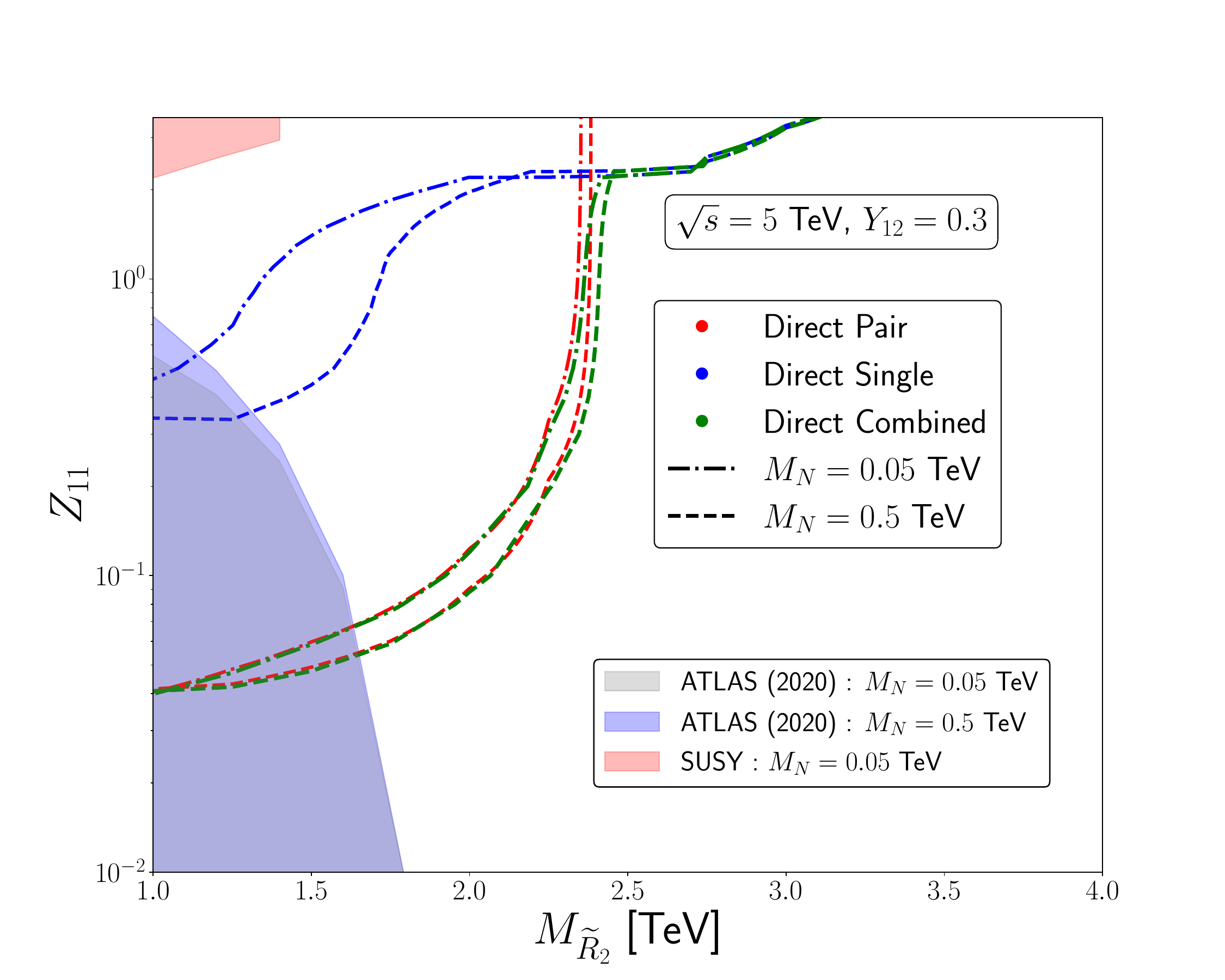}
\label{fig:Z11_vs_MLQ_all_5_TeV_Oct}}
\subfloat[]{

\includegraphics[scale=0.25]{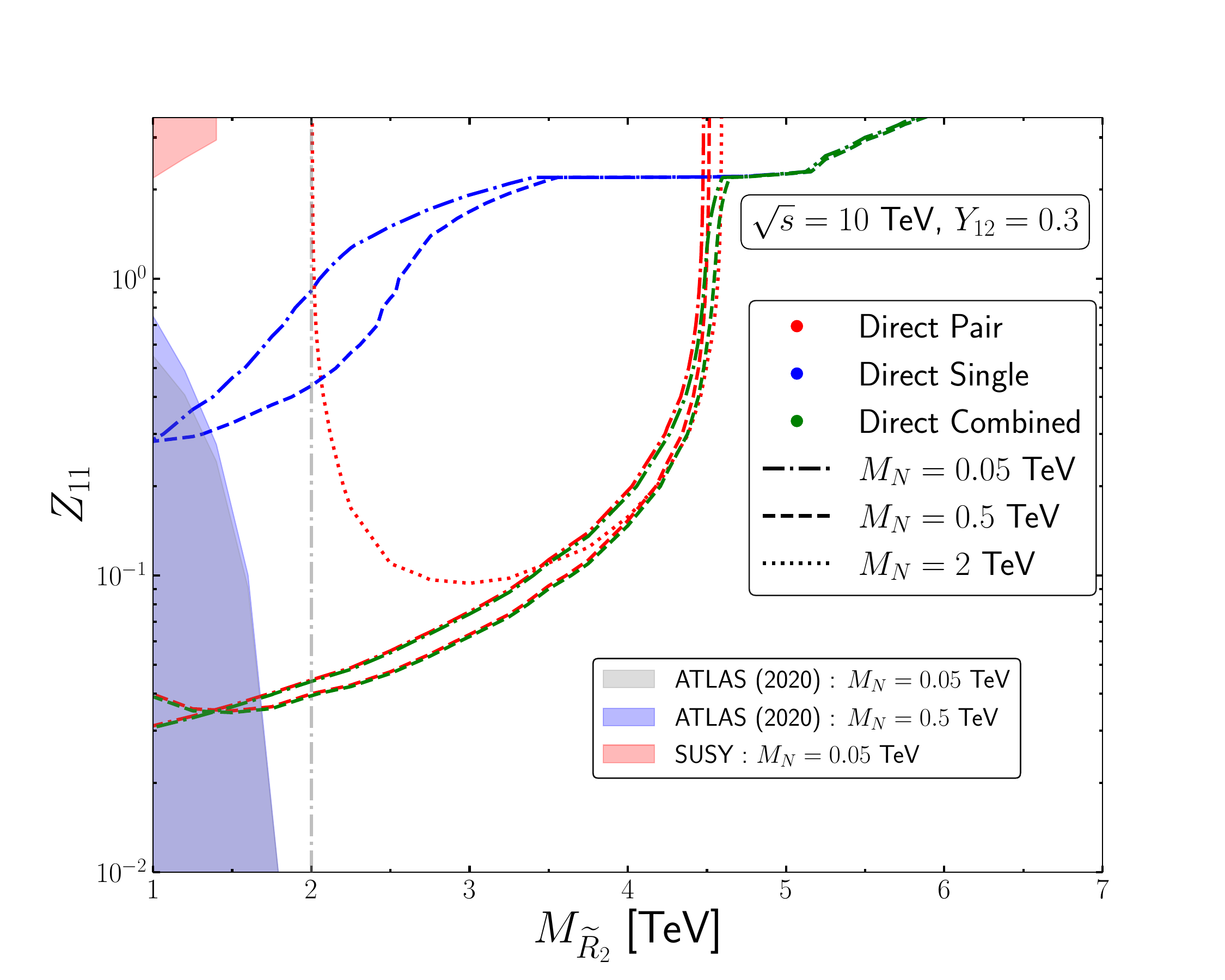}
\label{fig:Z11_vs_MLQ_all_10_TeV_Oct}}
\captionsetup{justification=raggedright,singlelinecheck=false} 
\caption{Projected sensitivity in the $Z_{11}$–$M_{\widetilde{R}_2}$ plane for direct and indirect sLQ search at the muon collider. Left and right panel correspond to C.O.M energy $5$ and $10$ TeV, respectively. The shaded regions indicate the exclusion bounds from existing ATLAS  and CMS  searches, as specified. See text for further details.}
\label{Fig:ZuNr_MR2_final}
\end{figure}


While for $M_N=50$ and 500 GeV, the projection contours demonstrate similar pattern, for $M_N=2.0$ TeV, there is a significant difference in the $5\sigma$ contour, as is evident from Fig.~\ref{fig:Y12_vs_MLQ_all_10_TeV_FINAL_17May}. The difference is most prominent for $M_N=2.0$ TeV around $M_{\widetilde{R}_2} \sim M_N$. For $M_{\widetilde{R}_2} <M_N$,  the decay $\widetilde{R}_2 \to N j$ is kinematically forbidden. The drastic loss of sensitivity for pair production (combined mode) is evident from the red (green) dotted vertical lines at $M_{\widetilde{R}_2}=2.0\ $TeV. For $M_{\widetilde{R}_2}> M_N$, as sLQ mass approaches $M_N$, the sensitivity in $Y_{12}$ for single production mode gradually reduces, as represented by the blue dotted line which includes contributions both from $\widetilde{R}_{2} \mu j$ and $\widetilde{R}_{2} N j$ production channels. Note that, the latter production mode is not subject to the stringent constraint  $M_{\widetilde{R}_2}> M_N$ as for this mode $ \widetilde{R}_2 \to \mu j$ decay can give the desired signal. Hence,  a finite sensitivity in Yukawa $Y_{12}$ is obtained from this mode alone as $M_{\widetilde{R}_2} \sim M_N$. In our analysis, the masses are chosen such that the sLQ mass is larger than the RHN mass; therefore, we do not present sensitivity projections in the region $M_{\widetilde{R}_2} < M_N$. In the case of $M_{N} = 2.0$~TeV, this boundary is indicated by a grey dot-dashed line. Notably, in this region significant deviation may arise, as $N \to \widetilde{R}_2 j$, and $ \widetilde{R}_2 \to N^* j$  decays will be open, and these contributions need to be taken into account while evaluating the sensitivity. For other RHN masses such as 50 GeV  (500 GeV), the plot will show similar behavior around the respective threshold $M_{\widetilde{R}_2} \sim 50 $ GeV ($500$ GeV). These low masses of sLQ  are anyway disfavored by the LHC. 

A comparison between the current LHC limits and the projected sensitivity for direct production at the muon collider highlights that  a heavier  $\widetilde{R}_2$ of mass as high as $6.0$ TeV and with $\mathcal{O}(1)$ Yukawa coupling, which is not constrained by the current LHC data can be probed  at a $\sqrt{s} =10$ TeV muon collider. The indirect search at the muon collider can improve the sensitivity by one order $Y_{12} \sim \mathcal{O}(0.1)$, for both $\sqrt{s}=5$ and 10 TeV configurations.

The various production and decay processes of sLQ at the muon collider also depend on the coupling $Z_{11}$. We show the sensitivity projection in the $Z_{11}-M_{\widetilde{R}_2}$ plane of  Figs.~\ref{fig:Z11_vs_MLQ_all_5_TeV_Oct} and \ref{fig:Z11_vs_MLQ_all_10_TeV_Oct} for $Y_{12}=0.3$. The  pair, single, and combined  production modes are represented by red, blue and green curves, respectively. Similar to Fig.~\ref{Fig:Ydl_MR2_final}, we show the excluded regions from the  ATLAS direct sLQ search~\cite{ATLAS:2020dsk}, for $M_{N} = 50$ GeV (grey shaded) and $500$ GeV (blue shaded). We follow the same method as outlined in Fig.~\ref{Fig:Ydl_MR2_final} to derive these bounds. In addition to this, we also include the bounds from the ATLAS gluino-pair SUSY search~\cite{ATLAS:2023afl} for $M_{N} = 50$ GeV denoted by the red-shaded region.  To derive this, we consider all relevant production modes of $\widetilde{R}_{2}$ at the LHC that can lead to same-sign dimuon and multi-jet final-state. Due to the Majorana nature of $N$ as mentioned above, the plausible production modes for this search are as follows: 
$\widetilde{R}_{2}^{+2/3}$ $\widetilde{R}_{2}^{-2/3}$, $\widetilde{R}_{2}^{1/3}$ $\widetilde{R}_{2}^{-1/3}$, $\widetilde{R}_{2}^{\pm2/3}$ $\widetilde{R}_{2}^{\pm2/3}$, $\widetilde{R}_{2}^{\pm1/3}$ $\widetilde{R}_{2}^{\pm1/3}$, $\widetilde{R}_{2}^{\pm2/3}$ $\widetilde{R}_{2}^{\mp1/3}$, $\widetilde{R}_{2}^{\pm2/3}$ $\widetilde{R}_{2}^{\pm1/3}$. The production is followed by the  decay modes $Nj\ \mu j$ and $Nj\ Nj$, followed by subsequently decays of $N \rightarrow \mu j j$ giving rise to either $4$ or 6 jets along with same-sign dimuon final-state. For every point in the $(M_{\widetilde{R}_2},\, M_{N})$ plane, the total predicted signal yield from all such channels is compared with the observed cross-section limits provided in the \texttt{SRLQD} signal region of Ref.~\cite{ATLAS:2023afl}, expressed in the $(M_{\widetilde{g}},\, M_{\widetilde{\chi}_1^{0}})$ mass plane \footnote{We explicitly checked that the cut-efficiency for our scenario matches with the cut-efficiency presented in the ATLAS search. Hence a re-interpretation of the cross-section limit is sufficient, instead of a full recast analysis.}. By varying the relevant couplings, we then derive the corresponding constraints. Figs.~\ref{fig:Z11_vs_MLQ_all_5_TeV_Oct} and \ref{fig:Z11_vs_MLQ_all_10_TeV_Oct} show that higher $Z_{11}$ values combined with lighter $M_{\widetilde R_2}$ are  constrained by this search. 

As is evident from Figs.~\ref{fig:Z11_vs_MLQ_all_5_TeV_Oct} and \ref{fig:Z11_vs_MLQ_all_10_TeV_Oct}, majority of the parameter space is unconstrained from the LHC searches, which can further be probed at muon collider. In particular, a $10$ TeV muon collider is sensitive to a very heavy sLQ with mass close to $6.0$ TeV, and $\mathcal{O}(1)$ Yukawa coupling.  In both left and right panels, the nature of pair production mode for various $M_N$ values show similar feature. In these figures, the sensitivity contours obtained from the combined production mode generically follow those from pair production as long as $M_{\widetilde{R}_2} < \sqrt{s}/2$. As $M_{\widetilde{R}_2}$ increases and approaches the kinematic limit of $\sqrt{s}/2$, the rapid suppression of the pair production cross-section causes the combined sensitivity to transition smoothly to that governed by single production. In the sLQ mass range  where the pair production is dominant, the qualitative behavior (dip of the contour and subsequent increment) is common to all benchmark values of $M_N$ and reflects the interplay between production of $\widetilde{R}_2$ and decay $\widetilde{R}_2 \to N j$. The sensitivity to pair production lowers as $M_{\widetilde{R}_2}$ approaches either the collider threshold, $M_{\widetilde{R}_2} \sim \sqrt{s}/2$, or the kinematic threshold, $M_{\widetilde{R}_2} \sim M_N$. For $M_N = 0.5~\text{TeV}$, this behavior is slightly visible as $M_{\widetilde{R}_2}$ approaches $\mathcal{O}(1)~\text{TeV}$. In contrast, for $M_N = 0.05~\text{TeV}$, the phase-space factor for $\widetilde{R}_2 \to N j$ is already saturated across the entire mass range considered. Hence, the dip in the contour behavior occurs at lower values of $M_{\widetilde{R}2}$, of order $\mathcal{O}(100),\text{GeV}$, which lie beyond the plotted range. For $M_N = 2.0~\text{TeV}$, this behavior is visible within the plotted range of $M_{\widetilde{R}_2}$.

The single production modes in both 
Fig.~\ref{fig:Z11_vs_MLQ_all_5_TeV_Oct} and Fig.~\ref{fig:Z11_vs_MLQ_all_10_TeV_Oct} are also very similar for RHN masses $50$ and $500$ GeV. For ${5}$ ($10$) TeV C.O.M energy, between sLQ mass $2.0-3.0$ TeV ($3.0-5.0$ TeV), the sensitivity contours are almost flat and independent of $Z_{11}$. This occurs, as for the chosen values of $Y_{12}$, the $Z_{11}$-independent single production process $\mu\mu\to\widetilde{R}_2\mu j$  can alone give a $5\sigma$ significance, although $\mu\mu\to\widetilde{R}_2 N j$ gives an additional sub-dominant contribution in the cross-section. Notably, in this region,  BR($\widetilde{R}_2 \to N j$) is almost constant with the increase of  sLQ mass, as $Z_{11}\gg Y_{12}$ and $M_{\widetilde{R}_2} \gg M_N$. Hence, no further suppression in total cross-section arises for the mode $\mu\mu\to\widetilde{R}_2\mu j$, followed by $\widetilde{R}_2 \to N j$. As $M_{\widetilde R_2}$ further increases, the $Z_{11}$-dependent single production mode $\mu\mu\to\widetilde{R}_2 N j$ becomes essential to achieve a $5\sigma$ sensitivity, as the dominant mode $\mu \mu \to \widetilde{R}_2 \mu j$ alone is insufficient. Due to a large phase space suppression in cross-section for a very heavy sLQ mass larger than $2.5$ ($5.0$) TeV for $\sqrt{s}=5$ (10) TeV, the required $Z_{11}$ to obtain a $5\sigma$ sensitivity thus increases. For $M_N=2.0$ TeV, the single production mode largely shows similar features as $M_N=50$ GeV and $500$ GeV, albeit with some minor differences near $M_{\widetilde{R}_2} \sim M_N$, evident from  Fig.~\ref{fig:Z11_vs_MLQ_all_10_TeV_Oct}.  In this region,  the contours closely resembles that in Fig.~\ref{fig:Y12_vs_MLQ_all_10_TeV_FINAL_17May}. A large value of $Z_{11}$ is required to obtain a $5\sigma$ significance, as in this region $\widetilde{R}_2 \mu j$ followed by $\widetilde{R}_2 \to N j $ is kinematically suppressed. The dominant contribution arises from $\widetilde{R}_2 Nj$. The grey dot-dashed vertical line at $M_{\widetilde{R}_2}=2.0$ TeV  indicates the kinematic threshold imposed by our assumption $M_{\widetilde{R}_2} > M_{N}$. We observe another interesting trend in the single production contour for $M_{N}=2.0$ TeV. There is a slight fall in $Z_{11}$ between $4.5<M_{\widetilde R_2}<6.0~\text{TeV}$, as the single production rate in this range is driven majorly by the $ \widetilde{R}_2\mu j$ channel (independent of $Z_{11}$), while $\mathrm{BR}(\widetilde{R}_2\to N j)$ continues to increase with sLQ mass, thus requiring a smaller $Z_{11}$ for the $5\sigma$ sensitivity.

Finally, we also present a qualitative comparison  of HL-LHC and muon collider sensitivity reach.  To compare our results with the HL-LHC discovery reach, we refer to the previous studies in  Refs.~\cite{Bhaskar:2023xkm, Duraikandan:2024kcy}. In Ref.~\cite{Bhaskar:2023xkm}, the authors investigate a dilepton plus jets final-state closely aligned with our signal topology, focusing on the $\widetilde{R}_2$ interacting exclusively with RHNs. In their analysis, the authors consider pair production of RHNs, arising from the decay of $\widetilde{R}_2$, with a fixed RHN mass of $M_N = 500$ GeV. For the HL-LHC with an integrated luminosity of $3$ ab$^{-1}$, their results indicate that to achieve a $5 \sigma$ discovery for $M_{\widetilde{R}_2} = 2.0$, $2.5$, and $3.0$ TeV, the required Yukawa coupling $Y$ is approximately $1.4$, $2.8$, and beyond $3.5$ in the case of up-type quark alignment, and $1.1$, $2.6$, and beyond $3.5$ for down-type alignment, respectively. In contrast, our results show that for the same mass points at a $5$ TeV muon collider, the required coupling values are significantly lower—being below $10^{-2}$ for $2.0$ TeV, approximately $0.8$ for $2.5$ TeV, and about $1.5$ for $3.0$ TeV. For a $10$ TeV muon collider, the required couplings remain below $10^{-2}$ for all the considered mass points. This comparison clearly demonstrates that the muon collider significantly outperforms the HL-LHC in terms of discovery potential for sLQs, especially in scenarios involving heavy final-state neutrinos. Our results quantitatively demonstrate that muon colliders provide exceptional sensitivity to the sLQ parameter space specially in the higher sLQ mass range.

\section{Conclusions}
\label{sec:9}

High-energy muon colliders provide a unique and powerful environment to probe LQ scenarios that are difficult or impossible to access at the LHC and HL-LHC. In this work, we have presented a comprehensive sensitivity study for the sLQ doublet $\widetilde{R}_2$ in scenarios where it couples to first-generation quarks, muons, and heavy RHNs. For completeness, we have also included the most relevant constraints derived from existing LHC searches. We consider two benchmark muon collider setups with center-of-mass energies of $\sqrt{s} = 5~\text{TeV}$ and 10~\text{TeV}, corresponding to integrated luminosities of $3~\text{ab}^{-1}$ and $10~\text{ab}^{-1}$, respectively. We have considered both indirect and direct probes of $\widetilde{R}_2$. The indirect search exploits deviations in dijet kinematics induced by $t$-channel exchange of $\widetilde{R}_2^{2/3}$, while the direct search includes all relevant pair and single production modes of $\widetilde{R}_2^{2/3}$ and $\widetilde{R}_2^{1/3}$. A common and minimal set of selection cuts on muon multiplicity, jet multiplicity, and leading-jet transverse momentum has been applied throughout. A key result of this study is that the indirect search channel provides a robust and largely mass-independent probe of $\widetilde{R}_2^{2/3}$. Owing to its coupling-driven nature, this channel remains sensitive even in regions where direct pair production becomes kinematically suppressed. As a result, indirect searches play an essential role in extending the reach of muon colliders into regions of parameter space that are inaccessible to direct production alone. For the direct searches, we find a clear hierarchy between pair and single production. Pair production dominates the sensitivity at lower LQ masses, whereas single production, governed by the sLQ–quark–muon and sLQ–quark–RHN Yukawa couplings, becomes increasingly important at higher masses. In particular, single production alone can achieve a $5\sigma$ discovery sensitivity for $M_{\widetilde{R}_2}\sim 3.0~\text{TeV}$ ($6.0~\text{TeV}$) at $\sqrt{s}=5~\text{TeV}$ ($10~\text{TeV}$), assuming $\mathcal{O}(1)$ Yukawa couplings. The transition between the pair-dominated and single-dominated regimes leads to characteristic features in the sensitivity contours. {At lower LQ masses, the required value of $Z_{11}$ displays a non-monotonic behavior due to the competing effects of production and decay: while the production cross-section decreases with mass $M_{\widetilde{R}_2}$, the branching ratio $\mathrm{BR}(\widetilde{R}_2 \to N j)$ increases as additional phase space opens up. At higher masses, however, the phase-space dependence of the decay saturates and the sensitivity becomes dominated by the kinematic suppression of the production cross-section, leading to a steadily increasing coupling requirement for $5\sigma$ discovery.} We find that the sensitivity effectively saturates for $M_{\widetilde{R}_2} \gtrsim 2.0~\text{TeV}$ ($3.0~\text{TeV}$) at $\sqrt{s}=5~\text{TeV}$ ($10~\text{TeV}$), where single production dominates. Beyond $M_{\widetilde{R}_2}\sim 3.0~\text{TeV}$ ($5.0~\text{TeV}$), Yukawa couplings of order unity are required to achieve discovery. 

Overall, the combination of indirect probes and single production enables the discovery of multi-TeV sLQ in regions of parameter space that remain inaccessible to the LHC. In contrast to hadron colliders, muon colliders retain strong sensitivity due to their clean experimental environment and their ability to exploit both coupling-driven and kinematics-driven production mechanisms. Our results demonstrate that future muon colliders can systematically explore a broad and theoretically well-motivated parameter space of sLQ–RHN models, establishing them as indispensable facilities for uncovering new physics at the multi-TeV scale. \\

\acknowledgments
\noindent
The authors acknowledge the use of SAMKHYA: High-Performance Computing Facility provided by the Institute of Physics (IOP), Bhubaneswar. The work of BD was partly supported by the U.S. Department of Energy under grant No. DE-SC0017987. BD thanks the participants of the 2025 GGI workshop on `Exploring the energy frontier with muon beams' for stimulating discussions. BD also acknowledges the local hospitality at IOP, Bhubaneswar during a visit in 2024 when a part of this work was done. 

\appendix
\section{Scalar LQ Decay Widths  \label{Appendix_LQ}}
The analytical expressions for the relevant partial decay widths of the $\widetilde{R}_2$ LQ components are provided below:
\begin{align}
	\Gamma(\widetilde{R}_{2}^{ 2/3}  \rightarrow \mu  d) & = \frac{3|Y_{12}|^2}{48 \pi M_{\widetilde{R}_{2}^{2/3}}}
	\left(   M_{\widetilde{R}_{2}^{2/3}}^2 - M_d^2  -  M_\mu^2 
	\right)\, 
	\lambda^{\frac{1}{2}}\left(1,y_d^2,y_\mu^2\right)
	,
	\label{eq3} \\
	\Gamma(\widetilde{R}_{2}^{2/3} \rightarrow u N) & = \frac{
		3|Z_{11}|^2}{48 \pi M_{\widetilde{R}_{2}^{2/3}}}\left(  M_{\widetilde{{R}}^{2/3}_{2}}^2  - M_{N}^2  -  M_u^2 
	\right)\, 
	\lambda^{\frac{1}{2}}\left(1,y_u^2,y_N^2\right)
	,
	\label{eq4} \\
	\Gamma(\widetilde{R}_{2}^{-1/3} \rightarrow d N) & = \frac{
		3|Z_{11}|^2}{48 \pi M_{\widetilde{R}_{2}^{-1/3}}}\left( 
	M_{\widetilde{R}_{2}^{-1/3}}^2 - M_{N}^2   -  M_d^2 
	\right)\, 
	\lambda^{\frac{1}{2}}\left(1,z_d^2,z_N^2\right)
	, 
	\label{eq5} \\
	\Gamma(\widetilde{R}_{2}^{-1/3} \rightarrow {\nu}_{\mu} d) & = \frac{3|Y_{12}|^2}{48 \pi M_{\widetilde{R}_{2}^{-1/3}}}
	\left(   M_{\widetilde{R}_{2}^{-1/3}}^2 - M_d^2  -  M_{\nu_{\mu}}^2 
	\right)\,
	\lambda^{\frac{1}{2}}\left(1,z_{d}^2,z_{\nu_{\mu}}^2\right)
	,
	\label{eq6}
\end{align}
\noindent
where 
$\lambda(a, b, c) = 
a^2 + b^{2} +c^{2} - 2 a b - 2 b c - 2 a c $, $y_i\equiv M_i/M_{\widetilde{R}_{2}^{2/3}}$, and $z_i\equiv M_i/M_{\widetilde{R}_{2}^{-1/3}}$. 
In the above expressions, $M_{\widetilde{R}_{2}^{2/3}}$ and $M_{\widetilde{R}_{2}^{-1/3}}$ are the masses of the $2/3$ and $-1/3$ components of $\widetilde{R}_{2}$ respectively. In the main text, we collectively denote them as $M_{\widetilde{R}_2}$, unless otherwise specified. Although the expressions above simplify in the limit $M_{\widetilde{R}_2}\gg M_f$, we retain their full forms here for completeness.
\section{RHN Decay Widths  \label{Appendix_n}}
Here we provide the analytic expressions of the partial widths for the two-body decays of a Majorana RHN $N \to \mu W, \,  \nu_{\mu} Z, \, \nu_{\mu} H$ and the relevant three-body decays
$N \to \mu u \bar{d},\ \nu_{\mu} d \bar{d},\ \nu_{\mu} u \bar{u},\ \nu_{\mu} \ell^+ \ell^-,\ \mu^{-} {\nu}_{\ell} \ell^{+},\ \nu_\mu {\mu}^{-} {\mu}^{+},\ \nu_{\mu} \nu_\ell \bar{\nu_\ell}$. 

\subsection{2-body decay widths}
The two-body partial decay widths of $N \rightarrow ab$ are as follows: 
\begin{align}
	\Gamma (N \rightarrow \mu^{+} W^{-}/ \mu^{-} W^{+}) &= \frac{g^{2}}{64\pi M_{N}M_{W}^{2}} |V_{\mu N}|^{2}   \left[ M_{W}^{2}(M_{{\mu}}^{2} + M_{N}^{2}) + (M_{\mu}^{2} - M_{N}^{2})^{2} - 2 M_{W}^{4} \right]  \lambda^{\frac{1}{2}}(1,x_{\mu}^{2},x_{W}^{2})\, , \\
	\Gamma (N \rightarrow \nu_{\mu} H / \bar{\nu}_{\mu} H) &= \frac{|V_{\mu N}|^{2}}{32 \pi M_{N} v^{2}} (M_{N}^{2} - M_{H}^{2})^{2} \, , \\ 
	\Gamma(N \rightarrow {\nu}_{\mu} Z / \bar{\nu}_{\mu} Z) & = \frac{(M_{N}^{2} - M_{Z}^{2})^2}{128 \pi c_{w}^{2}M_{Z}^{2}M_N^3} g^2 |V_{\mu N}|^{2}(M_N^2 + 2 M_{Z}^{2}) \, ,
\end{align}
where $x_i\equiv M_i/M_N$ and  
$c_w\equiv \cos\theta_w$ with $\theta_w$ being the weak mixing angle. As demonstrated, these decay modes are influenced directly by the active-sterile mixing parameter $V_{\mu N}$.

\subsection{3-body decay widths \label{threebody}}

The three-body partial decay widths of $N \rightarrow abc$ are as follows:
\begin{align}
	\Gamma(N \to \mu^{+} \bar{u}_{\alpha} {d}_{\beta} /\mu^{-} u_{\alpha} \bar{d}_{\beta}) =& \frac{M_N^5 N_c}{512 \pi^3}  
	|V_{\mu N}|^2 |V_{\text{CKM}}^{\alpha \beta}|^2 \left( \frac{g^2}{M_W^2} \right)^2  
	I_1(x_{\mu}, x_{u_{\alpha}}, x_{d_{\beta}})
	\nonumber  \\
	& + \frac{ N_c g^2 {\rm Re}(Z_{11} Y_{12}  V_{\mu N}^{*}V_{\text{CKM}}^{\alpha \beta *})}{512 \pi^3 M_N^3 M_W^2}  
	\int_{m_{23} = (M_{d_{\beta}} + M_\mu)^2}^{(M_N - M_{u_{{\alpha}}})^2} 
	\int_{m_{12} = (M_{u_{{\alpha}}} + M_{d_{\beta}})^2}^{(M_N - M_\mu)^2} dm_{12}dm_{23}\nonumber \\ 
	& \quad \times
	\frac{1}{(m_{23} - M_{\widetilde{R}_2^{2/3}}^2)(m_{12} - M_W^2)}  \Big[
	-M_{u_{\alpha}} M_{d_{\beta}} m_{12} (M_N^2 + M_\mu^2 - m_{12}) \nonumber \\ 
	& \quad + M_{u_{\alpha}} M_{d_{\beta}} (M_N^2 - M_\mu^2 - m_{12})(M_N^2 - M_\mu^2 + m_{12})  + 2 M_W^2 M_{u_{\alpha}} M_{d_{\beta}} (M_N^2 + M_\mu^2 - m_{12})
	\Big] 
	\nonumber \\
	&\quad  + N_c  \frac{|Z_{11}|^2  |Y_{12}|^2}{512 \pi^3 M_N^3}  
	\int_{m_{23} = (M_{d_{\beta}} + M_\mu)^2}^{(M_N - M_{u_{\alpha}})^2} 
	\int_{m_{12} = (M_{u_{\alpha}} + M_{d_{\beta}})^2}^{(M_N - M_\mu)^2}\, dm_{12} \, dm_{23}  \nonumber \\
	&\quad \quad \times 
	\frac{\left(M_N^2 - M_{u_{\alpha}}^2 - m_{23}\right)  \left(m_{23} - M_{d_{\beta}}^2 - M_\mu^2\right)}{\left(m_{23} - M_{\widetilde{R}_2^{2/3}}^2\right)^2}  ,
	\label{eq:A4} \\
	\Gamma(N \rightarrow \nu_\mu d_{\beta} \bar{d}_{\beta} / \bar{\nu}_\mu d_{\beta} \bar{d}_{\beta}) =& 
	\frac{M_N^5 N_c}{512 \pi^3} \Bigg[
	\frac{g^4}{4M_W^4} |V_{\mu N}|^2[(g_{L,d}^2 + g_{R,d}^2) I_1(x_{\nu_\mu}, x_{d_{\beta}}, x_{d_{\beta}}) 
	- 2g_{L,d} g_{R,d} G_3(x_{d_{\beta}}, x_{d_{\beta}}, x_{\nu_\mu})]\nonumber \\
	&\quad \quad + \frac{|Y_{\nu}|^2 Y_{d_{\beta}}^2}{2 M_H^4} 
	\left[ I_1(x_{d_{\beta}}, x_{d_{\beta}}, x_{\nu_\mu}) + 2 G_3(x_{d_{\beta}}, x_{d_{\beta}}, x_{\nu_\mu}) \right] \nonumber \\
	&\quad \quad + \frac{g^{2} {\rm Re}(V_{\mu N}^{*} Y_{\nu} Y_{d_{\beta}})}{2 M_{W}^{2} M_{H}^{2}} 
	(g_{L,d} - g_{R,d}) 
	I_2(x_{\nu_\mu}, x_{d_{\beta}}, x_{d_{\beta}}) \Bigg] \nonumber \\
	&\quad +  \frac{ |Z_{11}|^{2} \left(|(Y U_{\rm PMNS})_{12}|\right)^2 N_c}{512 \pi^3 M_N^3} 
	\int_{m_{23} = (M_{d_{\beta}} + M_{\nu_\mu})^2}^{(M_N - M_{d_{\beta}})^2} 
	\int_{m_{13} =(M_{d_{\beta}} + M_{\nu_\mu})^2}^{(M_N - M_{d_{\beta}})^2} \, dm_{13} \, dm_{23} \nonumber \\
	&\quad \times 
	\frac{\left(M_N^2 + M_{d_{\beta}}^2 - m_{23}\right) \left(m_{23} - M_{\nu_\mu}^2 - M_{d_{\beta}}^2\right)}{\left(m_{23} - M_{\widetilde{R}_2^{-1/3}}^2\right)^2}  \nonumber \\
	&\quad + \frac{N_c g^2 {\rm Re}( Z_{11}  (Y U_{\rm PMNS})_{12}    V_{\mu N}^{*})}{1024 \pi^3 M_N^3 M_Z^2 c_w^2} \int_{m_{23} = (M_{d_{\beta}} + M_{v\mu})^2}^{(M_N - M_{d_{\beta}})^2} \int_{m_{12} =4M_{d_{\beta}}^{2}}^{(M_N - M_\mu)^2}  \, dm_{12} \, dm_{23}\nonumber \\
	&\quad \times
	\frac{2 g_{L,d} M_{d_{\beta}}^2 \left(M_N^2 + M_\mu^2 - m_{12}\right) + 4 g_{R,d} \left(M_N^2 + M_{d_{\beta}}^2 - m_{23}\right) \left(m_{23} - M_{d_{\beta}}^2 - M_\mu^2\right)}{m_{23} - M_{\widetilde{R}_2^{-1/3}}^2} , \\
	\Gamma(N \rightarrow \nu_\mu u_{\alpha} \bar{u}_{\alpha} / \bar{\nu}_\mu u_{\alpha} \bar{u}_{\alpha}) =& 
	\frac{M_N^5 N_c}{512 \pi^3} \Bigg(
	\frac{g^4}{4M_W^4} |V_{\mu N}|^2 
	\left[ (g_{L,u}^2 + g_{R,u}^2) I_1(x_{\nu_\mu}, x_{u_{\alpha}}, x_{u_{\alpha}})  
	- 2 g_{L,u} g_{R,u} G_3(x_{u_{\alpha}}, x_{u_{\alpha}}, x_{\nu_\mu}) \right] \nonumber \\
	&\quad + \frac{g^{2} {\rm Re}(V_{\mu N}^{*} Y_{\nu} Y_{u_{\alpha}})}{2 M_{W}^{2} M_{H}^{2}} 
	(g_{L,u} - g_{R,u}) 
	I_2(x_{\nu_\mu}, x_{u_{\alpha}}, x_{u_{\alpha}}) \nonumber \\
	&\quad + \frac{|Y_{\nu}|^2 |Y_{u_{\alpha}}|^2}{2 M_H^4} 
	\left[ I_1(x_{u_{\alpha}}, x_{u_{\alpha}}, x_{\nu_\mu}) 
	+ 2 G_3(x_{u_{\alpha}}, x_{u_{\alpha}}, x_{\nu_\mu}) \right]
	\Bigg) \\
	\Gamma(N \rightarrow \nu_{\mu} \nu_\ell \bar{\nu_\ell} / \bar{\nu}_{\mu} \nu_\ell \bar{\nu_\ell}) =& 
	\frac{M_N^5}{512 \pi^3} \frac{g^4}{4M_W^4}|V_{\mu N}|^2 \left[
	(g_{L,\nu}^2 + g_{R,\nu}^2)  I_1(x_{\nu_\mu}, x_{\nu_\ell}, x_{\nu_\ell})
	- 2g_{L,\nu} g_{R,\nu}   G_3(x_{\nu_\ell}, x_{\nu_\ell}, x_{\nu_\mu})
	\right], 
	\end{align}
    
    \begin{align}
	\Gamma(N \rightarrow \nu_\mu \ell^{+} \ell^{-} / \bar{\nu}_\mu \ell^{+} \ell^{-}) =& 
	\frac{M_N^5}{512 \pi^3}  \Bigg(
	\frac{g^4}{4M_W^4}  |V_{\mu N}|^2[  (g_{L,\ell}^2 + g_{R,\ell}^2)  I_1(x_{\nu_\mu}, x_{\ell}, x_{\ell})  - 2g_{L,\ell} g_{R,\ell}   G_3(x_{\ell}, x_{\ell}, x_{\nu_\mu})] \nonumber \\
	&\quad + \frac{g^{2} {\rm Re}(V_{\mu N}^{*} Y_{\nu} Y_{\ell})}{2 M_{W}^{2} M_{H}^{2}} 
	(g_{L,\ell} - g_{R,\ell}) 
	I_2(x_{\nu_\mu}, x_{\ell}, x_{\ell}) \nonumber \\
	&\quad   + \frac{|Y_{\nu}|^2  Y_{\ell}^2}{ 2 M_H^4} \left[ I_1(x_{\ell}, x_{\ell}, x_{\nu_\mu}) + 2 G_3(x_{\ell}, x_{\ell}, x_{\nu_\mu}) \right] 
	\Bigg)\qquad \qquad  (\ell\neq \mu) \, \\
	\Gamma(N \rightarrow \nu_\mu \mu^{-} \mu^{+} / \bar{\nu}_\mu \mu^{-} \mu^{+}) =& 
	\frac{M_N^5}{512 \pi^3} \Bigg(
	\frac{g^4}{4M_W^4} |V_{\mu N}|^2 
	\left[(g_{L,\mu}^2 + g_{R,\mu}^2) I_1(x_{\nu_\mu}, x_{\mu}, x_{\mu}) 
	- 2g_{L,\mu} g_{R,\mu} G_3(x_{\mu}, x_{\mu}, x_{\nu_\mu}) \right] \nonumber \\
	&\quad + \frac{|Y_\nu|^2 Y_{\mu}^2}{2 M_H^4} 
	\left[I_1(x_{\mu}, x_{\mu}, x_{\nu_\mu}) 
	+ 2 G_3(x_{\mu}, x_{\mu}, x_{\nu_\mu})\right]  \nonumber \\
	&\quad + \frac{g^{2} {\rm Re}(V_{\mu N}^{*} Y_{\nu} Y_{\mu})}{2 M_{W}^{2} M_{H}^{2}} 
	(g_{L,\mu} - g_{R,\mu}) 
	I_2(x_{\nu_\mu}, x_{\mu}, x_{\mu}) + \frac{g^4 |V_{\mu N}|^2}{M_W^4} 
	I_1(x_{\mu}, x_{\nu_\mu}, x_{\mu})  \nonumber \\
	&\quad - 8 g_{L,\mu} C_{t2} \operatorname{Re}(V_{\mu N}^{*} C_{t3}) 
	I_1(x_{\mu}, x_{\nu_\mu}, x_{\mu})
	+ 8 g_{R,\mu} C_{t2} \operatorname{Re}(V_{\mu N}^{*} C_{t3}) 
	G_3(x_{\mu}, x_{\mu}, x_{\nu_\mu}) \nonumber \\
	&\quad - \frac{g^{2} {\rm Re}(V_{\mu N}^{*} Y_{\nu} Y_{\mu})}{M_{W}^{2} M_{H}^{2} } I_2(x_{\nu_\mu}, x_{\mu}, x_{\mu})
	\Bigg),\\
	\Gamma(N \rightarrow \mu^{-} {\nu}_{\ell} \ell^{+} / \mu^{+} \bar{\nu}_{\ell} \ell^{-}) &= 
	\frac{M_N^5 g^4}{512 \pi^3 M_W^4}  
	|V_{\mu N}|^2  I_1(x_{\mu}, x_{{\nu}_{\ell}}, x_{\ell}) \qquad \qquad  (\ell\neq \mu) 
	,
\end{align}
where $N_c$ denotes the number of colors. 
The kinematic functions are given by
\begin{align}
	I_1(x_a, x_b, x_c) & = \int_{(x_a + x_b)^2}^{(1 - x_c)^2} \frac{dz}{z}\, (z - x_a^2 - x_b^2) \, (1 + x_c^2 - z) \, \lambda^{\frac{1}{2}}(1, z, x_c^2)\, \lambda^{\frac{1}{2}}(z, x_a^2, x_b^2)  \, , \\ I_2(x_a, x_b, x_c) & = -\int_{(x_a + x_b)^2}^{(1 - x_c)^2} \frac{dz}{z}\, x_c\, (z - x_a^2 - x_b^2) \, \lambda^{\frac{1}{2}}(1, z, x_c^2) \, \lambda^{\frac{1}{2}}(1, x_a^2, x_b^2) \, , \\
	G_3(x_a, x_b, x_c) & = -\int_{(x_a + x_b)^2}^{(1 - x_c)^2} \frac{dz}{z}\, x_a \, x_b \, (1 + x_c^2 - z) \, \lambda^{\frac{1}{2}}(1, z, x_c^2) \, \lambda^{\frac{1}{2}}(z, x_a^2, x_b^2) \, .
\end{align}
The weak couplings are given by 
\begin{align}
	& g_{L,\nu} = 1, \, g_{R,\nu} = 0, \,
	g_{L,\ell} = 2s_w^2 - 1, \, g_{R,\ell} = 2s_w^2, \, 
	g_{L,d} = \frac{2}{3} s_{w}^{2} -1, \, g_{R,d} = \frac{2}{3} s_{w}^{2} , \, g_{L,u} = 1- \frac{4}{3} s_w^2, \, g_{R,u} = - \frac{4}{3} s_w^2 \, .
\end{align}
The Yukawa couplings are defined as
\begin{align}
	Y_{\nu} &= \frac{\sqrt{2}\, V_{\mu N}\, M_N}{v}, \qquad
	Y_{\ell} = \frac{\sqrt{2}\, M_{\ell}}{v}, \qquad
	Y_{d_{\beta}} = \frac{\sqrt{2}\, M_{d_{\beta}}}{v}, \qquad
	Y_{u_{\alpha}} = \frac{\sqrt{2}\, M_{u_{\alpha}}}{v}.
\end{align}
The coefficients $C_{t2}$ and $C_{t3}$ are given by
\begin{align}
	C_{t2} &= \frac{g^2}{2 M_W^2}, 
	\qquad
	C_{t3} = \frac{g^2 V_{\mu N}}{4 M_W^2}.
\end{align}
Here, $M_W$, $M_H$, and $M_Z$ denote the masses of the $W$ boson, Higgs boson, and $Z$ boson, respectively;  $M_{u_{\alpha}}$ and $M_{d_{\beta}}$ correspond to the masses of the up-type quark of generation $\alpha$ and the down-type quark of generation $\beta$, respectively;  
$M_{\ell}$ and $M_{\nu_\ell}$ denote the charged lepton and SM neutrino masses; $M_{\mu}$ and $M_{\nu_\mu}$ represent the muon and muon-neutrino masses. $N_c$ denotes the color factor.

\bibliographystyle{utphys}  
\bibliography{bibitem}  

@article{Minkowski:1977sc,
    author = "Minkowski, Peter",
    title = "{$\mu \to e\gamma$ at a Rate of One Out of $10^{9}$ Muon Decays?}",
    reportNumber = "Print-77-0182 (BERN)",
    doi = "10.1016/0370-2693(77)90435-X",
    journal = "Phys. Lett. B",
    volume = "67",
    pages = "421--428",
    year = "1977"
}

@article{Cowan:2010js,
    author = "Cowan, Glen and Cranmer, Kyle and Gross, Eilam and Vitells, Ofer",
    title = "{Asymptotic formulae for likelihood-based tests of new physics}",
    eprint = "1007.1727",
    archivePrefix = "arXiv",
    primaryClass = "physics.data-an",
    doi = "10.1140/epjc/s10052-011-1554-0",
    journal = "Eur. Phys. J. C",
    volume = "71",
    pages = "1554",
    year = "2011",
    note = "[Erratum: Eur.Phys.J.C 73, 2501 (2013)]"
}

@article{Desai:2023jxh,
    author = "Desai, Nishita and Sengupta, Amartya",
    title = "{Status of leptoquark models after LHC Run-2 and discovery prospects at future colliders}",
    eprint = "2301.01754",
    archivePrefix = "arXiv",
    primaryClass = "hep-ph",
    month = "1",
    year = "2023"
}

@article{Varzielas:2023qlb,
    author = "Varzielas, Ivo de Medeiros and Sengupta, Amartya",
    title = "{Constraining flavoured leptoquarks with LHC and LFV}",
    eprint = "2301.04119",
    archivePrefix = "arXiv",
    primaryClass = "hep-ph",
    doi = "10.1016/j.nuclphysb.2024.116495",
    journal = "Nucl. Phys. B",
    volume = "1001",
    pages = "116495",
    year = "2024"
}

@article{Padhan:2019dcp,
    author = "Padhan, Rojalin and Mandal, Sanjoy and Mitra, Manimala and Sinha, Nita",
    title = "{Signatures of $\tilde{R}_2$ class of Leptoquarks at the upcoming $ep$ colliders}",
    eprint = "1912.07236",
    archivePrefix = "arXiv",
    primaryClass = "hep-ph",
    reportNumber = "IP/BBSR/2019-11",
    doi = "10.1103/PhysRevD.101.075037",
    journal = "Phys. Rev. D",
    volume = "101",
    number = "7",
    pages = "075037",
    year = "2020"
}

@article{Georgi:1974sy,
    author = "Georgi, H. and Glashow, S.L.",
    title = "{Unity of All Elementary Particle Forces}",
    doi = "10.1103/PhysRevLett.32.438",
    journal = "Phys. Rev. Lett.",
    volume = "32",
    pages = "438--441",
    year = "1974"
}

@article{Fritzsch:1974nn,
    author = "Fritzsch, Harald and Minkowski, Peter",
    title = "{Unified Interactions of Leptons and Hadrons}",
    reportNumber = "CALT-68-467",
    doi = "10.1016/0003-4916(75)90211-0",
    journal = "Annals Phys.",
    volume = "93",
    pages = "193--266",
    year = "1975"
}

@article{Barbier:2004ez,
    author = "Barbier, R. and others",
    title = "{R-parity violating supersymmetry}",
    eprint = "hep-ph/0406039",
    archivePrefix = "arXiv",
    doi = "10.1016/j.physrep.2005.08.006",
    journal = "Phys. Rept.",
    volume = "420",
    pages = "1--202",
    year = "2005"
}

@article{Cai:2017jrq,
    author = "Cai, Yi and Herrero-Garc{\'\i}a, Juan and Schmidt, Michael A. and Vicente, Avelino and Volkas, Raymond R.",
    title = "{From the trees to the forest: a review of radiative neutrino mass models}",
    eprint = "1706.08524",
    archivePrefix = "arXiv",
    primaryClass = "hep-ph",
    reportNumber = "ADP-17-29-T1035",
    doi = "10.3389/fphy.2017.00063",
    journal = "Front. in Phys.",
    volume = "5",
    pages = "63",
    year = "2017"
}

@article{Mandal:2018qpg,
    author = "Mandal, Sanjoy and Mitra, Manimala and Sinha, Nita",
    title = "{Probing leptoquarks and heavy neutrinos at the LHeC}",
    eprint = "1807.06455",
    archivePrefix = "arXiv",
    primaryClass = "hep-ph",
    reportNumber = "IP/BBSR/2018-9",
    doi = "10.1103/PhysRevD.98.095004",
    journal = "Phys. Rev. D",
    volume = "98",
    number = "9",
    pages = "095004",
    year = "2018"
}

@article{Antusch:2014woa,
    author = "Antusch, Stefan and Fischer, Oliver",
    title = "{Non-unitarity of the leptonic mixing matrix: Present bounds and future sensitivities}",
    eprint = "1407.6607",
    archivePrefix = "arXiv",
    primaryClass = "hep-ph",
    reportNumber = "MPP-2014-313",
    doi = "10.1007/JHEP10(2014)094",
    journal = "JHEP",
    volume = "10",
    pages = "094",
    year = "2014"
}

@article{Dorsner:2016wpm,
    author = "Dor\v{s}ner, I. and Fajfer, S. and Greljo, A. and Kamenik, J.F. and Ko\v{s}nik, N.",
    title = "{Physics of leptoquarks in precision experiments and at particle colliders}",
    eprint = "1603.04993",
    archivePrefix = "arXiv",
    primaryClass = "hep-ph",
    doi = "10.1016/j.physrep.2016.06.001",
    journal = "Phys. Rept.",
    volume = "641",
    pages = "1--68",
    year = "2016"
}

@article{Buchmuller:1986zs,
    author = "Buchmuller, W. and Ruckl, R. and Wyler, D.",
    title = "{Leptoquarks in Lepton - Quark Collisions}",
    reportNumber = "DESY-86-150, ITP-UH-14-86",
    doi = "10.1016/0370-2693(87)90637-X",
    journal = "Phys. Lett. B",
    volume = "191",
    pages = "442--448",
    year = "1987",
    note = "[Erratum: Phys.Lett.B 448, 320--320 (1999)]"
}

@article{Duraikandan:2024kcy,
    author = "Duraikandan, Gokul and Khanna, Rishabh and Mandal, Tanumoy and Mitra, Subhadip and Sharma, Rachit",
    title = "{Right-handed neutrino production through first-generation leptoquarks}",
    eprint = "2412.19751",
    archivePrefix = "arXiv",
    primaryClass = "hep-ph",
    doi = "10.1103/PhysRevD.111.075032",
    journal = "Phys. Rev. D",
    volume = "111",
    number = "7",
    pages = "075032",
    year = "2025"
}

@article{CMS:2024bej,
    collaboration = "CMS",
    title = "{Search for $t$-channel scalar and vector leptoquark exchange in the high mass dimuon and dielectron spectrum in proton-proton collisions at $\sqrt{s} = 13~\mathrm{TeV}$}",
    reportNumber = "CMS-PAS-EXO-22-013",
    year = "2024",
    note = "\href{https://inspirehep.net/files/8f2b81ae49b033ab8794c9f5c22eb54a}{\bf CMS-PAS-EXO-22-013}."}

@article{ATLAS:2020dsk,
    author = "Aad, Georges and others",
    collaboration = "ATLAS",
    title = "{Search for pairs of scalar leptoquarks decaying into quarks and electrons or muons in $ \sqrt{s} $ = 13 TeV $pp$ collisions with the ATLAS detector}",
    eprint = "2006.05872",
    archivePrefix = "arXiv",
    primaryClass = "hep-ex",
    reportNumber = "CERN-EP-2020-084",
    doi = "10.1007/JHEP10(2020)112",
    journal = "JHEP",
    volume = "10",
    pages = "112",
    year = "2020"
}

@article{Babu:2019mfe,
    author = "Babu, K.S. and Dev, P. S. Bhupal and Jana, Sudip and Thapa, Anil",
    title = "{Non-Standard Interactions in Radiative Neutrino Mass Models}",
    eprint = "1907.09498",
    archivePrefix = "arXiv",
    primaryClass = "hep-ph",
    reportNumber = "FERMILAB-PUB-19-304-T, OSU-HEP-19-04",
    doi = "10.1007/JHEP03(2020)006",
    journal = "JHEP",
    volume = "03",
    pages = "006",
    year = "2020"
}

@article{Crivellin:2023zui,
    author = "Crivellin, Andreas and Mellado, Bruce",
    title = "{Anomalies in particle physics and their implications for physics beyond the standard model}",
    eprint = "2309.03870",
    archivePrefix = "arXiv",
    primaryClass = "hep-ph",
    reportNumber = "PSI-PR-23-34, ZU-TH 53/23, ICPP-73",
    doi = "10.1038/s42254-024-00703-6",
    journal = "Nature Rev. Phys.",
    volume = "6",
    number = "5",
    pages = "294--309",
    year = "2024"
}

@article{Fischer:2021sqw,
    author = "Fischer, Oliver and others",
    title = "{Unveiling hidden physics at the LHC}",
    eprint = "2109.06065",
    archivePrefix = "arXiv",
    primaryClass = "hep-ph",
    doi = "10.1140/epjc/s10052-022-10541-4",
    journal = "Eur. Phys. J. C",
    volume = "82",
    number = "8",
    pages = "665",
    year = "2022"
}

@article{Alwall:2014hca,
      author         = "Alwall, J. and Frederix, R. and Frixione, S. and Hirschi,
                        V. and Maltoni, F. and Mattelaer, O. and Shao, H. -S. and
                        Stelzer, T. and Torrielli, P. and Zaro, M.",
      title          = "{The automated computation of tree-level and
                        next-to-leading order differential cross sections, and
                        their matching to parton shower simulations}",
      journal        = "JHEP",
      volume         = "07",
      year           = "2014",
      pages          = "079",
      doi            = "10.1007/JHEP07(2014)079",
      eprint         = "1405.0301",
      archivePrefix  = "arXiv",
      primaryClass   = "hep-ph",
      reportNumber   = "CERN-PH-TH-2014-064, CP3-14-18, LPN14-066, MCNET-14-09,
                        ZU-TH-14-14",
      SLACcitation   = "%%CITATION = ARXIV:1405.0301;%%"
}

@article{Alloul:2013bka,
    author = "Alloul, Adam and Christensen, Neil D. and Degrande, C\'eline and Duhr, Claude and Fuks, Benjamin",
    title = "{FeynRules  2.0 - A complete toolbox for tree-level phenomenology}",
    eprint = "1310.1921",
    archivePrefix = "arXiv",
    primaryClass = "hep-ph",
    reportNumber = "CERN-PH-TH-2013-239, MCNET-13-14, IPPP-13-71, DCPT-13-142, PITT-PACC-1308",
    doi = "10.1016/j.cpc.2014.04.012",
    journal = "Comput. Phys. Commun.",
    volume = "185",
    pages = "2250--2300",
    year = "2014"
}

@article{Sjostrand:2006za,
    author = "Sjostrand, Torbjorn and Mrenna, Stephen and Skands, Peter Z.",
    title = "{PYTHIA 6.4 Physics and Manual}",
    eprint = "hep-ph/0603175",
    archivePrefix = "arXiv",
    reportNumber = "FERMILAB-PUB-06-052-CD-T, LU-TP-06-13",
    doi = "10.1088/1126-6708/2006/05/026",
    journal = "JHEP",
    volume = "05",
    pages = "026",
    year = "2006"
}

@article{deFavereau:2013fsa,
    author = "de Favereau, J. and Delaere, C. and Demin, P. and Giammanco, A. and Lema\^\i{}tre, V. and Mertens, A. and Selvaggi, M.",
    collaboration = "DELPHES 3",
    title = "{DELPHES 3, A modular framework for fast simulation of a generic collider experiment}",
    eprint = "1307.6346",
    archivePrefix = "arXiv",
    primaryClass = "hep-ex",
    doi = "10.1007/JHEP02(2014)057",
    journal = "JHEP",
    volume = "02",
    pages = "057",
    year = "2014"
}

@inproceedings{Soyez:2008pq,
    author = "Soyez, Gregory",
    title = "{The SISCone and anti-k(t) jet algorithms}",
    booktitle = "{16th International Workshop on Deep Inelastic Scattering and Related Subjects}",
    eprint = "0807.0021",
    archivePrefix = "arXiv",
    primaryClass = "hep-ph",
    doi = "10.3360/dis.2008.178",
    pages = "178",
    month = "7",
    year = "2008"
}

@article{Cacciari:2011ma,
    author = "Cacciari, Matteo and Salam, Gavin P. and Soyez, Gregory",
    title = "{FastJet User Manual}",
    eprint = "1111.6097",
    archivePrefix = "arXiv",
    primaryClass = "hep-ph",
    reportNumber = "CERN-PH-TH-2011-297",
    doi = "10.1140/epjc/s10052-012-1896-2",
    journal = "Eur. Phys. J. C",
    volume = "72",
    pages = "1896",
    year = "2012"
}

@article{Gell-Mann:1979vob,
    author = "Gell-Mann, Murray and Ramond, Pierre and Slansky, Richard",
    title = "{Complex Spinors and Unified Theories}",
    eprint = "1306.4669",
    archivePrefix = "arXiv",
    primaryClass = "hep-th",
    reportNumber = "PRINT-80-0576",
    journal = "Conf. Proc. C",
    volume = "790927",
    pages = "315--321",
    year = "1979"
}

@article{Mohapatra:1979ia,
    author = "Mohapatra, Rabindra N. and Senjanovic, Goran",
    title = "{Neutrino Mass and Spontaneous Parity Nonconservation}",
    reportNumber = "MDDP-TR-80-060, MDDP-PP-80-105, CCNY-HEP-79-10",
    doi = "10.1103/PhysRevLett.44.912",
    journal = "Phys. Rev. Lett.",
    volume = "44",
    pages = "912",
    year = "1980"
}

@article{Yanagida:1979as,
    author = "Yanagida, Tsutomu",
    editor = "Sawada, Osamu and Sugamoto, Akio",
    title = "{Horizontal gauge symmetry and masses of neutrinos}",
    reportNumber = "KEK-79-18-95",
    journal = "Conf. Proc. C",
    volume = "7902131",
    pages = "95--99",
    year = "1979"
}

@article{Li:2023tbx,
    author = "Li, Peiran and Liu, Zhen and Lyu, Kun-Feng",
    title = "{Heavy neutral leptons at muon colliders}",
    eprint = "2301.07117",
    archivePrefix = "arXiv",
    primaryClass = "hep-ph",
    doi = "10.1007/JHEP03(2023)231",
    journal = "JHEP",
    volume = "03",
    pages = "231",
    year = "2023"
}

@article{Kwok:2023dck,
    author = "Kwok, Tsz Hong and Li, Lingfeng and Liu, Tao and Rock, Ariel",
    title = "{Searching for heavy neutral leptons at a future muon collider}",
    eprint = "2301.05177",
    archivePrefix = "arXiv",
    primaryClass = "hep-ph",
    doi = "10.1103/PhysRevD.110.075009",
    journal = "Phys. Rev. D",
    volume = "110",
    number = "7",
    pages = "075009",
    year = "2024"
}

@article{Mekala:2023diu,
    author = {Mekala, Krzysztof and Reuter, Jurgen and Zarnecki, Aleksander Filip},
    title = "{Optimal search reach for heavy neutral leptons at a muon collider}",
    eprint = "2301.02602",
    archivePrefix = "arXiv",
    primaryClass = "hep-ph",
    doi = "10.1016/j.physletb.2023.137945",
    journal = "Phys. Lett. B",
    volume = "841",
    pages = "137945",
    year = "2023"
}

@article{Bessaa:2014jya,
    author = "Bessaa, Assia and Davidson, Sacha",
    title = "{Constraints on $t$ -channel leptoquark exchange from LHC contact interaction searches}",
    eprint = "1409.2372",
    archivePrefix = "arXiv",
    primaryClass = "hep-ph",
    doi = "10.1140/epjc/s10052-015-3313-0",
    journal = "Eur. Phys. J. C",
    volume = "75",
    number = "2",
    pages = "97",
    year = "2015"
}

@article{Babu:2020hun,
    author = "Babu, K. S. and Dev, P. S. Bhupal and Jana, Sudip and Thapa, Anil",
    title = "{Unified framework for $B$-anomalies, muon $g-2$ and neutrino masses}",
    eprint = "2009.01771",
    archivePrefix = "arXiv",
    primaryClass = "hep-ph",
    reportNumber = "OSU-HEP-20-12",
    doi = "10.1007/JHEP03(2021)179",
    journal = "JHEP",
    volume = "03",
    pages = "179",
    year = "2021"
}

@article{ATLAS:2020yat,
    author = "Aad, Georges and others",
    collaboration = "ATLAS",
    title = "{Search for new non-resonant phenomena in high-mass dilepton final states with the ATLAS detector}",
    eprint = "2006.12946",
    archivePrefix = "arXiv",
    primaryClass = "hep-ex",
    reportNumber = "CERN-EP-2020-066",
    doi = "10.1007/JHEP11(2020)005",
    journal = "JHEP",
    volume = "11",
    pages = "005",
    year = "2020",
    note = "[Erratum: JHEP 04, 142 (2021)]"
}

@article{CMS:2023qdw,
    author = "Hayrapetyan, Aram and others",
    collaboration = "CMS",
    title = "{Search for a third-generation leptoquark coupled to a {\ensuremath{\tau}} lepton and a b quark through single, pair, and nonresonant production in proton-proton collisions at $ \sqrt{s} $ = 13 TeV}",
    eprint = "2308.07826",
    archivePrefix = "arXiv",
    primaryClass = "hep-ex",
    reportNumber = "CMS-EXO-19-016, CERN-EP-2023-144",
    doi = "10.1007/JHEP05(2024)311",
    journal = "JHEP",
    volume = "05",
    pages = "311",
    year = "2024"
}

@article{Pati:1974yy,
    author = "Pati, Jogesh C. and Salam, Abdus",
    title = "{Lepton Number as the Fourth Color}",
    reportNumber = "IC-74-7",
    doi = "10.1103/PhysRevD.10.275",
    journal = "Phys. Rev. D",
    volume = "10",
    pages = "275--289",
    year = "1974",
    note = "[Erratum: Phys.Rev.D 11, 703--703 (1975)]"
}

@article{Weinberg:1975gm,
    author = "Weinberg, Steven",
    title = "{Implications of Dynamical Symmetry Breaking}",
    reportNumber = "PRINT-75-0804 (HARVARD)",
    doi = "10.1103/PhysRevD.19.1277",
    journal = "Phys. Rev. D",
    volume = "13",
    pages = "974--996",
    year = "1976",
    note = "[Addendum: Phys.Rev.D 19, 1277--1280 (1979)]"
}

@article{Susskind:1978ms,
    author = "Susskind, Leonard",
    title = "{Dynamics of Spontaneous Symmetry Breaking in the Weinberg-Salam Theory}",
    reportNumber = "SLAC-PUB-2142",
    doi = "10.1103/PhysRevD.20.2619",
    journal = "Phys. Rev. D",
    volume = "20",
    pages = "2619--2625",
    year = "1979"
}

@article{Mandal:2018kau,
    author = "Mandal, Tanumoy and Mitra, Subhadip and Raz, Swapnil",
    title = "{$R_{D^{(*)}}$ motivated $\mathcal{S}_1$ leptoquark scenarios: Impact of interference on the exclusion limits from LHC data}",
    eprint = "1811.03561",
    archivePrefix = "arXiv",
    primaryClass = "hep-ph",
    doi = "10.1103/PhysRevD.99.055028",
    journal = "Phys. Rev. D",
    volume = "99",
    number = "5",
    pages = "055028",
    year = "2019"
}

@article{Bhaskar:2021pml,
    author = "Bhaskar, Arvind and Das, Diganta and Mandal, Tanumoy and Mitra, Subhadip and Neeraj, Cyrin",
    title = "{Precise limits on the charge-2/3 U1 vector leptoquark}",
    eprint = "2101.12069",
    archivePrefix = "arXiv",
    primaryClass = "hep-ph",
    doi = "10.1103/PhysRevD.104.035016",
    journal = "Phys. Rev. D",
    volume = "104",
    number = "3",
    pages = "035016",
    year = "2021"
}

@article{Choi:2018stw,
    author = "Choi, Soo-Min and Kang, Yoo-Jin and Lee, Hyun Min and Ro, Tae-Gyu",
    title = "{Lepto-Quark Portal Dark Matter}",
    eprint = "1807.06547",
    archivePrefix = "arXiv",
    primaryClass = "hep-ph",
    doi = "10.1007/JHEP10(2018)104",
    journal = "JHEP",
    volume = "10",
    pages = "104",
    year = "2018"
}

@article{Bandyopadhyay:2016oif,
    author = "Bandyopadhyay, Priyotosh and Mandal, Rusa",
    title = "{Vacuum stability in an extended standard model with a leptoquark}",
    eprint = "1609.03561",
    archivePrefix = "arXiv",
    primaryClass = "hep-ph",
    reportNumber = "IITH-PH-0001-16, IMSC-2016-09-01",
    doi = "10.1103/PhysRevD.95.035007",
    journal = "Phys. Rev. D",
    volume = "95",
    number = "3",
    pages = "035007",
    year = "2017"
}

@article{Das:2017kkm,
    author = "Das, Debottam and Ghosh, Kirtiman and Mitra, Manimala and Mondal, Subhadeep",
    title = "{Probing sterile neutrinos in the framework of inverse seesaw mechanism through leptoquark productions}",
    eprint = "1708.06206",
    archivePrefix = "arXiv",
    primaryClass = "hep-ph",
    reportNumber = "HRI-RECAPP-2017-011, IP-BBSR-2017-10",
    doi = "10.1103/PhysRevD.97.015024",
    journal = "Phys. Rev. D",
    volume = "97",
    number = "1",
    pages = "015024",
    year = "2018"
}

@article{Bhaskar:2020kdr,
    author = "Bhaskar, Arvind and Das, Debottam and De, Bibhabasu and Mitra, Subhadip",
    title = "{Enhancing scalar productions with leptoquarks at the LHC}",
    eprint = "2002.12571",
    archivePrefix = "arXiv",
    primaryClass = "hep-ph",
    reportNumber = "IP-BBSR/2020-2",
    doi = "10.1103/PhysRevD.102.035002",
    journal = "Phys. Rev. D",
    volume = "102",
    number = "3",
    pages = "035002",
    year = "2020"
}

@article{Cottin:2021tfo,
    author = "Cottin, Giovanna and Fischer, Oliver and Mandal, Sanjoy and Mitra, Manimala and Padhan, Rojalin",
    title = "{Displaced neutrino jets at the LHeC}",
    eprint = "2104.13578",
    archivePrefix = "arXiv",
    primaryClass = "hep-ph",
    doi = "10.1007/JHEP06(2022)168",
    journal = "JHEP",
    volume = "06",
    pages = "168",
    year = "2022"
}

@article{AlAli:2021let,
    author = "Al Ali, Hind and others",
    title = "{The muon Smasher{\textquoteright}s guide}",
    eprint = "2103.14043",
    archivePrefix = "arXiv",
    primaryClass = "hep-ph",
    doi = "10.1088/1361-6633/ac6678",
    journal = "Rept. Prog. Phys.",
    volume = "85",
    number = "8",
    pages = "084201",
    year = "2022"
}

@article{InternationalMuonCollider:2025sys,
    author = "Accettura, Carlotta and others",
    collaboration = "International Muon Collider",
    title = "{The Muon Collider}",
    eprint = "2504.21417",
    archivePrefix = "arXiv",
    primaryClass = "physics.acc-ph",
    reportNumber = "FERMILAB-PUB-25-0309-AD-PPD-T",
    month = "4",
    year = "2025"
}

@article{Han:2025wdy,
    author = "Han, Tao and Low, Matthew and Wu, Tong Arthur and Xie, Keping",
    title = "{Colorful particle production at high-energy muon colliders}",
    eprint = "2502.20443",
    archivePrefix = "arXiv",
    primaryClass = "hep-ph",
    doi = "10.1007/JHEP06(2025)109",
    journal = "JHEP",
    volume = "06",
    pages = "109",
    year = "2025"
}

@article{Ghosh:2023xbj,
    author = "Ghosh, Nivedita and Rai, Santosh Kumar and Samui, Tousik",
    title = "{Search for a leptoquark and vector-like lepton in a muon collider}",
    eprint = "2309.07583",
    archivePrefix = "arXiv",
    primaryClass = "hep-ph",
    doi = "10.1016/j.nuclphysb.2024.116564",
    journal = "Nucl. Phys. B",
    volume = "1004",
    pages = "116564",
    year = "2024"
}

@article{Bandyopadhyay:2021pld,
    author = "Bandyopadhyay, Priyotosh and Karan, Anirban and Mandal, Rusa and Parashar, Snehashis",
    title = "{Distinguishing signatures of scalar leptoquarks at hadron and muon colliders}",
    eprint = "2108.06506",
    archivePrefix = "arXiv",
    primaryClass = "hep-ph",
    reportNumber = "IITH-PH-0002/21, SI-HEP-2021-22",
    doi = "10.1140/epjc/s10052-022-10809-9",
    journal = "Eur. Phys. J. C",
    volume = "82",
    number = "10",
    pages = "916",
    year = "2022"
}

@article{ATLAS:2022wcu,
    author = "Aad, Georges and others",
    collaboration = "ATLAS",
    title = "{Search for pair-produced scalar and vector leptoquarks decaying into third-generation quarks and first- or second-generation leptons in pp collisions with the ATLAS detector}",
    eprint = "2210.04517",
    archivePrefix = "arXiv",
    primaryClass = "hep-ex",
    reportNumber = "CERN-EP-2022-145",
    doi = "10.1007/JHEP06(2023)188",
    journal = "JHEP",
    volume = "2306",
    pages = "188",
    year = "2023"
}

@article{Bhaskar:2023xkm,
    author = "Bhaskar, Arvind and Chaurasia, Yash and Deka, Kuldeep and Mandal, Tanumoy and Mitra, Subhadip and Mukherjee, Ananya",
    title = "{Right-handed neutrino pair production via second-generation leptoquarks}",
    eprint = "2301.11889",
    archivePrefix = "arXiv",
    primaryClass = "hep-ph",
    doi = "10.1016/j.physletb.2023.138039",
    journal = "Phys. Lett. B",
    volume = "843",
    pages = "138039",
    year = "2023"
}

@article{Angelescu:2021lln,
    author = "Angelescu, Andrei and Be{\v{c}}irevi{\'c}, Damir and Faroughy, Darius A. and Jaffredo, Florentin and Sumensari, Olcyr",
    title = "{Single leptoquark solutions to the B-physics anomalies}",
    eprint = "2103.12504",
    archivePrefix = "arXiv",
    primaryClass = "hep-ph",
    reportNumber = "ZU-TH 12/21",
    doi = "10.1103/PhysRevD.104.055017",
    journal = "Phys. Rev. D",
    volume = "104",
    number = "5",
    pages = "055017",
    year = "2021"
}

@article{ATLAS:2023afl,
    author = "Aad, Georges and others",
    collaboration = "ATLAS",
    title = "{Search for pair production of squarks or gluinos decaying via sleptons or weak bosons in final states with two same-sign or three leptons with the ATLAS detector}",
    eprint = "2307.01094",
    archivePrefix = "arXiv",
    primaryClass = "hep-ex",
    reportNumber = "CERN-EP-2023-123",
    doi = "10.1007/JHEP02(2024)107",
    journal = "JHEP",
    volume = "02",
    pages = "107",
    year = "2024"
}

@article{CMS:2018lab,
    author = "Sirunyan, Albert M and others",
    collaboration = "CMS",
    title = "{Search for pair production of second-generation leptoquarks at $\sqrt{s}=$ 13 TeV}",
    eprint = "1808.05082",
    archivePrefix = "arXiv",
    primaryClass = "hep-ex",
    reportNumber = "CMS-EXO-17-003, CERN-EP-2018-218",
    doi = "10.1103/PhysRevD.99.032014",
    journal = "Phys. Rev. D",
    volume = "99",
    number = "3",
    pages = "032014",
    year = "2019"
}

\end{document}